\documentclass[11pt]{article}

\usepackage[utf8]{inputenc}
\usepackage{latexsym}
\usepackage{amsmath}
\usepackage{amssymb}
\usepackage{amsthm}
\usepackage{hyperref}
\usepackage{cite}
\usepackage{graphicx}
\usepackage{color}
\usepackage{enumitem}
\usepackage{comment}
\usepackage{bm}
\usepackage[left=1in,right=1in,top=1in,bottom=1in]{geometry}
\usepackage{xspace}
\setlist[description]{font=\normalfont\itshape\textbullet\space}
\usepackage[all]{xy}
\usepackage{doi, url}
\usepackage{arydshln}
\usepackage{xcolor} 
\usepackage{framed}
\usepackage{physics}
\usepackage{tikz}
\usepackage{tikz-cd}
\usepackage{mathdots}
\usepackage{yhmath}
\usepackage{cancel}
\usepackage{siunitx}
\usepackage{array}
\usepackage{multirow}
\usepackage{gensymb}
\usepackage{tabularx}
\usepackage{extarrows}
\usepackage{booktabs}
\usetikzlibrary{fadings}
\usetikzlibrary{patterns}
\usetikzlibrary{shadows.blur}
\usetikzlibrary{shapes}

\renewcommand{\paragraph}[1]{\vspace{6pt} \noindent \textbf{#1}\xspace}

\theoremstyle{plain}
\newtheorem{theorem}{Theorem}[section]

\newtheorem{corollary}[theorem]{Corollary}
\newtheorem{lemma}[theorem]{Lemma}

\newtheorem{claim}[theorem]{Claim}
\newtheorem{fact}[theorem]{Fact}

\theoremstyle{definition}
\newtheorem{problem}[theorem]{Problem}

\newtheorem{definition}[theorem]{Definition}

\newcommand{\GL}{\mathrm{GL}}
\newcommand{\F}{\mathbb{F}}
\newcommand{\Z}{\mathbb{Z}}
\newcommand{\Q}{\mathbb{Q}}

\newcommand{\poly}{\mathrm{poly}}

\newcommand{\M}{\mathrm{M}}

\newcommand{\Hom}{\mathrm{Hom}}

\newcommand{\End}{\mathrm{End}}

\DeclareMathOperator{\Rad}{Rad}

\DeclareMathOperator{\diag}{diag}

\renewcommand{\le}{\leqslant}
\renewcommand{\ge}{\geqslant}

\newcommand{\too}%
{\xrightarrow{\text{\raisebox{-3pt}{$\sim$}}\,}}

\mathchardef\mhyphen="2D

\usepackage[T1]{fontenc}
\def\DJ{{\hbox{D\kern-.8em\raise.15ex\hbox{--}\kern.35em}}}

\newcommand{\ord}{\mathrm{ord}}

\title{Faster Isomorphism for Groups of Class 2}

\author{Xiaorui Sun\footnote{Computer Science Department, University of Illinois at Chicago. {\tt xiaorui@uic.edu}}}

\date{}
\begin{document}

\maketitle


\thispagestyle{empty}

\thispagestyle{empty}

\begin{abstract}
The finite-group isomorphism problem asks whether two groups, given by their Cayley tables, are isomorphic. Despite decades of study, obtaining an $N^{o(\log N)}$-time algorithm for general groups of order $N$ remains a major open challenge.

Groups of nilpotency class $2$ have long been a central obstacle to achieving this goal. Recent breakthroughs have established $N^{o(\log N)}$-time algorithms for $p$-groups of class $2$ with odd $p$ (Sun, STOC '23; Ivanyos–Mendoza–Qiao–Sun–Zhang, FOCS '24; Grochow–Qiao–Stange–Sun, STOC '25) and for groups of Frattini class $2$ (Grochow–Ivanyos–Qiao–Sun, FOCS '26). However, the problem remained open for general groups of class $2$.

In this paper, we present an algorithm for testing isomorphism of groups of class $2$, with running time $N^{O((\log N)^{25/26} \log \log N)}$. 
Our algorithm combines a tensor-based restriction of the search space, an encoding of group relations by alternating matrix tuples, and an orbit computation to recover the remaining central power information.
\end{abstract}

\newpage

\thispagestyle{empty}
\tableofcontents
\newpage

\setcounter{page}{1}

\section{Introduction}
The \emph{group isomorphism problem} asks whether two finite groups, given by their Cayley tables, are isomorphic.
It is a well-studied problem in NP that is neither known to admit a polynomial-time algorithm nor known to be NP-complete~\cite{garey2002computers}.
The problem is closely connected to computational group theory, algebraic complexity theory, and cryptography~\cite{patarin1996hidden, petrank1997code, holt2005handbook, TI1, duong2025blind, blaserdigital, NQT}. 
Furthermore, the reduction from group isomorphism to graph isomorphism, together with Babai's quasipolynomial-time graph isomorphism algorithm~\cite{babai2016graph,babai2019canonical}, suggests that group isomorphism has become a bottleneck for developing a sub-quasipolynomial time algorithm for graph isomorphism.


The group isomorphism problem has been extensively studied since the 1970s~\cite{FN, Mil78, savage1980n2}. 
The classical $N^{\log N+O(1)}$-time algorithm is commonly attributed to Tarjan~\cite{Mil78}. 
The best-known algorithm for the general group isomorphism problem runs in time $N^{(1/4+o(1))\log N}$~\cite{rosenbaum2013bidirectional}. 
Despite decades of research, a substantial gap remains between this upper bound and the long-standing goal of achieving an $N^{o(\log N)}$-time algorithm.

Groups of (nilpotency) class $2$ have long been recognized as a major obstacle to faster group isomorphism algorithms~\cite{BCGQ11, LW12,brooksbank2012computing, rosenbaum2013bidirectional, brooksbank2015fast, LQ17, BGLQW}.
Groups of class $2$ have a lower central series of length $2$, so all their commutators lie in the center. Consequently, both $Z(G)$ and $G/Z(G)$ are abelian.
Despite this strong structural property, isomorphism testing for these groups still has a worst-case running time of the same general order as that for arbitrary groups.
This persistent difficulty has made groups of class $2$ a central focus of efforts to develop faster group isomorphism algorithms.

Significant progress has been made for special cases of groups of class $2$.
For $p$-groups of class $2$ and exponent $p$ with odd $p$, the first $N^{o(\log N)}$-time algorithm was obtained in~\cite{Sun23}, and the running time was subsequently improved to $N^{\widetilde O((\log N)^{1/2})}$ in~\cite{IMQSZ}.
This approach was later extended to all $p$-groups of class $2$ for odd primes~\cite{GQSS25}.
More recently, an $N^{O((\log N)^{1/2})}$-time algorithm was developed for $2$-groups of Frattini class $2$~\cite{grochow2026beyond}.
Although these breakthroughs demonstrate the potential for faster algorithms, obtaining an $N^{o(\log N)}$-time algorithm for general $2$-groups of class $2$ remained open.

Our main result resolves this problem by removing the restriction to odd primes and giving an algorithm for arbitrary finite $p$-groups of class $2$.

\begin{theorem}\label{thm:main_intro}
For every prime $p$, isomorphism of two $p$-groups of class $2$ of order $N$ can be decided in time $N^{O((\log N)^{25/26} \log \log N)}$.
\end{theorem}

Every finite group of class $2$ decomposes as a direct product of its Sylow $p$-subgroups, and this decomposition can be computed efficiently~\cite{KN, wilson2009finding, WilsonProductsII}.
Applying the theorem to the corresponding Sylow subgroups therefore yields the following corollary.

\begin{corollary}\label{cor:main_intro}
Isomorphism of two finite nilpotent groups of class $2$ and order $N$ can be decided in time $N^{O((\log N)^{25/26} \log \log N)}$.
\end{corollary}

\subsection{The significance of $2$-groups of class 2}

\paragraph{Finite class 2 groups.} A group is called nilpotent if its lower central series, obtained by repeatedly taking commutators, eventually becomes trivial. The nilpotency class measures the length of this series. In particular, a group has class at most $2$ if and only if $[G,G]\leq Z(G)$, meaning that all commutators are central.

Among nilpotent groups, $p$-groups of class $2$ are the simplest genuinely nonabelian cases, yet they already exhibit a wide range of commutator and power structures.
Their importance is also reflected in the enumeration of finite groups.
Pyber~\cite{pyber1993enumerating} showed that $p$-groups account for logarithmically almost all finite groups, in the sense that 
\[\lim_{N \to \infty} \frac{\log(\text{\# $p$-groups of order } \leq N)}{\log(\text{\# all groups of order} \leq N)} = 1\]
Recent results further establish that, as $N\to\infty$, 
almost all finite groups of order at most $N$, counted up to isomorphism, are $2$-groups of class 2~\cite{li2026almost, su2026almost}.
These results underscore the importance of understanding general $2$-groups of class $2$, both algorithmically and structurally.

\paragraph{Difficulty for $p = 2$.} 
The case $p=2$ presents several distinct challenges that have slowed progress, despite substantial advances for odd primes and groups of Frattini class $2$~\cite{Sun23,IMQSZ,GQSS25,grochow2026beyond}.

An important technique underlying previous progress is tensor isomorphism, which allows the commutator structures of groups of class $2$ to be compared. In particular, representing the commutator map as a tensor makes it possible to study group isomorphism using linear algebraic methods. However, the commutator tensor alone does not capture all the information needed to determine a group isomorphism, especially when $p=2$.

The prime~$2$ also introduces algebraic difficulties absent from several earlier approaches in odd characteristic. Every group of exponent~$2$ is abelian, so even the simplest nonabelian $2$-groups must have elements of order at least~$4$. In particular, the commutator pairing does not determine the squaring map. The dihedral and quaternion groups of order~$8$, for example, have equivalent commutator pairings on their quotients by the center but different square relations. More generally, the quadratic information carried by squaring is not determined by its polarization in characteristic~$2$.

Some algorithmic arguments for odd characteristic also rely on dividing by~$2$ when a commutator correction is absorbed into a power map. Likewise, structural uniqueness claims for decompositions of matrix tuples under congruence may require modification in characteristic~$2$. 
Recent work on groups of Frattini class $2$ addressed closely related obstacles by developing methods for quadratic form space isometry over $\mathbb F_2$~\cite{grochow2026beyond}. 
In the present problem, however, the analogous data lie over $\mathbb Z/2^\tau\mathbb Z$ for potentially large~$\tau$, 
so 
the extension is not merely a formal change of coefficient rings. 

For general $2$-groups of class $2$, the task is more demanding than recovering a quadratic form over $\mathbb F_2$. 
Elements may have orders that are arbitrarily large powers of $2$, and the relevant changes of generators take place over rings such as $\mathbb Z/2^\tau\mathbb Z$.
Thus, an algorithm must simultaneously handle the bilinear commutator  map, central power relations of generators with different orders, and the failure of those power relations to behave linearly. 

\subsection{Related work}

The group isomorphism problem has been investigated across various classes of groups. Polynomial-time algorithms have been developed for abelian groups~\cite{savage1980n2, vikas1996ano, Kavitha2007LinearAbelianIso}, groups arising as semidirect products of an abelian group with a cyclic group~\cite{wilson2009decomposing, wilson2009finding, LeGallRosenbaum}, groups containing normal Hall subgroups~\cite{qiao2012isomorphism}, groups with abelian Sylow towers~\cite{babai2012polynomial_1}, and groups with no abelian normal subgroups~\cite{babai2012polynomial}. Dietrich and Wilson demonstrated that the group isomorphism problem can be solved in nearly linear time for most orders~\cite{dietrich2022group}.

The isomorphism problem for nilpotent groups of class $2$ is closely related to tensor isomorphism through the bilinear maps induced by group commutators~\cite{TI1,GQSS25}.
Algorithms for tensor isomorphism and related matrix tuple problems have been studied extensively~\cite{IvanyosQiao19,Sun23,TI3,IMQSZ}.
The computational complexity of tensor isomorphism, including completeness results and reductions among algebraic isomorphism problems, has been investigated in~\cite{TI1,TI3,TI4,GQSS25}.
Its applications to cryptography, particularly pseudorandom group actions and post-quantum digital signatures, have also been explored~\cite{patarin1996hidden,ji2019general,MEDS2023}.
\subsection{Technique Overview}
\label{sec:technique_overview}

Our strategy uses tensor isomorphism \emph{to restrict the search for group isomorphisms}, rather than reducing the entire problem to a single tensor isomorphism instance as in previous approaches~\cite{Sun23,GQSS25,grochow2026beyond,IMQSZ}.  
We then encode most commutator and power relations in an alternating matrix tuple over a finite ring. The remaining information is confined to a small set of central power coefficients, which we recover through an explicit orbit computation.

\subsubsection{Beyond the commutator tensor}\label{sec:beyond_commutator_tensor}

Let $G$ be a $p$-group of class $2$. We use the convention $[g,h]=ghg^{-1}h^{-1}$ and write $\widetilde g=gZ(G)$. Since $[G,G]\leq Z(G)$, the commutator operation induces an alternating bilinear map on the abelian $p$-group $G / Z(G)$:
\[
  b_G:G/Z(G)\times G/Z(G)\longrightarrow [G,G]
  \text{ with } (\widetilde g,\widetilde h)\longmapsto[g,h].
\]
After choosing cyclic decompositions of the source and target, its coefficients can be arranged into an alternating tensor. Every group isomorphism induces an equivalence between the corresponding commutator maps. For groups of class 2 and exponent~$p$ with odd~$p$, Baer's correspondence~\cite{Bae38} makes this information sufficient; subsequent tensor-based reductions have also handled more general families in odd characteristic~\cite{Sun23,GQSS25,IMQSZ}.
For $2$-groups of Frattini class 2, a classical result going back to Higman~\cite{HigmanEnum} replaces the alternating tensor description by an isometry problem for a linear space of quadratic forms. The resulting problem is closely related to tensor isomorphism and can be attacked by extensions of tensor isomorphism techniques~\cite{grochow2026beyond}. 

What are the necessary and sufficient conditions for two groups of class $2$ to be isomorphic?
To answer this question, let $z_1,\dots,z_s$ be generators of a direct sum decomposition of $Z(G)$, and let $g_1,\dots,g_m$ be lifts to $G$ of generators of a direct sum decomposition of $G/Z(G)$. That is,
\[
Z(G)=\oplus_{i=1}^{s}\langle z_i\rangle,
\qquad G/Z(G)=\bigoplus_{i=1}^{m}\langle\widetilde g_i\rangle,
\qquad d_i=\ord(\widetilde g_i),
\]
where $\widetilde g_i=g_iZ(G)$, $\langle g\rangle$ denotes the subgroup generated by $g$, and $\ord(g)$ denotes the order of $g$.
Choose analogous decompositions for $H$, with central generators $y_1,\dots,y_s$ and lifts $h_1,\dots,h_m$ of quotient generators $\widetilde h_1,\dots,\widetilde h_m$.
If the corresponding cyclic orders agree, and
the coefficients of the pairwise commutators ($[g_i,g_j]$ and $[h_i,h_j]$) and central powers ($g_i^{d_i}$ and $h_i^{d_i}$) agree with respect to the central generators $z_1,\dots,z_s$ and $y_1,\dots,y_s$, respectively, then $G$ and $H$ are isomorphic.
Conversely, if $G$ and $H$ are isomorphic, then there exist choices of such decompositions and lifts for which the corresponding cyclic orders, commutator coefficients, and central power coefficients all agree.
This characterization gives necessary and sufficient conditions for two groups of class $2$ to be isomorphic. It is implicit in Section 8.9 of~\cite{holt2005handbook} (see also Lemma~\ref{lem:iso_condition}).

This commutator-and-power condition also explains why 
neither an isomorphism between the centers nor an isomorphism between the central quotients alone suffices.
For example, the dihedral and quaternion groups of order~$8$ illustrate the difference. Their quotient commutator forms are equivalent, yet the lifts of their noncentral cosets have different squares. In the Frattini setting these differences are captured by a quadratic form; in the general case, the corresponding relations may occur at several distinct powers of~$2$.
This example shows why central power relations must also be preserved for arbitrary groups of class $2$.

From the perspective of group isomorphism, preserving central power relations introduces an additional obstacle: these relations do not generally transform linearly under changes of generators.
In a group of class 2,
we have 
\begin{equation}\label{eq:overview_power_identity}
  (xy)^q=x^qy^q[y,x]^{\binom q2}
\end{equation}
for any $x, y$ in the group and any positive integer $q$. 
More generally, 
raising a product of generators to the $q$-th power gives
\[
 \left(\prod_{j=1}^{r}a_j^{e_j}\right)^q
 =\prod_{j=1}^{r}a_j^{qe_j}
 \prod_{1\leq j<k\leq r}[a_k,a_j]^{\binom q2e_je_k}.
\]
The displayed products are taken in increasing index order. 
Thus, a linear change of generators need not induce a linear change in their central powers, because additional commutator corrections arise from interactions between different generators. This is precisely what prevents us from identifying the group isomorphism problem with an unconstrained tensor isomorphism problem in characteristic~$2$.

If $x^q$ and $y^q$ are central for $q=p^s$, then $[x,y]^q=1$. For odd~$p$, the exponent $\binom{p^s}{2}$ is divisible by $p^s$, so the correction vanishes. For $p=2$, it need not vanish, but it always has order at most~$2$. This is the essential distinction: the nonlinear correction is not arbitrary, yet it cannot simply be ignored. Our algorithm exploits its small order rather than trying to represent the entire central power map linearly.

The key conceptual difference is that we do not require tensor isomorphism alone to certify group isomorphism. Instead, we use the tensor isomorphism algorithm to constrain the transformations that remain after choosing a small exceptional set of coordinates. These constraints guide our choice of group generators and isolate the part of the central power map that may fail to behave additively.
The commutator tensor therefore provides the initial restriction of the search space, while the subsequent matrix tuple and orbit stages complete the group isomorphism test.

\subsubsection{Using tensor isomorphism to restrict the search}

\begin{figure}[h]
\begin{center}
\vspace{.5cm}
\includegraphics[width=0.5\textwidth]{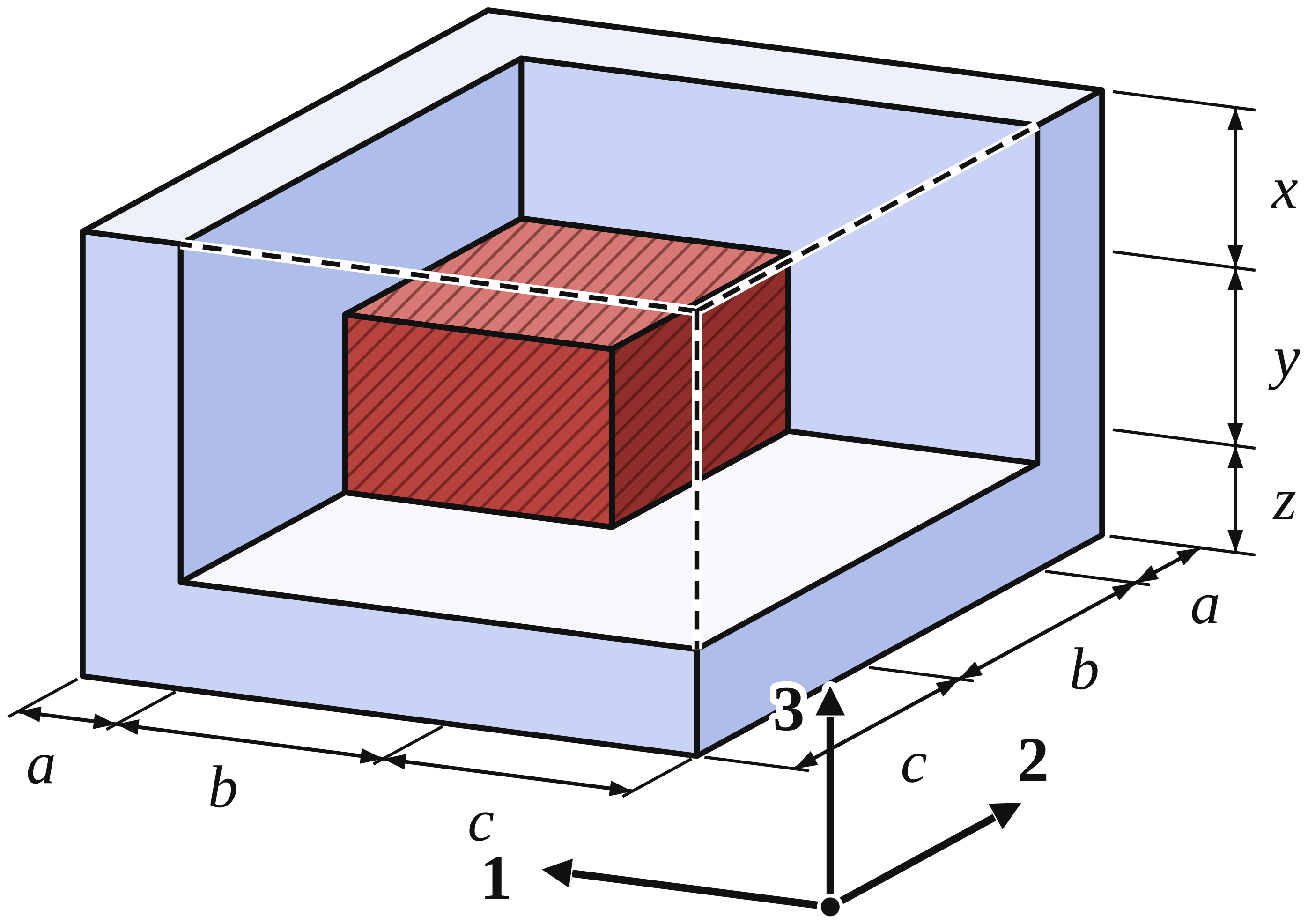}
\vspace{-.5cm}\caption{Restricted tensor}\label{fig:fig_1}
\end{center}
\end{figure}

Recent tensor isomorphism algorithms achieve speedups by restricting attention to transformations that respect a carefully chosen block structure. 
Roughly speaking, after enumerating partial information about the commutator tensor, one can arrange it so that large blocks are fixed, leaving only changes of basis within certain blocks and additions between blocks.
Figure~\ref{fig:fig_1} illustrates this restricted form: the $(a)$ and $(x)$ blocks are exceptional, while the large red region records information preserved by the remaining admissible transformations.

To describe the restricted form more concretely, partition each of the first two equivalent tensor directions into blocks $(a),(b),(c)$, and the third direction into $(x),(y),(z)$. For slices indexed by $(c)$ in either of the first two directions, the entries in the relevant $(b,c)\times(x, y)$ region vanish; the $(x)$-slices in the third direction also vanish on the $(b,c)\times(b,c)$ region. The exceptional blocks $(a)$ and $(z)$ are substantially smaller than their respective ambient directions. These zero conditions constrain the possible changes of basis.

An appropriate restricted form need not be unique. The tensor algorithm makes a bounded number of choices (or guesses based on a small amount of sampled tensor information) and enumerates the resulting possibilities. Once these choices are fixed, large blocks of the two tensors must already agree. In the remaining directions, the relevant freedom is, schematically, to add $(c)$-coordinates to $(b)$-coordinates in the first two directions, to add $(x)$-coordinates to $(y)$-coordinates in the third, and to make internal invertible changes of coordinates within $(c)$ and $(x)$. Consequently, the search need only consider the small exceptional region and the limited remaining transformations. The rigid region is represented by the red portion of the figure and the residual freedom by its blue portion.


We interpret this block structure in terms of the commutator tensor defined in Section~\ref{sec:beyond_commutator_tensor}. Informally, it identifies subgroups $Z(G) \leq C\leq B\leq G$, where $C$ corresponds to the $(c)$ block and $B$ to the union of the $(b)$ and $(c)$ blocks.
The structure in Figure~\ref{fig:fig_1} implies that $G$ is generated by $B$ together with a small number of distinguished elements corresponding to $(a)$, while $[C,B]$ lies in a small exceptional subgroup of the center corresponding to $(z)$, denoted by $Z$.

The red region of the restricted tensor, which is preserved by the remaining admissible transformations, also defines a labeling map $\rho$ on $B$.
Because the relevant tensor slices depend additively on the group elements, the resulting labeling map is additive, i.e., $\rho(xy) = \rho(x) + \rho(y)$, and satisfies $\rho(c) = 0$ for each $c \in C$.

If two groups are isomorphic and their commutator tensors have the same restricted form as in Figure~\ref{fig:fig_1} (up to changes in the blue region), then the tensors supply additive information about the groups, represented by the maps $\rho$, that an isomorphism must respect.
Our tensor isomorphism procedure enumerates candidate pairs of labeling maps so that, whenever an isomorphism exists, at least one pair is preserved by it.

This restriction is particularly useful when changing representatives. 
Suppose we replace a representative $g$ by $gu$, where $\rho(u)=0$. When $g$ belongs to the relevant subgroup $B$, the restriction ensures that $[u,g]$ lies in a small central subgroup $Z$. The class $2$ power formula then shows that the central powers of $g$ and $gu$ differ by an additional power term and a commutator correction in $Z$.
In particular, after projecting the center onto a complement $Y$ of~$Z$, the troublesome commutator term disappears. In this sense, the tensor labeling makes most central power information essentially linear while confining the residual nonlinear term to the small central subgroup $Z$.

More precisely, for an appropriate subgroup $F$ and exceptional central subgroup $Z$, 
the commutators between any element of the kernel of $\rho$ and elements of $F$ lie in $Z$, i.e., $[x, F] \leq Z$ whenever $\rho(x) = 0$. Thus, when a representative is multiplied by several elements with zero label, the nonlinear correction in its central power relation is confined to~$Z$. Tensor isomorphism therefore gives more than a necessary commutator invariant: it limits where the difficult central power information can move.

\subsubsection{Group restriction}

The restricted tensor identifies useful subgroups, whereas 
the commutator-and-power condition for group isomorphism  in Section~\ref{sec:beyond_commutator_tensor} is stated in terms of direct sum decompositions of the center and the quotient by the center. The bridge between these descriptions is a \emph{group restriction}. We choose
\[
  Z(G)=Z\oplus Y,
  \qquad Z=\oplus_{j=1}^{\zeta}\langle z_j\rangle,
\]
several distinguished elements $g_1,\ldots,g_\gamma$, and a subgroup $F$ whose image $\langle F,Z(G)\rangle/Z(G)$ is a direct summand of $G/Z(G)$, so that
\[
  G/Z(G)=\oplus_{i=1}^{\gamma}\langle\widetilde g_i\rangle\oplus \langle F, Z(G) \rangle / Z(G).
\]
The parameters are chosen so that the exceptional families $g_i$ and $z_j$ contain few elements, 
and the labeling map ensures that commutators involving elements of its kernel lie in $Z$. These conditions preserve the structural information obtained from the tensor reduction while giving the decompositions needed to test group isomorphism.

Passing from the tensor restriction to a group restriction requires more than identifying the relevant subgroups. The isomorphism commutator-and-power condition needs cyclic direct summands in both $Z(G)$ and $G/Z(G)$, while the tensor reduction initially supplies only subgroups and  
the information that certain commutators lie in a small subgroup. We therefore enlarge the exceptional central subgroup to a direct summand $Z$ and choose a complement~$Y$. Similarly, we select distinguished quotient generators $g_1,\ldots,g_\gamma$ and a complementary quotient subgroup represented by elements in~$F$.


A key subtlety is that a subgroup of an abelian $p$-group generated by a few elements need not lie in a direct summand generated by equally few cyclic factors. To keep the exceptional central subgroup small, we develop decomposition lemmas that take account of the orders of cyclic factors. This produces a bounded family of group restrictions such that, whenever the original groups are isomorphic, at least one pair of enumerated restrictions is compatible with an isomorphism. The remaining task is to test isomorphism between two fixed restrictions.

\subsubsection{Matching most group relations by matrix tuple isometry}

For a group restriction, a \emph{representative vector} records a cyclic decomposition of $Y$, the relevant exceptional central elements in $Z \cap F$, and elements of $F$ corresponding to an ordered cyclic decomposition of $\langle F, Z(G) \rangle / Z(G)$.

Because a representative vector records a central decomposition and lifts of a quotient decomposition, it determines commutator coefficients, central power coefficients, and labels.
If these coefficients and labels agree for representative vectors of two groups, then the groups are isomorphic. 
However, the same restriction admits many representative vectors, so enumerating them all would negate the advantage gained from the tensor-based search space reduction.
Moreover, the coefficients need not transform linearly under arbitrary changes of the chosen lifts. We therefore distinguish a weakened equivalence, called \emph{matching}, from a full group restriction isomorphism. The first stage matches the information preserved by suitable transformations; the second recovers the information deliberately omitted.

We call two representative vectors \emph{matching} when their commutator coefficients, their $Y$-power coefficients, and their labels agree, while their $Z$-power coefficients agree modulo the order-$p$ subgroup of each distinguished cyclic factor. Equivalently, we deliberately omit the highest base-$p$ digit of each relevant $Z$-power coefficient. Full equality of these coefficients would certify isomorphism of the restrictions; matching is the weaker condition that we can encode by linear-algebraic means.

The reason for this relaxation is the power identity~\eqref{eq:overview_power_identity}. For changes of generators selected by the order filtration and the labeling constraints, the commutator correction in the relevant powers has order at most~$p$. Hence it disappears from the portion of the $Z$-power data retained by matching. The $Y$-power data also remain linear when changes of generators are chosen to multiply an arbitrary element by several zero-label elements.

To match representative vectors, we encode the relations among their entries in an alternating matrix tuple over $\Z / p^\tau \Z$ for a suitable parameter $\tau$.
An alternating matrix tuple is a sequence of alternating matrices of the same dimension.
The goal of this encoding is to ensure that two group restrictions have matching representative vectors if and only if the matrix tuples constructed from any representative vectors of the two restrictions, denoted by $\mathbf{A}$ and $\mathbf{B}$, are congruent, i.e., $\mathbf{A} = P^T \mathbf{B} P$ for some invertible matrix $P$. 
Moreover, we want the admissible isometries to transform one representative vector into others that match the representative vector of the other group.

This task is not straightforward because the $Y$-components of the power coefficients are not always additive. 
For example, let $g, h$ be two elements of $F$ with $\ord(\widetilde{g}) = \ord(\widetilde{h}) = p$, where $p = 2$. Then $(g h)^p$ is central, and 
the $Y$-components of its power coefficients are guaranteed to agree with those of $g^p h^p$ only when at least one of $\rho(g)$ and $\rho(h)$ is zero, by the properties of the group restriction.
To enforce this compatibility condition when applying linear
transformations to the matrix tuple, we introduce auxiliary
group elements so that transformations between representative
vectors combine at most one nonzero $\rho$-value element 
 with elements whose $\rho$-values are zero.
This ensures that the $Y$-component of the commutator
correction vanishes.
This step requires careful design because multiplying
two elements with nonzero $\rho$-values can produce an
element with zero $\rho$-value.
We therefore need to encode additional information in
the matrix tuples to prevent such multiplications,
even when the resulting element has zero $\rho$-value.
At the same time, we must ensure that every element
with zero $\rho$-value can be expressed as a product
of representative and auxiliary elements with zero
$\rho$-values.
Furthermore, the auxiliary elements must be chosen
carefully to prevent an excessive increase in the
dimension of the matrix tuples.

With these carefully constructed matrix tuples, we use a matrix tuple isometry algorithm over the finite ring $\Z / p^\tau \Z$ to 
find the required congruence. 
For odd $p$, an algorithm over $\F_p$ was developed in~\cite{IvanyosQiao19}, followed by an algorithm over $\Z / p^\tau \Z$ in~\cite{GQSS25}. 
For $p=2$, an algorithm over $\F_2$ was developed in~\cite{grochow2026beyond}. 
In this work, we extend the algorithm from $\F_2$ to $\Z / 2^\tau \Z$, obtaining a running-time bound that depends on $\tau$ (Theorem~\ref{thm:matrix_tuple_isometry_overall}). 
We present this matrix tuple congruence algorithm in Appendix~\ref{sec:matrix_tuple_isometry}; it may be of independent interest. 
The algorithm also computes a generating set for the automorphism group of an alternating matrix tuple.

The subgroup of matrix tuple automorphisms relevant to the group restriction acts on the representative vectors, subject to the compatibility conditions. Thus the matrix routine supplies both a matching witness and a finite generating set from which the candidate matching vectors can be explored. 

\subsubsection{Recovering the missing power information}

Matching extended representative vectors does not yet guarantee a group restriction isomorphism: the omitted highest base-$p$ digits of the $Z$-power coefficients must also agree.
We collect these digits into a matrix in $\mathbb F_p^{\phi\times\zeta}$, where $\phi$ is the length of the representative vector and $\zeta$ is the number of exceptional central factors.
Our goal is to enumerate the possible highest-digit matrices arising from representative vectors that match a fixed representative vector of the other group restriction.
There are at most $p^{\phi\zeta}$ such matrices, and our final parameter choices make their enumeration affordable; in particular, $\phi$ is at most $\log_p N$ and $\zeta$ is sublinear in $\log_p N$.

The key is to enumerate the possible highest-digit matrices associated with matching representative vectors, rather than the vectors themselves. This is feasible because the matching conditions determine all the relevant information except for the highest base-$p$ digits of the $Z$-power coefficients.


The matrix tuple automorphisms induce transformations on these missing digits. Their action is not ordinary matrix multiplication: changing generators uses ordered products in a nonabelian group, so the update formula must account for commutator corrections arising from~\eqref{eq:overview_power_identity} and carries from the lower base-$p$ digits. We derive an explicit update rule and compute the resulting orbit of possible matrices.

We then test whether the highest-digit data required by the second group restriction belong to this orbit.  
Orbit membership provides the missing full power relations, and the commutator-and-power condition then yields an isomorphism between the group restrictions.

\paragraph{Acknowledgments}
The author thanks Joshua Grochow for fruitful discussions regarding Lemma~\ref{lem:iso_condition}.

\paragraph{AI Methodology}
The author used ChatGPT 6 and Claude Opus 5.5 to improve the writing and structure of the proofs, verify correctness, and create figures. 
All algorithms, lemmas, and proofs were developed by the author, with the exception of the proofs for Lemma~\ref{lem:upper_tri_generating_set} and Lemma~\ref{lem:upper_triangularize} in Appendix~\ref{sec:matrix_tuple_isometry}, which were generated by ChatGPT 6 under the direct guidance of the author. 
The author verified the correctness and originality of all content, including references.

\section{Preliminary}

Throughout the paper, we suppose $p$ is a prime, which is assumed to be 2 unless specified, but can be any positive prime. 
While the result of this paper improves the result for $p = 2$, 
we present the results for general $p$ for the purpose of presenting a general algorithm that works for all the possible $p$-groups of class 2. 
We use $R$ to denote $\Z / p^\tau \Z$ for some parameter $\tau$ to be decided later. 

\subsection{Matrix and matrix tuple}

Throughout the paper, we assume all the vectors are column vectors unless specified. Let $M(m \times n, R)$ be the $R$-module of all the matrices over $R$, and $\GL(n, R)$ be the group of $n \times n$ invertible matrices over $R$. 
We use $e_i$ to denote the column vector such that the $i$-th coordinate is 1, and all the other coordinates are zero. 
For a matrix in $\M(m \times n, R)$, the rank of a matrix is the inner rank of the matrix, which is the smallest integer $r$ such that the matrix can be written as $\sum_{i = 1}^r u_i v_i^T$ for some $u_i \in R^m$ and $v_i \in R^n$.  
The inner rank satisfies the condition that $\rank(AB) \leq \min\{\rank(A), \rank(B) \}$.

An $n \times r$ matrix tuple $\mathbf{A}$ of length $m$ over $R$ is an ordered sequence $(A_1, \dots, A_m) \in \M(n \times r, R)^m$. 
A square matrix tuple $\mathbf{A} = (A_1, \dots, A_m) \in \M(n \times n, R)^m$ is an \emph{alternating} matrix tuple if $A_i[j, j]$ = 0 for each $1 \leq i \leq m, 1 \leq j \leq n$, and $A_i^T = - A_i$ for each $1 \leq i \leq m$.
For an $n \times r$ matrix tuple $\mathbf{A}$ of length $m$ and a column vector $v \in R^r$, we use $\mathbf{A}v$ to denote the $n \times 1$ column vector tuple $(A_1 v, \dots, A_m v)$.

Throughout this paper, unless specified, we assume that all matrix tuples are non-degenerate, i.e., 
there are no invertible matrices $X \in \GL(n,  R)$ and $Y \in \GL(r,  R)$ such that some row of $X \mathbf{A}Y$ is a zero row or some column of $X \mathbf{A}Y$ is a zero column.

We consider two different actions on matrix tuples. For a tuple $\mathbf{A} = (A_1, \dots, A_m) \in \M(n \times r, R)^m$, the left-right action by matrices $U\in \GL(n,  R)$ and $W \in \GL(r, R)$ is given by \[U^T \mathbf{A} W := (U^T A_1 W, \dots, U^T A_m W).\] In the congruence action, we require that the matrices in $\mathbf{A}$ are square; under this action, a matrix $U \in \GL(n, R)$ acts on $\mathbf{A}$ by conjugation via $U^T \mathbf{A} U$. 

For a matrix tuple $\mathbf{A} = (A_1, \dots, A_m) \in \M(n \times r, R)^m$, the $(n \cdot m) \times r$  matrix $C_\mathbf{A}$ is defined as 
\[
C_\mathbf{A} := \begin{pmatrix}
A_1 \\ \vdots \\ A_m
\end{pmatrix}. 
\]

The alternating matrix-tuple isometry problem asks whether two matrix tuples $\mathbf{A}$ and $\mathbf{B}$ are congruent, i.e., whether there exists an invertible matrix $P$ such that $P^T\mathbf{A}P=\mathbf{B}$. For odd $p$, polynomial-time algorithms are known over both $\F_p$~\cite{IvanyosQiao19} and $\Z/p^\tau\Z$~\cite{GQSS25}. For $p=2$, an $n^{O(\sqrt{n})}$-time algorithm over $\F_2$ was developed in~\cite{grochow2026beyond}. We extend this result to $\Z/2^\tau\Z$, with the proof provided in Appendix~\ref{sec:matrix_tuple_isometry}.

\begin{theorem}[c.f.~Theorem 1.2 of~\cite{grochow2026beyond}]\label{thm:matrix_tuple_isometry_overall}
Let $R = \Z / p^\tau \Z$ for any prime $p$ and positive integer $\tau$. 
Let $\mathbf{A}$ and $\mathbf{B}$ be two alternating matrix tuples in $\M(n, R)^m$ for some positive integers $n, m, \tau$. 
There is an algorithm to determine if there is a matrix $P \in \GL(n, R)$ such that $P^T \mathbf{A} P = \mathbf{B}$ in time \[p^{O(\tau n^{7/4} \log (\tau n))} \cdot \poly(m),\] and 
an algorithm to output a generating set of the invertible automorphism group in the same time of size $p^{O(\tau n^{7/4} \log (\tau n))} \cdot \poly(m)$.
\end{theorem}

\subsection{Group theory}

Throughout the paper, 
for a group $G$ and an element $g\in G$, we use $\ord(g)$ to denote the order of $g$ and $\widetilde g$ to denote $g Z(G)$, the coset of $Z(G)$ containing $g$. 
For two elements of the group $g, h$, we define $[g, h] := g h g^{-1} h^{-1}$. 

For an $m \times n$ matrix $Q$ over $R$ and a column vector $\vec y = (y_1, \dots, y_n)^T$ of group elements whose orders divide $p^\tau$, 
we write $Q\vec y$ for the vector $(y_1',\dots,y_m')^T$ with $y_i' = \prod_{j=1}^n y_j^{Q[i,j]}$, the product taken in increasing order of $j$. 
We remark that if the group is not abelian, then the composition law does not hold, i.e., $Q_1 (Q_2 \vec y)$ is not equal to $(Q_1 Q_2) \vec y$ in general. 

For a finite abelian $p$-group $G$, the rank of $G$ is the integer $r$ such that $G \cong \oplus_{i=1}^r (\Z/p^{\alpha_i}\Z)$ for some positive integers $\alpha_1,\dots,\alpha_r$. 
For an abelian $p$-group $G$ with rank $k$, suppose $g_1, \dots, g_k$ form a decomposition of $G$, i.e., $G = \oplus_{i = 1}^k \langle g_i \rangle$. 
For an element $g$ of $G$, 
we call $\alpha_i$ the coefficient of $g$ with respect to $g_i$, relative to $g_1,\dots,g_k$, if $g = \prod_{i = 1}^k g_i^{\alpha_i}$ and $0 \leq \alpha_i < \ord(g_i)$ for each $1 \leq i \leq k$.

The following theorem is from Hillar and Rhea’s description of the endomorphism ring and automorphism group of abelian $p$-groups ~\cite{HillarRhea} (originally proved by Ranum nearly a century earlier~\cite{Ranum}).

\begin{theorem}\label{thm:group_decomposition}
Let $G$ be an abelian $p$-group with $G = \oplus_{i = 1}^r \langle g_i\rangle $ and $\vec g = (g_1, \dots, g_r)^T$. 
Let $Q$ be a matrix in $\M(r, R)$ such that every element of $G$ is of order at most $p^\tau$. 
$(g_1', \dots, g_r')^T = Q \vec g$ forms a direct sum decomposition of $G$ with $\ord(g_i') = \ord(g_i)$ for every $i$ if and only if 
$Q$ is invertible and $Q[i, j]$ is a multiple of $\ord(g_j) / \ord(g_i)$ if $\ord(g_i) \leq \ord(g_j)$.
\end{theorem}

\subsection{Tensor and tensor isomorphism}
A $\ell \times n \times m$ tensor $\mathcal{T}$ over $R$ is $(t_{i, j, k})$ where $i \in [\ell], j \in [n], k \in [m]$ and $t_{i, j, k} \in R$. 
We use $\mathcal{T}_{i, j, k}$ for some $1 \leq i \leq \ell, 1 \leq j \leq n, 1 \leq k \leq m$ to denote the $(i, j, k)$-th entry of the tensor. 
Given matrices $X \in \M(\ell, R), Y \in \M(n, R), Z \in \M(m, R)$, $(X, Y, Z)$ sends $\mathcal{T}$ to $\mathcal{T}' = (t_{i, j, k}')$, also denoted as $(X, Y, Z) \circ \mathcal{T}$, such that 
\[t'_{i, j, k} = \sum_{i', j', k'} X_{i, i'} Y_{j, j'} Z_{k, k'} t_{i', j', k'}.\]

A tensor $\mathcal{T} = (t_{i, j, k}) \in R^{\ell \times n \times m}$ is an alternating tensor if $\ell = n$, and for each $i, j, k$, $t_{i, j, k} = - t_{j, i, k}$ and $t_{i, i, k} = 0$ for each $i, k$.
In the alternating tensor isomorphism problem, we are given two $n \times n \times m$ alternating matrix tensors $\mathcal{T}$ and $\mathcal{T}'$ and want to determine if there exist 
$X \in \GL(n, R)$ and $Y \in \GL(m, R)$ such that $\mathcal{T}' = (X, X, Y) \circ \mathcal{T}$. 

For any $1 \leq i \leq n$, 
Let $N_{\mathcal{T}, i}$ be the $n \times m$ matrix over $R$ such that $N_{\mathcal{T}, i}[j, k] = t_{i, j, k}$.
More generally, for any column vector $\vec x = (x_1, \dots,  x_n)^T\in R^n$,
$N_{\mathcal{T}, \vec x}$ is the $n \times m$ matrix 
\[\sum_{i = 1}^n x_i \cdot N_{\mathcal{T}, i}.\]
For any $1 \leq k \leq m$, 
Let $M_{\mathcal{T}, k}$ be the $n \times n$ matrix over $R$ such that $M_{\mathcal{T}, k}[i, j] = t_{i, j, k}$.
For any column vector $\vec y = (y_1, \dots, y_m)^T \in R^m$,
$M_{\mathcal{T}, \vec y}$ is the $n \times n$ matrix \[\sum_{k = 1}^m y_k \cdot M_{\mathcal{T}, k}.\] 
Let $c_{\mathcal{T}, 0}$ be the integer such that $p^{c_{\mathcal{T}, 0}}$ is the number of different $N_{\mathcal{T}, \vec x}$ matrices for all the possible $\vec x$, and 
$c_{\mathcal{T}, 1}$ be the integer such that $p^{c_{\mathcal{T}, 1}}$ is the number of different $M_{\mathcal{T}, \vec y}$ matrices for all the possible $\vec y$. 
For an alternating tensor over $R$, define $c_{\mathcal{T}} :=c_{\mathcal{T}, 0} + c_{\mathcal{T}, 1}$ to be the complex parameter of the tensor $\mathcal{T}$.

In~\cite{GQSS25}, the alternating tensor isomorphism problem over a ring $R$ is reduced to the following restricted alternating tensor isomorphism problem. 
In the restricted alternating tensor isomorphism problem, with two integer parameters $d$ and $w$, 
given two alternating tensors $\mathcal{T}, \mathcal{T}' \in R^{n \times n \times m}$, and two matrices $P, P' \in \M(d \times n, R)$  such that 
\begin{enumerate}
\item for any $\vec x \in R^n$ satisfying $P \vec x = 0$, the $(j, k)$-th entry of $N_{\mathcal{T}, \vec x}$ is zero for any $j, k > w$;
\item for any $\vec x \in R^n$ satisfying $N_{\mathcal{T}, \vec x} = 0$, $P\vec x = 0$,
\end{enumerate}
decide if there are $X \in \GL(n, R)$ and $Y \in \GL(m, R)$ such that $(X, X, Y) \circ \mathcal{T} = \mathcal{T}'$ and $P \cdot X^T = P'$. 

In~\cite{GQSS25}, the following theorem was proved. 

\begin{theorem}[c.f. Lemma 5.16 of \cite{GQSS25}]\label{thm:tensor_iso}
Let $\mathcal{T}$ and $\mathcal{T}'$ be two alternating tensors over $R$ with the same complex parameter $c_{\mathcal{T}}$. 
There is an algorithm with running time $p^{O(c_{\mathcal{T}}^{17 / 9} \cdot (\log p)^{8/9})}$ to reduce the tensor isomorphism problem to $p^{O(c_{\mathcal{T}}^{17 / 9} \cdot (\log p)^{8/9})}$ instances of the restricted alternating tensor isomorphism problem with parameters $d \leq c_{\mathcal{T}}^2$ and $w = O(\tau^2 \cdot c_{\mathcal{T}}^{4/9} \cdot (\log p)^2 )$ in the following sense:
\begin{enumerate}
\item For each instance of the restricted alternating tensor isomorphism problem, the algorithm produces $U, U' \in \GL(n, R)$, $W, W' \in \GL(m, R)$ and $P, P' \in \M(d \times n, R)$ such that $(U, U, W) \circ \mathcal{T}$, $(U', U', W') \circ \mathcal{T}'$, $P$ and $P'$ form an input instance of the restricted alternating tensor isomorphism problem. 
\item For any $X$ and  $Y$ such that $(X, X, Y) \circ \mathcal{T} = \mathcal{T}'$,
 there are produced $U, U', W, W', P$, and $P'$ such that 
The pair $U' X U^{-1}$ and $W' Y W^{-1}$ is a solution for the restricted alternating tensor isomorphism problem with  $(U, U, W) \circ \mathcal{T}$ and $ (U', U', W') \circ \mathcal{T}'$, $P,$ and  $P'$ as input. 
 
 \end{enumerate}
\end{theorem}

\section{Necessary and sufficient condition for class 2 isomorphism}

We establish the necessary and sufficient condition for group isomorphism. 
The following result was implicitly implied by Section 8.9 of the Handbook of Computational Group Theory~\cite{holt2005handbook}. 
For completeness, we give a self-contained proof.

\begin{lemma}\label{lem:iso_condition}
Let $G$ and $H$ be finite $p$-groups of nilpotent class $2$ such that both $Z(G)$ and $Z(H)$ have rank $k$, and $G/Z(G)$ and $H/Z(H)$ have rank $k'$. Let $\ell=k+k'$. Then $G$ and $H$ are isomorphic if and only if there exist $g_1,\dots,g_\ell\in G$ and $h_1,\dots,h_\ell\in H$ satisfying the following conditions:
\begin{enumerate}
\item $Z(G)=\oplus_{i=1}^k\langle g_i\rangle$ and $Z(H)=\oplus_{i=1}^k\langle h_i\rangle$, with $\ord(g_i)=\ord(h_i)$ for every $i\in[k]$.

\item $G/Z(G)=\oplus_{i=k+1}^{\ell}\langle {\widetilde{g}_i}\rangle$ and $H/Z(H)=\oplus_{i=k+1}^{\ell}\langle {\widetilde{h}_i}\rangle$, with $\ord({\widetilde{g}_i})=\ord({\widetilde{h}_i})$ for every $k+1\leq i\leq\ell$.

\item Let $t_i=\ord(g_i)$ for $1\leq i\leq k$ and $t_i=\ord({\widetilde{g}_i})$ for $k+1\leq i\leq\ell$. Define $f:G\to H$ by
\[
f(g_1^{m_1}\cdots g_\ell^{m_\ell})=h_1^{m_1}\cdots h_\ell^{m_\ell}.
\]
Then
\begin{enumerate}
\item $f(g_i^{t_i})=f(g_i)^{t_i}$ for every $k+1\leq i\leq\ell$;
\item $f([g_i,g_j])=[f(g_i),f(g_j)]$ for every $i,j\in[\ell]$.
\end{enumerate}
\end{enumerate}
Furthermore, whenever $g_1,\dots,g_\ell$ and $h_1,\dots,h_\ell$ satisfy these three conditions, the resulting map $f$ is an isomorphism.
\end{lemma}

\begin{proof}
By the first two conditions of the lemma, every element of $G$ has a unique representation
$g=g_1^{m_1}\cdots g_\ell^{m_\ell}$ with $ 0\leq m_i<t_i,$
and every element of $H$ has the analogous representation using the $h_i$.

The forward implication is immediate. Suppose that $f_0:G\to H$ is an isomorphism. Choose $g_1,\dots,g_k$ such that $Z(G)=\oplus_{i=1}^k\langle g_i\rangle$, and choose $g_{k+1},\dots,g_\ell$ whose images form a direct-sum decomposition of $G/Z(G)$. Set $h_i=f_0(g_i)$. Then the first two conditions  hold, and the map $f$ in the third condition is exactly $f_0$. Hence the third condition holds.

Conversely, suppose that $g_1,\dots,g_\ell$ and $h_1,\dots,h_\ell$ satisfy the three conditions. Because $G$ has class $2$, all commutators are central, and for all $x,y\in G$, $xy=[x,y]yx$.
Thus any product of powers of the $g_i$ can be collected into the ordered form $g_1^{a_1}\cdots g_\ell^{a_\ell}$ by repeatedly combining adjacent powers of the same generator, 
replacing $g_i^{t_i}$ by empty if $i\leq k$, or by its unique ordered central representation if $i>k$, and interchanging adjacent out-of-order generators, i.e., replacing $g_i^a g_j^b$ by
$[g_i,g_j]^{ab}g_j^b g_i^a$  for some $i > j$. 
The same collection process applies to the $h_i$. Condition (3a) guarantees that a power $g_i^{t_i} \in Z(G)$, for $i>k$, is replaced by the corresponding ordered central factor (involving only $g_1,\dots,g_k$) after applying $f$, and for $i \leq k$, is replaced by empty; condition (3b) guarantees the same for commutator factors. 

Now take $x,y\in G$ and denote
$
x=g_1^{a_1}\cdots g_\ell^{a_\ell}, y=g_1^{b_1}\cdots g_\ell^{b_\ell}$.
Apply the collection process to $xy$, obtaining $g_1^{c_1}\cdots g_\ell^{c_\ell}$. Applying the same sequence of collection steps to $f(x)f(y)$ gives $f(g_1)^{c_1}\cdots f(g_\ell)^{c_\ell}$. Therefore
\[
f(xy)=f(g_1)^{c_1}\cdots f(g_\ell)^{c_\ell}=f(x)f(y),
\]
so $f$ is a homomorphism. The first two conditions imply that the ordered normal forms in $G$ and $H$ have the same exponent ranges, so $f$ is bijective. Hence $f$ is an isomorphism.
\end{proof}

We also prove the following lemma for the power central coefficients. 

\begin{lemma}\label{lem:high_bit}
Let $G$ be a finite $p$-group of class 2. Let $a_1, \dots, a_k$ be a set of elements of order at most $p^t$ for some non-negative integer $t$.
Let $g = \prod_{i = 1}^k a_i^{m_i}$. 
Then \[g^{p^t} = \prod_{i = 1} a_i^{m_i p^t} \cdot z\] for some $z$ of order at most $p$.  
\end{lemma}
\begin{proof}
Since $G$ has class 2, we have
\[
g^{p^t}
=
\prod_{i=1}^k a_i^{m_i p^t}
\cdot
\prod_{1\leq i<j\leq k}
[a_j,a_i]^{m_i m_j\binom{p^t}{2}}.
\]
Let
\[
z=
\prod_{1\leq i<j\leq k}
[a_j,a_i]^{m_i m_j\binom{p^t}{2}}.
\]
Since $\ord(a_i)\leq p^t$, every commutator
$[a_j,a_i]$ has order at most $p^t$.
Moreover, $p^t$ divides $p\binom{p^t}{2}$.
Thus $z^p=1$, proving the lemma.
\end{proof}

\section{Commutator tensor and group restriction}

In this section, we reduce the group isomorphism problem to the group restriction isomorphism problem. 
The purpose of the group restriction is to applying the results from tensor isomorphism to decompose the center $Z(G)$ by 
\[\langle z_1 \rangle \oplus \dots \oplus \langle z_\zeta \rangle \oplus Y\] and $G / Z(G)$ by \[\langle {\widetilde{g}_1}\rangle  \oplus \dots \oplus \langle {\widetilde{g}_\gamma} \rangle\oplus \langle F, Z(G) \rangle / Z(G)\] so that elements within $F$ zero labels have commutators with other elements of $F$ in $\langle z_1 \rangle \oplus \dots \oplus \langle z_\zeta \rangle$. 
And thus, when multiplying elements with zero labels, the power projected onto $Y$ does not change.

In Section~\ref{sec:group_restriction}, we define group restriction and group restriction isomorphism. 
In Section~\ref{sec:group_decomposition}, we present several lemmas to decompose a group into a direct sum decomposition under certain constraints. 
In Section~\ref{sec:tensor_iso}, we use tensor isomorphism  techniques to create labels for group elements. 
In Section~\ref{sec:restriction_reduction}, we reduce the group isomorphism problem to restricted group isomorphism problem. 
\subsection{Group restriction and group restriction isomorphism}\label{sec:group_restriction}

We formally define group restriction, which is to encode the restrictions obtained from the tensor isomorphism algorithm and other conditions to make the elements in the restricted
group ($F$ in the definition) behave well. 

\begin{definition}[Group restriction]\label{def:restricted_group}

Let $G$ be a finite $p$-group of class $2$. 
A group restriction of $G$ is $(z_1, \dots, z_\zeta, Y, g_1, \dots, g_\gamma, F, \rho)$ for positive integers $\zeta, \gamma, \kappa, \tau$ and $\delta$,  satisfying the following conditions: 
\begin{enumerate}
\item $z_1, \dots, z_\zeta \in Z(G)$ and a subgroup $Y \leq Z(G)$ satisfy the following conditions:
\begin{enumerate}
\item $Z(G) = \langle z_1 \rangle \oplus \dots \oplus \langle z_\zeta \rangle \oplus Y$; 
\item every element  $z \in Y$ is of order at most $p^\tau$. 
\end{enumerate}

\item elements $g_1, \dots, g_\gamma$ and a subgroup $F \leq G$ satisfy the following conditions:
\begin{enumerate}
\item[(c)] there exists a $K \leq G / Z(G)$ such that $G / Z(G) = \langle {\widetilde{g}_1} \rangle \oplus \dots \oplus \langle {\widetilde{g}_\gamma} \rangle \oplus K$, every element of $K$ is of order at most $p^{\delta}$, and every ${\widetilde{g}} \in K$ has a $g' \in {\widetilde{g}}$ satisfying $\ord(g') \leq p^\tau$;
\item[(d)] $F$ is the subgroup generated by $\{g \in G: {\widetilde{g}} \in K, \ord(g) \leq p^{\tau}\}$, and $\langle F, Z(G) \rangle / Z(G) = K$. 
\end{enumerate}
\item A function $\rho: F \rightarrow R^\kappa$ satisfies the following conditions:
\begin{enumerate}
\item[(e)] $\rho(g) = (0, \dots, 0)$ for every $g \in Z(G) \cap F$;
\item[(f)] for every $g \in F$ such that $\rho(g) = (0, \dots, 0)$, $[g, F] \leq \langle z_1 \rangle \oplus \dots \oplus \langle z_\zeta \rangle$, where $[g, F]$ is the subgroup generated by $\{[g, x] : x \in F\}$;
\item[(g)] for every $g, h \in F$, $\rho(g\cdot h) = \rho(g) + \rho(h)$. 
\end{enumerate}
\end{enumerate}
We also use $\psi$ to denote the rank of $Y$, $\eta$ to denote the rank of $\langle F, Z(G) \rangle / Z(G)$, $Z$ to denote $\oplus_{i = 1}^\zeta \langle z_i\rangle$, and $Z_0 = Z \cap F$.
\end{definition}

The following fact shows that all the elements of $F$ have an order at most $p^\tau$.

\begin{fact}\label{fact:f_order}
For a group restriction with parameter $\tau$ and $\delta$ such that $\tau \geq \delta + 1$, 
every element of $F$ is of order at most $p^\tau$. 
\end{fact}
\begin{proof}
Since every element $g \in F$ satisfies $\ord({\widetilde{g}}) \leq p^{\delta}$, 
$\ord([g, g']) \leq p^{\delta}$ for any $g' \in F$. 
Hence, for any $g, g' \in F$ of order at most $p^\tau$, we have 

\[(g \cdot g')^{p^\tau} = g^{p^\tau} g'^{p^\tau} [g', g]^{\binom{p^\tau}{2}} = g^{p^\tau} g'^{p^\tau} ([g', g]^{p^\delta})^{p^{\tau -\delta}(p^\tau - 1) / 2} = 1,\]
where the last equality uses the fact that $p^\tau - 1$ is a multiple of 2 for all the odd primes, and $p^{\tau - 1}$ is a multiple of 2 if $p = 2$. 
By induction, any element in $F$ is of order at most $p^\tau$. 
\end{proof}

We define the group restriction isomorphism problem, which is to determine if two groups under the imposed restrictions are isomorphic with respect to the restriction. 

\begin{problem}[Group restriction isomorphism]
Given two groups $G$ and $H$ with their restrictions \[(z_1, \dots, z_\zeta, Y, g_1, \dots, g_\gamma, F, \rho)  \text{ for } G  \text{ and } (z_1', \dots, z_\zeta', X, g_1', \dots, g_\gamma', E, \rho') \text{ for }H,\] decide whether there exists a group restriction isomorphism $f:G\rightarrow H$, i.e., an isomorphism satisfying the following conditions:
\begin{enumerate}
\item $f(z_i) = z_i'$ for every $1 \leq i \leq \zeta$; 
\item $f(g_i) = g_i'$ for every $1 \leq i \leq \gamma$;  
\item $f(Y) = X$ and $f(F) = E$; 
\item for every $g \in F$, $\rho'(f(g)) = \rho(g)$.
\end{enumerate}
\end{problem}

Without loss of generality, when studying the group restriction isomorphism problem, 
we may assume that the following conditions hold, since they are necessary and can be verified efficiently.
\begin{enumerate}
\item the parameters $\zeta, \gamma, \tau, \kappa, \psi$ and $\eta$ are the same for the two restrictions;
\item $Y$ and $X$ are isomorphic;
\item $\langle F , Z(G) \rangle / Z(G)$ and $\langle E, Z(H) \rangle / Z(H)$ are isomorphic; 
\item $\ord(z_i) = \ord(z_i')$ for every $i \in [\zeta]$; 
\item $\ord({\widetilde{g}_i}) = \ord({\widetilde{g}_i'})$ for every $i \in [\gamma]$. 
\item $\alpha_{i, j, k}$ for $G$ and $H$ are the same for any $1\leq i, j \leq \gamma$, where (taking $G$ as an example)
$\alpha_{i,j,k}$ for each $i,j\in[\gamma]$ is defined by
$
[g_i,g_j]
=\left(\prod_{k=1}^{\zeta}
z_k^{\alpha_{i,j,k}}\right)y_{i,j}$ for some 
$y_{i,j}\in Y$, where
$0\leq\alpha_{i,j,k}<\ord(z_k)$.

\end{enumerate} 

Based on  Lemma~\ref{lem:iso_condition},
we give necessary and efficient conditions for the group restriction isomorphism.

\begin{lemma}\label{lem:restricted_group_iso_condition}
Let   
\[(z_1, \dots, z_\zeta, Y, g_1, \dots, g_\gamma, F, \rho) \text{ and }(z_1', \dots, z_\zeta', X, g_1', \dots, g_\gamma', E, \rho')\] be group restrictions for $G$ and $H$ respectively. 
Then there is a group restriction isomorphism from $G$ to $H$ if and only if 
there are \[z_{\zeta + 1}, \dots, z_{\zeta + \psi} \in Y, z_{\zeta + 1}', \dots, z_{\zeta + \psi}' \in X, 
g_{\gamma + 1}, \dots, g_{\gamma + \eta} \in F, \text{ and } g_{\gamma + 1}', \dots, g_{\gamma + \eta}' \in E\] satisfying all of the following conditions:
\begin{enumerate}

\item 
\[Y = \oplus_{i = 1}^\psi \langle z_{\zeta + i} \rangle \text{ and }
X = \oplus_{i = 1}^\psi \langle z_{\zeta + i}' \rangle\] with $\ord(z_{\zeta + i}) = \ord(z_{\zeta + i}')$ for every $i \in [\psi]$; 
\item 
\[\langle F , Z(G)\rangle / Z(G) =\oplus_{i = 1}^\eta \langle {\widetilde{g}_{\gamma + i}} \rangle  \text{ and } 
\langle E , Z(H)\rangle / Z(H) = \oplus_{i = 1}^\eta \langle {\widetilde{g}_{\gamma + i}'} \rangle \]
with 
$\ord({\widetilde{g}_{\gamma + i}}) = \ord({\widetilde{g}_{\gamma + i}'})$ for every $i \in [\eta]$; 

\item For every $i \in [\gamma + \eta]$, the coefficients of $g_i^{\ord({\widetilde{g}_i})}$ with respect to $z_1, \dots, z_{\zeta + \psi}$ equal the coefficients of $g_i'^{\ord({\widetilde{g}_i'})}$ with respect to $z_1', \dots, z_{\zeta + \psi}'$. 

\item For every $i, j \in [\gamma + \eta]$, the coefficients of $[g_i, g_j]$ with respect to $z_1, \dots, z_{\zeta + \psi}$ equal the coefficients of $[g_i', g_j']$ with respect to $z_1', \dots, z_{\zeta + \psi}'$.

\item $\rho(g_i) = \rho'(g_i')$ for every $\gamma + 1 \leq i \leq \gamma + \eta$. 
\end{enumerate}
\end{lemma}
\begin{proof}
Suppose first that $f:G\to H$ is a group restriction isomorphism. Choose $z_{\zeta+1},\dots,z_{\zeta+\psi}\in Y$ such that
\[
Y=\oplus_{i=1}^{\psi}\langle z_{\zeta+i}\rangle,
\]
and choose $g_{\gamma+1},\dots,g_{\gamma+\eta}\in F$ such that
\[
\langle F,Z(G)\rangle/Z(G)=\oplus_{i=1}^{\eta}\langle {\widetilde{g}_{\gamma+i}}\rangle.
\]
Set $z_{\zeta+i}'=f(z_{\zeta+i})$ and $g_{\gamma+i}'=f(g_{\gamma+i})$. Since $f(Y)=X$, $f(F)=E$, and $f(Z(G))=Z(H)$, the first two conditions of the lemma follow. The order equalities follow because $f$ is an isomorphism. The third and fourth conditions follow from preservation of powers and commutators under an isomorphism, and the fifth condition is exactly the defining property of a group restriction isomorphism.

Conversely, suppose that the elements in the statement satisfy all five conditions. Consider the ordered generating systems
\[
z_1,\dots,z_{\zeta+\psi},g_1,\dots,g_{\gamma+\eta}
\]
and
\[
z_1',\dots,z_{\zeta+\psi}',g_1',\dots,g_{\gamma+\eta}'.
\]
By conditions (1) and (2), every element of $G$ and $H$ has a unique ordered representation with respect to these generators, with the same exponent ranges on the two sides. Define $f:G\to H$ by
\[
f\left(z_1^{\alpha_1}\cdots z_{\zeta+\psi}^{\alpha_{\zeta+\psi}}g_1^{\beta_1}\cdots g_{\gamma+\eta}^{\beta_{\gamma+\eta}}\right)
=z_1'^{\alpha_1}\cdots z_{\zeta+\psi}'^{\alpha_{\zeta+\psi}}g_1'^{\beta_1}\cdots g_{\gamma+\eta}'^{\beta_{\gamma+\eta}},
\]
where $0\leq\alpha_i<\ord(z_i)$ and $0\leq\beta_i<\ord({\widetilde{g}_i})$. The first four conditions of this lemma imply the hypotheses of Lemma~\ref{lem:iso_condition}, so $f$ is a group isomorphism.

It remains to verify the additional restriction conditions. By definition, $f(z_i)=z_i'$ for $i\leq\zeta$ and $f(g_i)=g_i'$ for $i\leq\gamma$. The first condition gives $f(Y)=X$.

Let $K_G$ and $K_H$ denote the subgroups $K$ defined in Definition~\ref{def:restricted_group} for $G$ and $H$ respectively. 
The induced isomorphism $G/Z(G)\to H/Z(H)$ for $f$ maps $K_G$ onto $K_H$ by $f(g_{\gamma + i}) = g'_{\gamma + i}$ for any $1 \leq i \leq \eta$. Since $f$ preserves element orders, it maps
\[
S=\{g\in G:{\widetilde{g}}\in K_G,\ \ord(g)\leq p^\tau\}
\]
onto the analogous set $S'$ for $H$. By condition (d) in the definition of a restriction, $F=\langle S\rangle$ and $E=\langle S'\rangle$. Hence $f(F)=E$.

Finally, let $g\in F$. Since the images of $g_{\gamma+1},\dots,g_{\gamma+\eta}$ form a direct-sum decomposition of $\langle F,Z(G)\rangle/Z(G)$, there are integers $m_1,\dots,m_\eta$ and an element $c\in F\cap Z(G)$ such that
\[
g=c\,g_{\gamma+1}^{m_1}\cdots g_{\gamma+\eta}^{m_\eta}.
\]
Condition (e) in the definition of a restriction gives $\rho(c)=0$, and condition (5), together with additivity, gives
\[
\rho(g)=\sum_{i=1}^{\eta}m_i\rho(g_{\gamma+i}).
\]
Since $f(c)\in E\cap Z(H)$, condition (e) for the restriction of $H$ gives $\rho'(f(c))=0$. Therefore
\[
\rho'(f(g))=\sum_{i=1}^{\eta}m_i\rho'(g_{\gamma+i}')=\sum_{i=1}^{\eta}m_i\rho(g_{\gamma+i})=\rho(g).
\]
Thus $f$ is a group restriction isomorphism.
\end{proof}


\subsection{Group decomposition}\label{sec:group_decomposition}

The main issue is not simply to find generators for a subgroup: a subgroup of an abelian $p$-group need not be a direct summand.  The following lemmas allow a bounded number of cyclic factors to be separated while keeping the remaining summand inside a prescribed subgroup.


\begin{lemma}\label{lem:direct_sum_decomposition}
Let $G$ be a finite abelian $p$-group. 
There are at most $|G|^2$ pairs $(g, U)$ with $g \in G$ and $U \leq G$ such that $G = \langle g \rangle \oplus U$ and $\rank(U) = \rank(G) - 1$. 
Furthermore, there is an algorithm to compute all these pairs in time $O(\poly(|G|))$. 
\end{lemma}

\begin{proof}
We first bound the number of different $(g, U)$ pairs by showing that there are at most $|G|$ different pairs for a fixed $g$. 
Suppose we have two subgroups $U, U' \leq G$ such that $G = \langle g \rangle \oplus U = \langle g \rangle \oplus U'$. 
We have $U \cong U'$. 
Let $f : U \rightarrow U'$ be the function such that $f(u) = b_u$, where $b_u$ is the element in $U'$ such that $u = a_u \cdot b_u$ for some $a_u \in \langle g \rangle$ and  $b_u \in U'$. 
The function $f$ must be injective, because otherwise there exist different $x, y \in U$ such that $f(x) = f(y)$, and consequently $xy^{-1} \in \langle g \rangle$, contradicting the fact that $xy^{-1}$ is a non-identity element of $U$, and $G = \langle g \rangle \oplus U$. 
$f$ is also surjective because $|U| = |U'|$, and thus $f$ is bijective. 
In addition, $f$ is a homomorphism as for any $u, u' \in U$, \[u \cdot u'= a_u \cdot b_u \cdot a_{u'} \cdot b_{u'} = (a_u \cdot a_{u'}) \cdot (b_u \cdot b_{u'})\] and thus \[f(u \cdot u') = b_u \cdot b_{u'} = f(u) \cdot f(u').\]
Hence, $f$ is an isomorphism between $U$ and $U'$.

Let $u_1, \dots, u_k \in G$ be arbitrary elements such that $U = \langle u_1 \rangle \oplus \dots \oplus \langle u_k\rangle$. 
Since $G$ is abelian, once $f(u_1), \dots, f(u_k)$ is determined, the entire $f$ is determined. 

On the other hand, $f(u_1), \dots, f(u_k)$ are determined if $a_{u_1}, \dots, a_{u_k}$ are determined. 
Let $\alpha_i$ denote the order of $u_i$. 
We have \[u_i^{\alpha_i} = (a_{u_i} \cdot b_{u_i})^{\alpha_i} = a_{u_i}^{\alpha_i} \cdot b_{u_i}^{\alpha_i} = 1.\] 
Since $a_{u_i} \in \langle g\rangle$, $b_{u_i} \in U'$, and $G = \langle g \rangle \oplus U'$, we have \[a_{u_i}^{\alpha_i} = b_{u_i}^{\alpha_i} = 1.\]
Since $G$ is a $p$-group, the order of $u_i$ must be a multiple of the order of $a_{u_i}$, and thus the number of possible different $a_{u_i}$ is upper bounded by the order of $u_i$. 
So, the number of different $f$ is upper bounded by the product of the order of $u_i$ for all the $i$, which is upper bounded by $|U| \leq |G|$. 
On the other hand, $U'=f(U)$ is determined by $f$, so there are at most as many $U'$ as there are $f$.
The bound is obtained by $|G|$ possible different $g$.

Here, we describe an algorithm to compute all the $(g, U)$ pairs for a given $g$, and the overall algorithm is obtained by enumerating all the possible $g$. 
To compute all the possible $U$, 
we first compute an arbitrary direct sum decomposition $G = \oplus_{i = 1}^{r} \langle g_i \rangle$ for a rank $r$ group $G$.
Then we find an $i \in [r]$ such that if $g_i$ is replaced by $g$, $G$ still decomposed by $g_1, \dots, g_{r}$. 
If there is no such an $i$, then there is no $(g, U)$ pair. 
Let $u_1, \dots, u_{r - 1}$ be $g_1, \dots, g_{i-1}, g_{i + 1}, \dots, g_r$, and $U = \oplus_{j = 1}^{r - 1} \langle u_j \rangle$. 
To compute all the other $U'$ such that $G = \langle g\rangle \oplus U'$, one only needs to compute $f(u_1), \dots, f(u_{r-1})$. 
The algorithm enumerates all the possible $f(u_1), \dots, f(u_{r-1})$, and verifies if each enumeration forms a direct sum decomposition of $G$ with $g$.

By Theorem~\ref{thm:group_decomposition}, replacing $g_i$ by $g$ for some $i$ is available if $G$ can be decomposed as $\langle g \rangle \oplus U$ for some $U$. 
Hence, the algorithm is correct. 
It takes an $O(\poly(|G|))$ time to compute an arbitrary direct sum decomposition by Smith normal form. Since there are $|G|$ possible $f(u_1), \dots f(u_k)$ as discussed above, 
the overall running time is $O(\poly(|G|))$. 
\end{proof}

By applying Lemma~\ref{lem:direct_sum_decomposition} recursively, we have the following corollary.

\begin{corollary}\label{cor:enumerate_decomposition}
Let $G$ be a finite abelian $p$-group. 
There are at most $|G|^{2k}$ tuples $(g_1, \dots, g_k, U)$ with $g_1, \dots, g_k \in G$ and $U \leq G$ such that $G = \langle g_1 \rangle \oplus \dots \oplus \langle g_k \rangle \oplus U$ and $\rank(U) = \rank(G) - k$. 
Furthermore, there is an algorithm to compute all these tuples in time $|G|^{O(k)}$. 
\end{corollary}

The following lemma suggests that given a subgroup $H$ of $G$ such that $G$ is generated by $H$ and a few other elements, 
we can find a direct sum of $G$ such that a subgroup of $H$ is a summand of the direct sum with a bounded number of other cyclic components. 

\begin{lemma}\label{lem:decomposition_subgroup}
Let $G$ be a finite abelian $p$-group of exponent $p^b$, and let $H\leq G$.
Suppose that $G=\langle H,g_1,\ldots,g_k\rangle$.
Then there exist $m\leq bk$ elements $x_1,\ldots,x_m\in G$ and a subgroup
$K\leq H$ such that
$G=\langle x_1\rangle\oplus\cdots\oplus\langle x_m\rangle\oplus K$.
\end{lemma}
\begin{proof}
By the classification of finite abelian $p$-groups, let
$
G=\oplus_{i=1}^b G_i$ with 
$G_i\cong (\mathbb Z/p^i\mathbb Z)^{r_i}$ for some $r_i$.
Let $H_i=H\cap G_i$. The natural map $G_i\to G/H$ has kernel $H_i$,
so $G_i/H_i$ embeds into $G/H$. Since $G/H$ is a finite abelian
$p$-group generated by at most $k$ elements, so is its subgroup
$G_i/H_i$.

By Smith normal form for subgroups of the homocyclic group $G_i$,
there is a basis $e_{i,1},\ldots,e_{i,r_i}$ of $G_i$ and integers
$0\leq a_{i,j}\leq i$ such that
\[
H_i=\oplus_{j=1}^{r_i}\langle p^{a_{i,j}}e_{i,j}\rangle.
\]
Consequently,
\[
G_i/H_i\cong
\oplus_{\substack{1\leq j\leq r_i, a_{i,j}>0}}
\mathbb Z/p^{a_{i,j}}\mathbb Z.
\]
It follows that at most $k$ of the $a_{i,j}$ are positive. Let
\[
C_i=\oplus_{a_{i,j}>0}\langle e_{i,j}\rangle \text{ and }
K_i=\oplus_{a_{i,j}=0}\langle e_{i,j}\rangle.
\]
Then $G_i=C_i\oplus K_i$, the group $C_i$ is a direct sum of at most
$k$ cyclic groups, and $K_i\leq H_i\leq H$.
Taking direct sums over $i$ gives
\[
G=\left(\oplus_{i=1}^b C_i\right)
\oplus \left(\oplus_{i=1}^b K_i\right). 
\]
Let  $K = \oplus_{i=1}^b K_i \leq K \leq H$. 
The first summand is a direct sum of at most $bk$ cyclic groups,
which proves the claim.
\end{proof}

The following lemma suggests that given a set of elements of $G$, one can decompose $G$ into a direct sum decomposition such that the given elements have non-zero projections only on a few cyclic components.

\begin{lemma}\label{lem:decomposition_subgroup_new}
Let $G$ be a finite abelian $p$-group of exponent $p^b$, and let $g_1, \dots, g_k$ be $k$ elements of $G$.
Then there exist $m \leq bk$ elements $x_1, \dots, x_m \in G$ and a subgroup
$K\leq G$ such that $G=\langle x_1\rangle\oplus\cdots\oplus\langle x_m\rangle\oplus K$
and $g_i \in \langle x_1\rangle\oplus\cdots\oplus\langle x_m\rangle$. 
\end{lemma}
\begin{proof}
We prove the case of $k = 1$. The general case can be obtained by first decomposing $G$ as $\langle x_1 \rangle \oplus \dots \oplus \langle x_m \rangle \oplus K$ with $m \leq b$ and $g_1 \in \langle x_1 \rangle \oplus \dots \oplus \langle x_m \rangle$, and then recurse on $K$ with $g_2', \dots, g_k'$, where 
$g_i'$ is the projection of $g_i$ onto $K$. 

Let $y_1, \dots, y_r$ be an arbitrary direct sum decomposition of $G$ with $\ord(y_1) \leq \dots \leq \ord(y_r)$. 
Let $0 \leq a_i < \ord(y_i)$ be the integer such that $g_1 = \prod_{i = 1}^{r} y_i^{a_i}$. 
Let $c_i$ be the $\log_p \ord(y_i^{a_i})$. 
We partition $[r]$ into $b$ groups such that \[C_j = \{i \in [r] : c_i = j\}\] for $1 \leq j \leq b$. 
For each $1 \leq j \leq b$ with $C_j \neq \emptyset$, let $i_j$ be the smallest integer such that in $C_j$.
By Theorem~\ref{thm:group_decomposition}, one can replace $y_{i_j}$ by another element $y_{i_j}' $ such that 
the new $y_1, \dots, y_r$ still form a direct sum decomposition of $G$, and the $a_i$ with respect to the new $y_i$ satisfies the following condition: 
\begin{enumerate}
\item $a_i$ does not change for every $i \notin C_j$. 
\item $a_{i_j}$ does not change. 
\item $a_i = 0$ for every $i \in C_j$ not equal to $i_j$.  
\end{enumerate}
Repeat above process for each $j \neq 0$, then we get the desirable decomposition. 
\end{proof}

\begin{lemma}\label{lem:order_refinement_final}
Let $G$ be a $p$-group of class 2, and $U$ be a subgroup of $G$ containing $Z(G)$ such that every element of $U / Z(G)$ has an order at most $p^\delta$ for some positive integer $\delta$. 
Let $F$ be the subgroup generated by all the elements of order at most $p^{m + \delta + 1}$ in $U$ for some positive integer $m$. 
Then every element of $F$ is of order at most $p^{m + \delta + 1}$, and 
there is an integer $k \leq (\log_p |G|) \cdot \delta / m$ such that $U$ is $\langle {\widetilde{u}_1} \rangle \oplus \dots \langle {\widetilde{u}_k} \rangle \oplus L$ such that $L \leq \langle F , Z(G) \rangle / Z(G)$. 
\end{lemma}

\begin{proof}

Let $q = p^{m+\delta+1}$. Since every element of $[U,G]$ has order at most $p^\delta$
and $p^\delta$ divides $\binom{q}{2}$, for all $x, y \in U$ we have
\[
(xy)^q = x^q y^q [y,x]^{\binom{q}{2}} = x^q y^q ,
\]
so $x \mapsto x^q$ is a homomorphism on $U$.
Write $Z(G) = \langle z_1 \rangle \oplus \dots \oplus \langle z_\ell \rangle$, where
$\ord(z_j) > p^m$ exactly for $j \leq t$. Then $t \leq (\log_p |G|)/m$.
For $j \in [\ell]$, let $\pi_j : Z(G) \to \langle z_j \rangle$ be the coordinate projection.

For every $x \in U$, we have $x^{p^\delta} \in Z(G)$. Hence $x^{p^{\delta+1}} \in Z(G)$, and
$x^q = \big(x^{p^{\delta+1}}\big)^{p^m}$.
Thus $\pi_j(x^q) = 1$ for every $j > t$, and consequently
\begin{equation}\label{eq:order_test}
x^q = 1 \iff \pi_j(x^q) = 1 \text{ for every } j \leq t .
\end{equation}

\begin{claim}\label{claim:order_refinement}
Let $1 \leq e \leq \delta$, and let $g_1, \dots, g_s \in U$ satisfy
$\ord({\widetilde{g}_i}) = p^e$ for every $i$ and
\[\langle g_1, \dots, g_s, Z(G)\rangle / Z(G) = \oplus_{i=1}^s \langle {\widetilde{g}_i} \rangle.\]
Then there are $h_1, \dots, h_s \in \langle g_1, \dots, g_s \rangle$ such that
\begin{enumerate}
\item $\oplus_{i=1}^s \langle {\widetilde{h}_i} \rangle = \oplus_{i=1}^s \langle {\widetilde{g}_i} \rangle$
and $\ord({\widetilde{h}_i}) = p^e$ for every $i$;
\item $h_i^q = 1$ for every $i > \min(s, t)$.
\end{enumerate}
\end{claim}
\begin{proof}
We process $j = 1, \dots, t$ in order. We maintain a list $h_1, \dots, h_s$ that
satisfies condition~1, together with an index $r_j \leq \min(s,t)$, such that
$\pi_{j'}(h_i^q) = 1$ for every $j' < j$ and every $i > r_j$.
Initially $h_i = g_i$ and $r_1 = 0$.

At step $j$, if $\pi_j(h_i^q) = 1$ for every $i > r_j$, set $r_{j+1} = r_j$.
Otherwise, choose $i_0 > r_j$ such that $\ord(\pi_j(h_{i_0}^q))$ is maximal among the
indices $i > r_j$. Since $\langle z_j \rangle$ is cyclic, maximality gives
$\pi_j(h_i^q) \in \langle \pi_j(h_{i_0}^q) \rangle$ for every $i > r_j$.
So there are integers $\alpha_i$ with $\pi_j(h_i^q) = \pi_j(h_{i_0}^q)^{\alpha_i}$.
For every $i > r_j$ with $i \neq i_0$, replace $h_i$ by $h_i h_{i_0}^{-\alpha_i}$.
Then swap $h_{i_0}$ with $h_{r_j + 1}$, and set $r_{j+1} = r_j + 1$.

Each replacement multiplies one generator by a power of another generator whose
coset has the same order $p^e$. This is an invertible change of basis that
preserves the orders of the cosets, so condition~1 is maintained.
Because $x \mapsto x^q$ is a homomorphism on $U$, every new $h_i$ satisfies
$\pi_j(h_i^q) = \pi_j(h_i^q)\,\pi_j(h_{i_0}^q)^{-\alpha_i} = 1$.
For $j' < j$, both factors already had trivial $\pi_{j'}$, so $\pi_{j'}(h_i^q) = 1$
still holds.

After step $t$, every $i > r_{t+1}$ satisfies $\pi_j(h_i^q) = 1$ for all $j \leq t$.
By~\eqref{eq:order_test}, this gives $h_i^q = 1$.
Since $r_{t+1} \leq \min(s,t)$, condition~2 holds.
\end{proof}

Fix a decomposition $U / Z(G) = \oplus_{i} \langle {\widetilde{u}_i} \rangle$.
For each $1 \leq e \leq \delta$, apply Claim~\ref{claim:order_refinement} to the $u_i$
with $\ord({\widetilde{u}_i}) = p^e$, replacing them by the resulting $h_i$.
This preserves the direct sum decomposition.

In each order class, at most $t$ of the new elements fail to satisfy $h^q = 1$.
Let $u_1, \dots, u_k$ be these elements, collected over all classes; then
$k \leq \delta t \leq (\log_p |G|) \cdot \delta / m$.
Every remaining element $h$ satisfies $h^q = 1$, so $h \in F$.
Their cosets therefore generate a subgroup $L \leq \langle F, Z(G) \rangle / Z(G)$, and
\[
U / Z(G) = \langle {\widetilde{u}_1} \rangle \oplus \dots \oplus \langle {\widetilde{u}_k} \rangle \oplus L .
\]
\end{proof}

\subsection{Tensor isomorphism for commutator map tensors}\label{sec:tensor_iso}

We apply the tensor isomorphism results to the commutator maps of two groups. As a result, we can assign each element $g$ (of a group) a label, denoted as $\rho(g)$ so that the label is additive.

Because $G$ has class $2$, its commutators are central and the commutator operation is bilinear on $G/Z(G)$.  After choosing cyclic coordinates, this operation can therefore be recorded by an alternating tensor.  The tensor describes commutators, but not by itself the central powers of the chosen lifts.

For a $p$-group  $G$ of class 2 and a positive integer $\delta$, 
$D_{p^\delta}(G)$ be the subgroup of $G$ generated by all the elements $g \in G$ such that ${\widetilde{g}} \in (G /  Z(G))_{p^\delta}$, where $ (G /  Z(G))_{p^\delta}$ is the subgroup of $G / Z(G)$ generated by all the elements in $G / Z(G)$ of order at most $p^\delta$. 
Suppose $(G / Z(G))_{p^\delta}$ is an abelian $p$-group that can be decomposed into a direct sum of $\gamma_0$ cyclic $p$-groups, and $Z(G)$ is an abelian $p$-group can be decomposed into a direct sum of $\zeta_0$ cyclic $p$-groups.

Let $g_1, \dots, g_{\gamma_0}$ be elements of $D_{p^\delta}(G)$ such that 
\[(G / Z(G))_{p^\delta} = \langle {\widetilde{g}_1} \rangle \oplus \dots \oplus \langle {\widetilde{g}_{\gamma_0}} \rangle \text{ with } \ord({\widetilde{g}_1}) \leq \dots \leq \ord({\widetilde{g}_{\gamma_0}}), \]
and $z_1, \dots, z_{\zeta_0}$ be elements of $Z(G)$ such that \[Z(G) = \langle z_1 \rangle \oplus \dots \langle z_{\zeta_0} \rangle \text{ with } \ord(z_1) \leq \dots \leq\ord(z_{\zeta_0}).\] 
The commutator tensor for $D_{p^{\delta}(G)}$ is an ${\gamma_0} \times {\gamma_0} \times {\zeta_0}$ alternating tensor over $R' = \Z / p^\delta \Z$ constructed as follows: for each $1 \leq i, j \leq {\gamma_0}$, suppose \[[g_i, g_j] =\prod_{k = 1}^{\zeta_0} z_k^{w_{i, j, k}}\] for some $0 \leq w_{i, j, k} < \ord(z_k)$. 
Then the tensor is $T = (t_{i, j, k})$ such that $t_{i, j, k} = w_{i, j, k}  \cdot p^\delta / \ord(z_k)$. 
Since $\ord({\widetilde{g}_i}) \leq p^\delta$, $w_{i, j, k} \cdot p^\delta$ is always a multiple of $\ord(z_k)$.

Before we prove the tensor isomorphism for tensors constructed from isomorphic groups, we first prove some useful lemma to establish the equivalence. 

\begin{lemma}\label{lem:inv_with_basis}
Let $G$ be an abelian $p$-group of rank $r$ whose exponent divides $p^\tau$. 
Let $g_1, \dots, g_r$ be elements of $G$ such that $G = \oplus_{i = 1}^r \langle g_i \rangle$. 
Let $W$ be the matrix $\diag( p^\tau / \ord(g_1), \dots, p^\tau / \ord(g_r))$ over $R$.  
Let $P, Q \in \M(r, R)$ be matrices such that $P  W   Q = W$. The following properties hold:
\begin{enumerate}
\item Let $\vec v = (v_1, \dots, v_r)^T = P (g_1, \dots, g_r)^T$. Then $G = \oplus_{i = 1}^r \langle v_i \rangle$ with $\ord(v_i) = \ord(g_i)$. 
\item $P$ and $Q$ are invertible.
\item For any $\vec s = (s_1, \dots, s_r)^T \in R^r$ such that  $s_i$ is a multiple of $p^\tau / \ord(g_i)$, we have
\[\prod_{i = 1}^{r} g_i^{s_i / (p^\tau / \ord(g_i))} = \prod_{i = 1}^r v_i^{t_i / (p^\tau / \ord(g_i))},\] where $(t_1, \dots, t_r)^T = Q^T \vec s $.
\end{enumerate}
\end{lemma}
\begin{proof}
For the first property, let $M$ be the $R$-module submodule spanned by the columns of $W$. 
Then the underlying abelian addition group of $M$ is isomorphic to $G$ with the $i$-th column of $W$ corresponding to $g_i$. 

Let $M'$ be the $R$-submodule spanned by the columns of $Q^TW$.
Since $Q^TWP^T=W$, we have $M\subseteq M'$. As $M$ is finite, $|M'| \leq |M|$, and hence $M'=M$.
On the other hand, $Q^T$ induces an isomorphism $f: M \rightarrow M'$ defined by $f(\vec w) = Q^T \vec w$. 

For $1\leq i\leq r$, let $\vec w_i$ denote the $i$-th column of $W$.
Then $M = \oplus_{i = 1}^r\langle \vec{w_i} \rangle = \oplus_{i = 1}^r \langle Q^T\vec{w}_i \rangle$ as an abelian group. 
Since for any $1 \leq i \leq r$, 
\[
\vec w_i=\sum_{j=1}^r P[i, j]Q^T\vec w_j,\]
$P$ is a transition matrix between two direct sum decompositions of the same abelian group. 
In addition, since $f$ is an isomorphism, the order of $w_i$ equals the order of $Q^T \vec w_i$ for the underlying abelian addition group of $M$. 
By the correspondence between $w_i$ and $g_i$, the first property holds.

Since $P$ is a transition matrix between two direct sum decompositions of the same abelian group, by Theorem~\ref{thm:group_decomposition}, $P$ is invertible. 
Using the same argument for $P W Q = W$, $Q^T$ is invertible, and thus $Q$ is invertible. 
So the second property holds.

To prove the third property, choose
$\vec a=(a_1,\ldots,a_r)^T$ such that
$\vec s=W\vec a$.
Since $P$ is invertible and $PWQ=W$, we have \[Q^TW=WP^{-T}.\]
Let $\vec b=P^{-T}\vec a$. Then
\[\vec t=Q^T\vec s=Q^TW\vec a=WP^{-T}\vec a=W\vec b.\]
Since $\vec v=P\vec g$, the change-of-generators identity gives \[\prod_{i=1}^r g_i^{a_i} = \prod_{i=1}^r v_i^{b_i}.\]
This proves the third property.
Here division by $p^\tau/\ord(g_i)$ is
interpreted modulo $\ord(g_i)$.
\end{proof}

\begin{lemma}\label{lem:inv_with_basis_new}
Let $G$ be an abelian $p$-group of rank $r$ whose exponent divides $p^\tau$, and  $\vec g = (g_1, \dots, g_r)^T$ be elements of $G$ such that $G = \oplus_{i = 1}^r \langle g_i \rangle$. 
Let $P$ be a matrix such that $\vec v = (v_1, \dots, v_r)^T := P \vec g$ is a direct sum decomposition of $G$ with $\ord(v_i) = \ord(g_i)$. 
Let $W = \diag(p^\tau / \ord(g_i))$. 
Then there is an invertible matrix $Q$ such that $PWQ = W$. 
\end{lemma}
\begin{proof}
Let $M=WR^r$
be the $R$-submodule of $R^r$ spanned by
the columns of $W$.
Since $\vec v=P\vec g$ is a direct-sum
decomposition of $G$ with
$\ord(v_i)=\ord(g_i)$ for every $i$,
Theorem~\ref{thm:group_decomposition}
implies that $P$ is invertible and
satisfies the corresponding divisibility
conditions.

These divisibility conditions imply
$P(M)\subseteq M$. Since $P$ is
invertible, its restriction to $M$ is
injective. As $M$ is finite, we obtain
$P(M)=M$.
For each $i\in[r]$, let $\vec w_i$
denote the $i$-th column of $W$.
Since $\vec w_i\in P(M)$, there exists
$\vec q_i\in R^r$ such that
\[ PW\vec q_i=\vec w_i.\]
Let \[Q=(\vec q_1,\ldots,\vec q_r).\]
Then $PWQ=W$. By
Lemma~\ref{lem:inv_with_basis},
$Q$ is invertible.
\end{proof}

We show that the tensors constructed from two isomorphic groups are isomorphic. 

\begin{lemma}\label{lem:commutator_tensor_iso}
Let $G$ and $H$ be two $p$-groups of class 2. 
Let $\mathcal{T}$ and $\mathcal{T}'$ be commutator tensors of $D_{p^\delta}(G)$ and $D_{p^\delta}(H)$ respectively. 
If $G$ and $H$ are isomorphic, then there exist invertible matrices $X$ and $Y$ such that $\mathcal{T}' = (X, X, Y) \circ \mathcal{T}$. 
\end{lemma}

\begin{proof}
Let $f$ be an arbitrary isomorphism from $G$ to $H$. 
Then $f(Z(G)) = Z(H)$ and $f(D_{p^\delta}(G)) = D_{p^\delta}(H)$. 
Suppose $\mathcal{T}$ is constructed from \[g_1, \dots, g_{\gamma_0} \in D_{p^{\delta}}(G) \text{ and }z_1, \dots, z_{\zeta_0} \in Z(G),\] and $\mathcal{T}'$ is constructed from \[h_1, \dots, h_{\gamma_0} \in D_{p^{\delta}}(H) \text{ and } v_1, \dots, v_{\zeta_0} \in Z(H).\]

Let $X$ be the matrix in $\M(\gamma_0, R')$ such that \[h_i = z \cdot \prod_{j = 1}^{\gamma_0} f(g_j)^{X[i, j]}\] for some $z \in Z(H)$.
By Theorem~\ref{thm:group_decomposition}, $X$ is invertible. 
By the construction of the tensor, \[(X, X, I) \circ \mathcal{T} = \mathcal{T}'',\] where $\mathcal{T}''$ is the tensor constructed from 
$h_1, \dots, h_{\gamma_0}$ and $f(z_1), \dots, f(z_{\zeta_0})$.

Now we show that for two tensors $\mathcal{T}'$ and $\mathcal{T}''$ constructed from different direct sum decompositions of $Z(H)$, there is an invertible matrix $Y$ such that $(I, I, Y)\circ \mathcal{T}'' = \mathcal{T}'$. 

Let $s$ be an arbitrary integer such that $\delta \leq s$ and every element of $Z(G)$ is of order at most $p^s$. 
Define $\gamma_0 \times \gamma_0 \times \zeta_0$ tensors $\mathcal{M}'$ and $\mathcal{M}''$ over $\Z / p^s \Z$ such that the $(i, j, k)$-th entry of  $\mathcal{M}'$ equals $(i, j, k)$-th entry of  $\mathcal{T}'$ multiplying $p^{s - \delta}$, and  
the $(i, j, k)$-th entry of  $\mathcal{M}''$ equals $(i, j, k)$-th entry of  $\mathcal{T}''$ multiplying $p^{s - \delta}$. 
Then $\mathcal{M}'$ and $\mathcal{M}''$  correspond to tensors constructed from  $v_1, \dots, v_{\zeta_0}$  and $f(z_1), \dots, f(z_{\zeta_0})$ respectively. 

Let $U$ be the matrix over $\Z / p^s \Z$ such that \[v_i = \prod_{j = 1}^{\zeta_0} f(z_j)^{U[i, j]},\] and 
$Q$ be the matrix over $\Z / p^s \Z$ such that \[Q := \diag(p^s / \ord(z_1), \dots, p^s / \ord(z_{\zeta_0})).\] 
Let $Y_0$ be the invertible matrix over $\Z / p^s \Z$ such that $U Q Y_0^T = Q$ promised by Lemma~\ref{lem:inv_with_basis_new}. 
We have \[(I, I, Y_0) \circ \mathcal{M}'' = \mathcal{M}'\] by the construction of $\mathcal{M}', \mathcal{M}''$, and Lemma~\ref{lem:inv_with_basis}.
By the correspondence between $\mathcal{T}', \mathcal{T}''$ and $\mathcal{M}', \mathcal{M}''$, 
let $Y$ be the matrix over $R'$ obtained by $Y_0$ modulo $p^{\delta}$.
$Y$ is still invertible, and satisfies 
$(I, I, Y) \circ \mathcal{T}'' = \mathcal{T}'$. 
Hence, $(X, X, Y) \circ \mathcal{T} = \mathcal{T}'$. 
\end{proof}

The purpose of the labeling algorithm is to make the subsequent search uniform over all possible group isomorphisms.  Each output pair is a candidate for the $\rho$ function of the group elements. 
In addition, the algorithm also identify elements whose commutators can be confined to the exceptional central summand.

\begin{framed}
\vspace{.2cm} \noindent \textbf{Labeling Algorithm}

\noindent \textbf{Input:} $p$-groups $G$ and $H$ of class 2, and positive integer $\delta$. 

\noindent \textbf{Output:} $\{(\lambda_i, \lambda_i')\}$ with \[\lambda_i: D_{p^\delta}(G)\rightarrow (\Z / p^{\delta} \Z)^{\kappa_i}  \text{ and }\lambda_i': D_{p^\delta}(H)\rightarrow (\Z / p^{\delta} \Z)^{\kappa_i}\] for some positive integer $\kappa_i$.
\begin{enumerate}
\item Construct the commutator tensors $\mathcal{T}$ for $D_{p^\delta}(G)$ from elements $g_1, \dots, g_{\gamma_0}$ and $z_1, \dots, z_{\zeta_0}$, and $\mathcal{T}'$ for $D_{p^\delta}(H)$ from elements $h_1, \dots, h_{\gamma_0}$ and $v_1, \dots, v_{\zeta_0}$. 
\item  Run the tensor isomorphism algorithm (Theorem~\ref{thm:tensor_iso}), and for each returned $U, U', W, W', P$, and $P'$, construct function $\lambda$ and $\lambda'$ as follows:
\begin{enumerate}
\item For each $g \in D_{p^\delta}(G)$, let $\lambda(g) = P U^{-T} (c_1, \dots, c_{\gamma_0})^T$
where $0 \leq c_i <\ord({\widetilde{g}_i})$ are  integers such that \[g = \prod_i g_i^{c_i} \cdot g' \text { with } g' \in Z(G).\] 
\item For each $h \in D_{p^\delta}(H)$, let $\lambda'(h) = P' U'^{-T} (c_1, \dots, c_{\gamma_0})^T$ 
where $0 \leq c_i < \ord({\widetilde{h}_i})$ are  integers such that \[h = \prod_i h_i^{c_i} \cdot h' \text{  with } h' \in Z(H). \]
\end{enumerate}
\item Return all the pairs $(\lambda, \lambda')$ constructed. 
\end{enumerate}

\end{framed}

\begin{lemma}\label{lem:labeling_function}
Let 
$G$ and $H$ be two $p$-groups of class 2 with $Z(G) \cong Z(H)$ and $G / Z(G) \cong H / Z(H)$. 
Let $R' = \Z / p^\delta \Z$. 
There is a labeling algorithm with running time \[p^{O( (\log_p |G|)^{17 / 9} \cdot (\log p)^{8/9})}\] to create a set $\{(\lambda_i, \lambda'_i) \}$ of \[p^{O(( \log_p |G|)^{17 / 9} \cdot (\log p)^{8/9})}\] pairs with $\lambda_i : D_{p^\delta}(G) \rightarrow (R')^{\kappa_i}$ and $\lambda_i' : D_{p^\delta}(H) \rightarrow (R')^{\kappa_i}$ for some integer $\kappa_i \leq (\log_p |G|)^2$ satisfying the following conditions:
\begin{enumerate}
\item $\lambda_i(g g') = \lambda_i(g) + \lambda_i(g')$ 
for any $g, g' \in D_{p^\delta}(G)$
and $\lambda_i'(h h') = \lambda_i'(h) + \lambda_i'(h')$ for any $h, h' \in D_{p^\delta}(H)$.
\item 
For each $i$, there is a subgroup $J_i \leq D_{p^\delta}(G)$ satisfying the following conditions:
\begin{enumerate}
\item $Z(G) \leq J_i$. 
\item $D_{p^\delta}(G)$ is generated by $J_i$ and at most $O(\delta^2 \cdot ( \log_p |G|)^{4/9} \cdot (\log p)^2 )$ other elements of $D_{p^\delta}(G)$. 
\item The group generated by $[g, J_i]$ for all the $g \in D_{p^\delta}(G)$ with $\lambda_i(g) = 0$ has a generating set of size at most $O(\delta^2 \cdot ( \log_p |G|)^{4/9} \cdot (\log p)^2 )$. 
\end{enumerate}
\item If $G$ and $H$ are isomorphic, then there is a pair $(\lambda_i, \lambda_i')$ and an isomorphism $f$ between the two groups such that \[f(D_{p^\delta}(G)) = D_{p^\delta}(H) 
\text{ and } \lambda_i(g) = \lambda_i'(f(g))\] for every $g \in D_{p^\delta}(G)$.
\end{enumerate}
\end{lemma}

\begin{proof}
By the construction of the commutator tensor, 
we have $c_{\mathcal{T}, 0} \leq \log_p |G / Z(G)|$, and $c_{\mathcal{T}, 1} \leq \log_p |Z(G)|$,
and thus 
$c_\mathcal{T}$ is upper bounded by $\log_p |G|$. 

For the first condition, for any $g, g'\in D_{p^\delta}(G)$, let $c_{g, 1}, \dots, c_{g, \gamma_0}$ and $c_{g', 1}, \dots, c_{g', \gamma_0}$ be integers so that 
\[\lambda(g) = PU^{-T} (c_{g, 1}, \dots, c_{g, \gamma_0})^T \text{ and }\lambda(g') = PU^{-T} (c_{g', 1}, \dots, c_{g', \gamma_0})^T.\]
Let $c_i$ be $(c_{g, i} + c_{g', i}) \pmod{\ord({\widetilde{g}_i})}$. 
We have \[\lambda(g g') = PU^{-T} (c_1, \dots, c_{\gamma_0})^T.\]
To prove the first condition, it is sufficient to show that 
\begin{equation}\label{equ:labeling_function}PU^{-T}(c_{g, 1} + c_{g', 1}, \dots, c_{g, \gamma_0} + c_{g', \gamma_0})^T = PU^{-T} (c_1, \dots, c_{\gamma_0})^T.\end{equation}
Since $c_{g, i} + c_{g', i} \equiv c_i \pmod{\ord({\widetilde{g}_i})}$, it is sufficient to show that $PU^{-T} \vec a_i = 0$ with 
\[\vec a_i = (0, \dots, 0, \ord({\widetilde{g}_i}), 0, \dots, 0)^T \text{ for every }i.\] 
Since $[g_i,g_j]^{\ord({\widetilde{g}_i})}=1$, it follows that 
$w_{i,j,k}\ord({\widetilde{g}_i})$ is a multiple of $\ord(z_k)$ for every $j$ and $k$, and therefore $\ord({\widetilde{g}_i})\,t_{i,j,k}\equiv0\pmod{p^\delta}$.
Thus, $N_{\mathcal{T}, \vec a_i} = 0$, and consequently 
\[N_{(U, U, I) \circ \mathcal{T}, U^{-T}\vec a_i} = 0, \text{ and thus } N_{(U, U, W) \circ \mathcal{T}, U^{-T}\vec a_i} = 0.\] 
Hence, $P U^{-T} \vec a_i = 0$ for every $i$ by the definition of restricted tensor isomorphism, and thus Equation~\eqref{equ:labeling_function} holds. 
The same argument also applies to $\lambda'$, and thus the first condition holds.


For the second condition, for each $i$, 
let \[(u_1, \dots, u_{\gamma_0})^T = U^{-T} (g_1, \dots, g_{\gamma_0})^T,\] and $J_i$ be the group generated by $u_{w + 1}, \dots u_{{\gamma_0}}$ and $Z(G)$. 
By the correspondence between the group and its commutator tensor and Theorem~\ref{thm:tensor_iso},  $J_i$ satisfies the second condition. 

For the last condition, 
we show that if $G$ and $H$ are isomorphic, then there exists a constructed pair $(\lambda_i, \lambda_i')$ satisfying the requirements of the lemma. 
Suppose $f$ is an isomorphism between the two groups. 
By Lemma~\ref{lem:commutator_tensor_iso}, there exist invertible matrices $X$ and $Y$ such that $(X, X, Y) \circ \mathcal{T} = \mathcal{T}'$. 
By Theorem~\ref{thm:tensor_iso}, 
there are  $U, U', W, W', P$, and $P'$ produced by the algorithm such that 
$U' X U^{-1}$ and $W' Y W^{-1}$ form a solution for the restricted alternating tensor isomorphism problem with  \[(U, U, W) \circ \mathcal{T}, (U', U', W') \circ \mathcal{T}', P, \text{ and }P'\] as input. 
By the correspondence between the group and its commutator tensor, 
if $\lambda_i$ and $\lambda_i'$ are constructed with these $U, U', W, W', P$ and $P'$, then we have $\lambda_i(g) = \lambda_i'(f(g))$.

The running time of the algorithm and the bound on the number of the pairs constructed is obtained by Theorem~\ref{thm:tensor_iso}. 
\end{proof}

\subsection{Reduce group isomorphism to restricted group isomorphism}\label{sec:restriction_reduction}
We first give an algorithm to construct a group restriction.

\begin{framed}
\noindent \textbf{Group Restriction Construction Algorithm}

\noindent \textbf{Input}: $p$-group $G$ of class 2 of order $p^n$ and function $\lambda: D_{p^\delta}(G) \rightarrow (\Z / p^\delta \Z)^d$, and a parameter $\tau$ such that $\tau \geq \delta + 1$.

\noindent \textbf{Output}: A set of group restrictions.
\begin{enumerate}
\item Let $\Gamma = \emptyset$, 
and $\rho : D_{p^\delta}(G) \rightarrow R^d$ s.t. $\rho(h) = \lambda(h) \cdot p^{\tau - \delta}$. 

\item For each $g_1, \dots, g_\gamma \in G$, $K \leq (G / Z(G))_{p^\delta}$, $z_1, \dots, z_\zeta \in Z(G)$, and $Y \leq Z(G)$ with both $\gamma$ and $\zeta$ being $O(n / \delta + n \delta / \tau + \delta^3 \cdot n^{4/9} \cdot (\log p)^2)$ s.t. (a), (b), and (d) of Definition~\ref{def:restricted_group}  are satisfied, 

\begin{enumerate}
\item Let $H \leq G$ be the subgroup generated by all the $g$ with ${\widetilde{g}} \in K$ and $\ord(g) \leq p^{\tau}$. 
\item 
If Definition~\ref{def:restricted_group} is satisfied,  add  $(z_1, \dots, z_\zeta, Y, g_1, \dots, g_\gamma, H, \rho)$ to $\Gamma$ (with $\rho$ restricted to $H$) .  

\end{enumerate}
\item Return $\Gamma$.
\end{enumerate}
\end{framed}

\begin{lemma}\label{lem:restricted_group_construction}
For a $p$-group $G$ of class 2 with the order $p^n$, two positive integers $\delta, \tau$ with $\tau \geq  \delta + 2$, and a function $\lambda : D_{p^\delta}(G) \rightarrow (\Z / p^\delta \Z)^d$ for some $d = \mathrm{poly}(n)$ satisfying the second condition of Lemma~\ref{lem:labeling_function}, 
the restricted group construct algorithm outputs all the group restrictions $(z_1, \dots, z_\zeta, Y, g_1, \dots, g_\gamma, H, \rho)$ for $G$ with \[\zeta, \gamma= O(n / \delta + n \delta / \tau+ \delta^3 \cdot n^{4/9} \cdot (\log p)^2), 
\ \ \ d = \poly(n),\] and 
$\rho$ obtained from $\lambda$ by extending to the proper domain. 
Furthermore, the running time of the algorithm is \[p^{O(n \cdot (n / \delta + n \delta / \tau + \delta^3 \cdot n^{4/9} \cdot (\log p)^2))}\] and the number of group restrictions returned by the algorithm is at most $p^{O(n \cdot (n/ \delta +n \delta / \tau + \delta^3 \cdot n^{4/9} \cdot (\log p)^2))}$.
\end{lemma}

\begin{proof}
We first prove the existence of the required group restriction  of the lemma. 
We show that there are $g_1, \dots, g_\gamma$ and $K \leq (G / Z(G))_{p^\delta}$ satisfying the following conditions:
\begin{enumerate}
\item $G / Z(G) = \langle {\widetilde{g}_1} \rangle \oplus \dots \oplus \langle {\widetilde{g}_\gamma} \rangle \oplus K$. 
\item For every ${\widetilde{g}} \in K$, there is a $g' \in {\widetilde{g}}$ s.t. $\ord(g') \leq p^\tau$.
\item The group generated by $[g, h]$ for all the $g, h \in G$ with ${\widetilde{g}}, {\widetilde{h}} \in K$ and $\lambda(g) = 0$ has a generating set of size at most $O(\tau^3 \cdot n^{4/9} \cdot (\log p)^2)$. 
\end{enumerate}
Take an arbitrary $P \leq G / Z(G)$ and $g_1, \dots, g_k$ 
 such that $G / Z(G) = \oplus_{i = 1}^{k} \langle {\widetilde{g}_i} \rangle \oplus P$, where $k$ is the number of elements of order greater than $p^\delta$ for an arbitrary direct sum decomposition of $G/ Z(G)$. 
 We have $k \leq \log_p |G| / \delta$. 
Such a $P$ can be obtained by taking an arbitrary direct sum decomposition of $G/ Z(G)$ and letting $P$ be the group generated by elements of order at most $p^\delta$ within the quotient group.

Let $J \leq D_{p^e}(G)$ be a subgroup satisfying the second condition of Lemma~\ref{lem:labeling_function}. 
Let $Q = (J / Z(G)) \cap P$. 
Then $P$ can be generated by $Q$ and at most other \[O(\delta^2 \cdot ( \log_p |G|)^{4/9} \cdot (\log p)^2 )\] elements in $P$ by Lemma~\ref{lem:labeling_function}.
By Lemma~\ref{lem:decomposition_subgroup}, there exist $g_{k + 1}, \dots, g_{k + \ell} \in G$ 
and $K' \leq Q$
such that $P =\oplus_{i = 1}^\ell \langle {\widetilde{g}_{k + i}}\rangle \oplus K'$ with $\ell = O(\delta^3 \cdot ( \log_p |G|)^{4/9} \cdot (\log p)^2 )$.
Finally, we obtained $K$ by decomposing $K'$ into $\oplus_{i = k + \ell + 1}^\gamma \langle {\widetilde{g}_{i}} \rangle \oplus K$ such that for every ${\widetilde{g}} \in K$, there is a $g' \in {\widetilde{g}}$ such that $\ord(g) \leq p^\tau$. 
By Lemma~\ref{lem:order_refinement_final}, $\gamma - k - \ell$ can be upper bounded by $\log_p |G|\cdot  \delta / \tau$. 



Once $K$ is determined, $F$ is determined by Definition~\ref{def:restricted_group}.
Next, we show that there are $z_1, \dots, z_{\zeta}$ and $Y$ satisfying the following conditions:
\begin{enumerate}
\item $Z(G)  = \langle z_1\rangle \oplus \dots \oplus \langle z_\zeta \rangle \oplus Y$. 
\item The group generated by $[g, F]$ for all $g \in F$ with $\lambda(h) = 0$ is a subgroup of $\langle z_1\rangle \oplus \dots \oplus \langle z_\zeta \rangle$. 
\item All the elements of $Y$ are of order at most $p^\tau$. 
\end{enumerate}
We first decompose $Z(G)$ as $\langle z_{1} \rangle \oplus \dots \oplus \langle z_s \rangle \oplus Y'$ such that every element of $Y'$ is of order at most $p^\delta$ for some $s = O(n / \delta)$. 
Such a $Y'$ can be obtained by taking an arbitrary direct sum decomposition of $Z(G)$ and letting $Y'$ be the group generated by elements of order at most $p^\delta$ in the decomposition. 
Since  the group generated by $[g, F]$ for all $g \in F$ with $\lambda(g) = 0$ has a generating set of size at most \[O(\delta^2 \cdot (\log_p |G|)^{4/9} \cdot (\log p)^2),\] 
by Lemma~\ref{lem:decomposition_subgroup_new}, 
we then decompose $Y'$ by $\langle z_{s+1}\rangle  \oplus \dots  \oplus \langle z_\zeta \rangle \oplus Y$ for some $\zeta = O(n / \delta + \delta^3 \cdot (\log_p |G|)^{4/9} \cdot (\log p)^2)$ such that $[g, F]\leq \langle z_1\rangle  \oplus \dots  \oplus \langle z_\zeta \rangle$.


Then the correctness of the algorithm is obtained by the description of the algorithm. 
By Corollary~\ref{cor:enumerate_decomposition}, the number of group restrictions returned by the algorithm is at most 
\[p^{O(n \cdot (n/ \delta +n \delta / \tau + \delta^3 \cdot n^{4/9} \cdot (\log p)^2))}.\] 
The running time of the algorithm is obtained by Corollary~\ref{cor:enumerate_decomposition}. 
\end{proof}

Finally, we present our group isomorphism algorithm to reduce the group isomorphism problem to restricted group isomorphism.
\begin{framed}
\noindent \textbf{Group Isomorphism Algorithm}

\noindent \textbf{Input}: Two $p$-groups $G$ and $H$ of class 2, and parameter 

\noindent \textbf{Output}: Yes or no. 

\begin{enumerate}
\item Return no if $|G|\neq |H|$, $Z(G) \ncong Z(H)$, or $G/Z(G) \ncong H / Z(H)$. 
\item Run the labeling algorithm to get functions $(\lambda_i, \lambda_i')$. 
\item For every $(\lambda_i, \lambda_i')$, 
\begin{enumerate}
\item Run the restricted group construction algorithm for $G$ with $\lambda_i$ and for $H$ with $\lambda_i'$. Let $\Gamma_1$ and $\Gamma_2$ denote the output for $G$ and $H$ respectively.
\item For every two group restrictions from $\Gamma_1$ and $\Gamma_2$ respectively, 
if the parameters of the two restrictions are the same, $|Y| = |X|$, and $|F| = |E|$, then run the restricted group isomorphism algorithm for the two group restrictions and return yes if the output is yes. 
\end{enumerate}
\item Output no. 
\end{enumerate}
\end{framed}

We prove the following lemma for the correctness of the algorithm. 

\begin{lemma}\label{lem:group_iso_reduction}
Given two $p$-groups of class 2 with order $p^n$ and two positive integers $\delta, \tau$ such that $\tau \geq \delta + 2$, there is an algorithm to reduce the group isomorphism problem for the two groups to  \begin{equation}\label{equ:time_bound_reduction} p^{O(n^{17 / 9} \cdot (\log p)^{8/9} + n \cdot (n / \delta + n \delta / \tau + \delta^3 \cdot n^{4/9} \cdot (\log p)^2))}\end{equation}instances of restricted group isomorphism with parameters 
$\zeta, \gamma = O(n / \delta + n \delta / \tau + \delta^3 \cdot n^{4/9} \cdot (\log p)^2)$ and $\kappa = \poly(n)$. 
The running time of the algorithm is also \eqref{equ:time_bound_reduction}. 
\end{lemma}

\begin{proof}

We first prove the correctness of the algorithm. 
If the two groups are not isomorphic, then the algorithm always outputs no, as one necessary condition for the group restriction isomorphism algorithm outputting yes is the two groups are isomorphic. 
If the two groups are isomorphic, then 
by Lemma~\ref{lem:labeling_function}, there is a pair $(\lambda, \lambda')$ such that the three conditions of Lemma~\ref{lem:labeling_function} are satisfied. 
Let $f$ be an isomorphism between $G$ and $H$ satisfying the last condition of Lemma~\ref{lem:labeling_function}. 
By Lemma~\ref{lem:restricted_group_construction}, 
for one group restriction $(z_1, \dots, z_\zeta, Y, g_1, \dots, g_\gamma, F, \rho)$ constructed from $\lambda$ and $G$, 
then $(z_1', \dots, z_\zeta', X, g_1', \dots, g_\gamma', E, \rho')$ for $H$ with 
\[z_i' = f(z_i), X = f(Y), g_i' = f(g_i), E = f(F),\] and $\rho'$ restricted to $E$ is a valid group restriction for $H$. 
Furthermore, $\rho(g) = \rho'(f(g))$ for any $g \in F$. 
By Lemma~\ref{lem:restricted_group_construction}, $(z_1', \dots, z_\zeta', X, g_1', \dots, g_\gamma', E, \rho')$ is produced when running the group restriction construction algorithm for $\lambda'$ and $H$. 
Hence, 
when running group restriction isomorphism algorithm for $(z_1, \dots, z_\zeta, Y, g_1, \dots, g_\gamma, H, \rho)$ and $(z_1', \dots, z_\zeta', X, g_1', \dots, g_\gamma', E, \rho')$, the output is yes, and then the overall algorithm output yes. 
Hence, if two groups are isomorphic, the algorithm outputs yes. 
If two groups are not isomorphic, then by the definition of restricted group isomorphism, no two group restrictions are restriction isomorphic. So the algorithm always outputs no. 

The bound on the number of restricted group isomorphism problem instances and the running time of the algorithm is obtained by Lemma~\ref{lem:labeling_function} and Lemma~\ref{lem:restricted_group_construction}. 
\end{proof}


\section{Restricted group isomorphism}

In this section, we present an algorithm to solve the group restriction isomorphism problem and prove the following theorem.

\begin{theorem}\label{thm:restricted_group_iso_algo}
For two group restrictions of two finite $p$-groups of class 2 with order $p^n$ for some positive integer $n$, 
there is an algorithm to solve the group restriction isomorphism problem in time 
\[p^{O\left(    \max\left\{ \tau n^{7 / 4} \cdot \delta^{7/ 2} \log(\tau \cdot n) ,  \tau \cdot n^{3/2} \delta^7 \log(\tau \cdot n),  n\cdot \zeta \right\}\right)}.\]
\end{theorem}

Together with Lemma~\ref{lem:group_iso_reduction}, Theorem~\ref{thm:main_intro} is obtained by taking $\delta = \lceil n^{1/ 26} \rceil$ and $\tau = \lceil  n^{1 / 13} \rceil$. 

In Section~\ref{sec:representative_vector}, we first define the representative vector for a group restriction and give an algorithm to construct a representative vector. 
In Section~\ref{sec:extended_vector}, we define the extended representative vector by adding some auxiliary group elements so that transformations on the representative vector are compatible to our constraints. 
In Section~\ref{sec:encoding}, we encode the information of an extended representative vector by a matrix tuple. 
In Section~\ref{sec:doubling}, we construct another matrix tuple to ensure certain conditions of the solution of the matrix tuple isometry so that we can extract the required transformations for the representative vector. 
In Section~\ref{sec:group_restriction_iso}, we solve the restricted group isomorphism problem by an orbit computation.

\subsection{Representative vector}\label{sec:representative_vector}

We define the representative vector. 
A representative vector provides a common coordinate system for the variable part of the restriction.  Its central coordinates distinguish the $Y$-part from specified central generators, while its noncentral coordinates describe the quotient subgroup.  Allowing controlled central modifications is useful because different lifts of the same quotient element need not have identical powers.

The intuition of the representative vector is to encode a direct sum decomposition of $Y$, denoted as $y_1, \dots, y_\psi$, 
and other elements from $F$, denoted as $s_1, \dots, s_\eta$ such that $\langle F, Z(G) \rangle / Z(G) = \oplus_{i = 1}^\zeta \langle \widetilde s_i \rangle$. 
However, since we want to preserve some central power information regarding the $Z = \oplus_{i = 1}^\zeta \langle z_i \rangle$ component, we also need to add some elements from $Z$, and give some freedom of these elements in the representative vector.

Let $\zeta_0$ be the number of elements in $z_1, \dots, z_\zeta$ of order greater than $p$. 
After reindexing if necessary, $z_1, \dots, z_{\zeta - \zeta_0}$ are of order $p$. 

\begin{definition}[Representative vector]\label{def:rep_vec}

For a group restriction $(z_1, \dots, z_\zeta, Y, g_1, \dots, g_\gamma, F, \rho)$ of a $p$-group of class 2 with parameters $\zeta, \gamma, \tau$ and $\kappa$, 
a vector 
\[\vec {r} = \begin{pmatrix}r_1 \\ \vdots \\ r_{\zeta_0 + \psi + \eta} \end{pmatrix}\]
 is a representative vector (with respect to the group restriction) if the following conditions are satisfied: 
\begin{enumerate}
\item For $1 \leq i \leq \psi$, $r_i$ equals $y_i \cdot y_{i}'$ such that 
$Y = \oplus_{i = 1}^\psi \langle y_{i} \rangle$ with $\ord(y_1) \leq \dots \leq \ord(y_{\psi})$,
and $y_i'$ is an element in \[\oplus_{i =\zeta - \zeta_0 + 1}^\zeta \langle z_i\rangle\] of order at most $p$. 
\item $r_{\psi + i}$ is $x_{\psi + 1} \cdot x_{\psi + 2}$, where 
$x_{\psi + 1}$ is 
a power of \[z_{i + (\zeta - \zeta_0)}^{\max\{1,  \ord(z_{i + \zeta - \zeta_0}) / p^\tau\}}
\text{ such that } \ord(r_{\psi + i}) = \ord(z_{i + (\zeta - \zeta_0)}^{\max\{1,  \ord(z_{i + \zeta - \zeta_0}) / p^\tau\}})\]
for every $i \in [\zeta_0]$, and $x_{\psi + 2}$ is an element in $\oplus_{i =\zeta - \zeta_0 + 1}^\zeta \langle z_i\rangle$ of order at most $p$. 
\item $\langle F,  Z(G) \rangle / Z(G) = \oplus_{i = 1}^\eta \langle {\widetilde{r}_{\psi + \zeta_0 + i}} \rangle$ with $\ord({\widetilde{r}_{\psi + \zeta_0 + 1}}) \leq  \dots \leq \ord({\widetilde{r}_{\psi + \zeta_0 +  \eta}})$. 
\item The vector \[\begin{pmatrix} \rho\left(r_{\psi + \zeta_0 + 1}^{p^{\tau - 1}}\right) \\ \vdots  \\ \rho\left(r_{\psi + \zeta_0 + 1}\right)  \\ \vdots \\  \rho\left(r_{\psi +  \zeta_0 + \eta}^{p^{\tau - 1}}\right) \\ \vdots \\ \rho(r_{\psi + \zeta_0 +\eta})\end{pmatrix}  \in R^{\eta \cdot \tau \cdot \kappa}\]  is lexicographically smallest among all the possible $r_1, \dots, r_{\psi + \zeta_0 + \eta}$ satisfying the first three conditions. 
\end{enumerate}
\end{definition}

We prove some basic properties of the representative vector.

\begin{fact}\label{fact:representative_vector_to_denote_element}
Let $\vec r$ be a representative vector.
Then every $g\in F$ can be written as
\[
g=\prod_{i=1}^{\psi+\zeta_0+\eta}r_i^{\alpha_i} \cdot z 
\]
for some integers
$0\leq\alpha_i<p^\tau$ and $z \in \langle z_1 \rangle \oplus \dots \langle z_{\zeta - \zeta_0}\rangle $.
\end{fact}
\begin{proof}
By Definition~\ref{def:rep_vec},
there are integers
$\alpha_{\psi+\zeta_0+1},\dots,
\alpha_{\psi+\zeta_0+\eta}$ such that
\[
{\widetilde{g}}=
\prod_{i=\psi+\zeta_0+1}^{\psi+\zeta_0+\eta}
{\widetilde{r}_i^{\alpha_i}}.
\]
The difference belongs to $F\cap Z(G)$,
which is generated by
$r_1,\dots,r_{\psi+\zeta_0}$ and $z_1, \dots, z_{\zeta - \zeta_0}$.
The fact follows.
\end{proof}

The minimality condition in the representative-vector definition has an algebraic consequence: a zero-label element cannot use a nonzero-label coordinate of the same order with a unit coefficient. This restriction later becomes a triangularity condition on compatible matrices.

\begin{lemma}\label{lem:representative_vector_decomposition}
For a representative vector $\vec r$ for a group restriction $(z_1, \dots, z_\zeta, Y, g_1, \dots, g_\gamma, F, \rho)$ of a group $G$ with parameter $\tau$ (and other parameters),
let $g$ be an element of $F$ with $\rho(g) = 0$ and $g \notin Z(G)$. 
Let $\alpha_{1}, \dots, \alpha_{\psi + \zeta_0 + \eta}$ be integers such that $g = z \cdot \prod_{i = 1}^{\psi + \zeta_0 +\eta} r_{i}^{\alpha_{i}}$ for some $z \in Z_0$. 
For any  $\psi + \zeta_0 + 1 \leq i \leq \psi + \zeta_0 + \eta$ such that $\ord({\widetilde{r}_{i}}) = \ord({\widetilde{g}})$, either $\rho(r_{i}) = 0$ or $\alpha_{ i}$ is a multiple of $p$.

\end{lemma}

\begin{proof}
We prove by contradiction. Suppose there is an index $i$ with $\psi + \zeta_0 + 1 \leq i \leq \psi + \zeta_0 + \eta$ satisfying the following three conditions: 
\begin{enumerate}
\item $\ord({\widetilde{r}_{ i}}) = \ord({\widetilde{g}})$.
\item $\rho(r_{ i}) \neq 0$.
\item $\alpha_i$ is not a multiple of $p$. 
\end{enumerate}
Since $g$ is not in $Z(G)$, $i$ must be greater than $\psi + \zeta_0$. 
Then for any $1 \leq j \leq \psi + \zeta_0 +\eta$, let $r_j'$ be $r_j$ if $j \neq i$ and \[r_j' = \prod_{k = 1}^{\psi + \zeta_0 + \eta} r_k^{\alpha_k}\] for $j = i$. 
Then $\ord({\widetilde{r}_j'}) = \ord({\widetilde{r}_j})$ for each $1 \leq j \leq \psi + \zeta_0 +\eta$, and thus by Theorem~\ref{thm:group_decomposition}, $r_1', \dots, r_{\psi + \zeta_0 +\eta}'$ satisfy the first three conditions of Definition~\ref{def:rep_vec}. 
But the lexicographic order of 
\[\begin{pmatrix}\rho(r_{\psi + \zeta_0 +1}'^{p^{ \tau - 1}}) \\ \vdots \\ \rho(r_{\psi + \zeta_0 +1}') \\ \vdots  \\ \rho(r_{\psi + \zeta_0 +\eta}'^{p^{\tau - 1}}) \\ \vdots \\ \rho(r_{\psi + \zeta_0 + \eta}')\end{pmatrix}\] 
is smaller than the lexicographic vector associated with $\vec r$
because \[\begin{pmatrix}\rho(r_{i}'^{p^{\tau-1}}) \\ \vdots \\ \rho(r_{i}')\end{pmatrix}\]
is a zero vector and 
\[\begin{pmatrix}\rho(r_{i}^{p^{\tau-1}}) \\ \vdots \\ \rho(r_{i})\end{pmatrix}\]
is non-zero. 
Hence, $\vec{r}$ is not a representative vector. Contradiction. 
\end{proof}

We use the following algorithm to construct a representative vector. 
\begin{framed}
\noindent \textbf{Representative Vector Algorithm}
\begin{enumerate}
\item Compute arbitrary $r_{1}, \dots, r_{\psi}$ such that $Y = \oplus_{i = 1}^\psi \langle r_{i} \rangle$ and $\ord(r_1) \leq  \dots \leq \ord(r_{\psi})$. 
\item Let $r_{\psi + i} = z_{i + \zeta - \zeta_0} ^{\max\{1,  \ord(z_{i + \zeta - \zeta_0}) / p^\tau\}}$ for any $1 \leq i \leq \zeta_0$. 
\item For $i = 1, \dots, \eta$, 
compute any $r_{\psi + \zeta_0 + i} \in F$ satisfying the following conditions:
\begin{enumerate}
\item There is no $a, b \in F$ such that ${\widetilde{r}_{\psi +\zeta_0 +  i}}= {\widetilde{ab}}$, 
${\widetilde{a}} \in \oplus_{j = 1}^{i-1} \langle {\widetilde{r}_{\psi + \zeta_0 + j}} \rangle$, and ${\widetilde{b}}$ has a $p$-th root in $\langle F, Z(G)\rangle / Z(G)$. 
\item $\ord({\widetilde{r}_{\psi + \zeta_0 + i}})$ is smallest among all the possible candidates satisfying the condition (a).
\item $(\rho(r_{\psi + \zeta_0 + i}^{p^{\tau - 1}})^T, \dots, \rho(r_{\psi + \zeta_0 + i})^T)^T$ has the smallest lexicographic order among all the candidates satisfying the conditions (a) and (b). 
\end{enumerate}
\item Return $r_1, \dots, r_{\psi + \zeta_0 + \eta}$.
\end{enumerate}
\end{framed}

\begin{lemma}\label{lem:representative_vec_construction}
The representative vector algorithm constructs a representative vector for a given group restriction $G$ of a $p$-group of class 2 in time $\poly(|G|)$ if all the parameters of the group restriction are polylogarithmic in $|G|$. 
\end{lemma}

\begin{proof}
We prove that the third condition of the definition of the representative vector  is satisfied by the algorithm.
Let $t = \psi + \zeta_0$. 
In the first iteration of Step 3 of the algorithm, by Theorem~\ref{thm:group_decomposition}, for the selected $r_{t + 1}$, there exist $r_{t + 2}', \dots, r_{t + \eta}'$ such that ${\widetilde{r}_{t + 1}}, {\widetilde{r}_{t + 2}'}, \dots, {\widetilde{r}_{t + \eta}'}$ is a decomposition of $\langle F, Z(G) \rangle / Z(G)$. 

Now suppose after the $i$-th iteration of the Step 3 of the algorithm, with produced $r_{t + 1}, \dots, r_{t + i}$, 
there exist $r_{t + i + 1}', \dots, r_{t + \eta}'$ such that 
\[\langle F, Z(G) \rangle / Z(G) = {\widetilde{r}_{t + 1}}, \dots, {\widetilde{r}_{t + i}}, {\widetilde{r}_{t + i + 1}'} \dots, {\widetilde{r}_{t + \eta}'}\].
So, $r_{t + i + 1}$ exists because one of $r_{t + i + 1}', \dots, r_{t + \eta}'$ satisfies the requirement. 
We show that after the $(i + 1)$-th iteration, 
there still exist $r_{t + i + 2}', \dots, r_{t + \eta}'$ such that 
\[\langle F, Z(G) \rangle / Z(G) = {\widetilde{r}_{t + 1}}, \dots, {\widetilde{r}_{t + i + 1}}, {\widetilde{r}_{t + i + 2}'} \dots, {\widetilde{r}_{t + \eta}'}.\]
Let $a_{t + 1}, \dots, a_{t + \eta}$ be arbitrary integers over $R$ such that \[\prod_{j = 1}^{i} r_{t + j}^{a_{t + j}} \prod_{j = i + 1}^\eta {\widetilde{r}_{t + j}'^{a_{t + j}}} = {\widetilde{r}_{t + i + 1}}.\] 
By Theorem~\ref{thm:group_decomposition}, there must exist $j > i$ such that  $\ord({\widetilde{r}_{t + j}'}) = \ord({\widetilde{r}_{t + i+1}})$ and $a_{t + j}$ is not a multiple of $p$. 
Without loss of generality, we can assume $j = i + 1$. 
Then replacing $r_{t + i + 1}'$ by $r_{t + i + 1}$ still forms a decomposition of $\langle F, Z(G) \rangle / Z(G)$ by Theorem~\ref{thm:group_decomposition}. 

Hence, $r_1, \dots, r_{t + \eta}$ satisfy the first three conditions of the definition of a representative vector. 
We show that the last condition of a representative vector is also satisfied. 
Let $s_1, \dots, s_{t + \eta}$ be an arbitrary representative vector, and \[B(g) = (\rho(g^{p^{\tau - 1}})^T, \dots, \rho(g)^T)^T.\]
Suppose, for a contradiction, that $i$ is the smallest integer such that $B(s_{t + i}) < B(r_{t + i})$. 
Then there must exist a $j \leq i$ such that $s_{t + j}$ satisfying $\ord({\widetilde{s}_{t + j}}) = \ord({\widetilde{r}_{t + i}})$ and ${\widetilde{s}_{t + j}} \notin \oplus_{k = 1}^{i-1} \langle {\widetilde{r}_{t + k}} \rangle$ is a valid candidate for $r_{t + i}$ in the $i$-th iteration of the algorithm.
The existence of such a $j$ is from $\oplus_{k = 1}^i \langle {\widetilde{s}_{t + k}} \rangle \cong \oplus_{k = 1}^i \langle {\widetilde{r}_{t + k}} \rangle $ and 
Theorem~\ref{thm:group_decomposition}. 
Since
\[\ord({\widetilde{s}_{t + j}}) = \ord({\widetilde{r}_{t + i}}) = \ord({\widetilde{s}_{t + i}}),\] by the definition of representative vector, we have 
$B(s_{t + j}) \leq B(s_{t + i}) < B(r_{t + i})$, 
contradicting the choice of $r_{t + i}$. Hence, the algorithm produces a representative vector.

The running time is $\poly(|G|)$ because one can enumerate candidate group elements, test the subgroup and root conditions, calculate their quotient orders, and compare their $\rho$-value blocks in polynomial time in $|G|$.
\end{proof}

\begin{definition}\label{def:parameters}
Let $\vec r$ be a representative vector for group restriction $(z_1, \dots, z_\zeta, Y, g_1, \dots, g_\gamma, F, \rho)$ with parameters $\zeta, \gamma, \kappa, \tau, \psi$ and $\eta$. 
For notational convenience, set $g_{\gamma+s}=r_s$ for every $1\leq s \leq \psi+\zeta_0 + \eta$. 
Let $s_j$ be the element of $Y$ such that $r_j = s_j \cdot c_j$ for some $c_j \in Z$.  
For $1\leq i,j\leq\gamma+\psi+ \zeta_0 + \eta$, $1\leq k \leq\zeta$, and $1\leq\ell\leq\psi$, define the following coefficients:
\begin{enumerate}
\item let $0 \leq \alpha_{i, j, k} < \ord(z_k)$ and $0 \leq \beta_{i, j, \ell} < \ord(r_\ell)$ be the unique integers in the indicated ranges such that 
\begin{equation}\label{equ:match_1}[g_i, g_j] = \prod_{k = 1}^\zeta z_k^{\alpha_{i, j, k}} \prod_{\ell = 1}^{\psi} s_\ell^{\beta_{i, j, \ell}};\end{equation}
\item let $0 \leq \xi_{i, k} < \ord(z_k)$ and $0 \leq \pi_{i, \ell} < \ord(r_\ell)$ be the unique integers in the indicated ranges such that 
\begin{equation}\label{equ:match_2}g_i^{\ord({\widetilde{g}_i})} = \prod_{k = 1}^\zeta z_k^{\xi_{i, k}} \prod_{\ell = 1}^{\psi} s_\ell^{\pi_{i, \ell}}.\end{equation}
\end{enumerate}
\end{definition}

We define the matching relation of two representative vectors.
The matching relation deliberately compares less than the full group presentation: commutator and $Y$-power data agree exactly, but the selected $Z$-power coefficients are compared only modulo $\ord(z_k)/p$.  Thus at most one base-$p$ digit of each such coefficient is deferred to the final stage of the algorithm.

\begin{definition}\label{def:rep_vec_match}

Let $\vec{r} = (r_1, \dots, r_{\psi + \zeta_0 +\eta})^T$ and $\vec{r'} = (r_1', \dots, r_{\psi + \zeta_0 +\eta}')^T$ be representative vectors of two group restrictions 
\[(z_1, \dots, z_\zeta, Y, g_1, \dots, g_\gamma, F, \rho) \text{ for } G \text{ and } (z_1', \dots, z_\zeta', X, g_1', \dots, g_\gamma', E, \rho') \text{ for } H\]
with the same parameters. 
Let $\alpha_{i, j, k}, \beta_{i, j, k}, \xi_{i, k}, \pi_{i, k}$ and $\alpha_{i, j, k}', \beta_{i, j, k}', \xi_{i, k}', \pi_{i, k}'$ be the integers defined in Definition~\ref{def:parameters} for $\vec r$ and $\vec{r'}$ respectively.

We say that $\vec r$ and $\vec r'$ match if 
the following conditions are satisfied: 
\begin{enumerate}
\item $\alpha_{i, j, k} = \alpha_{i, j, k}'$  and $\beta_{i, j, \ell} = \beta_{i, j, \ell}'$ for any $1 \leq i, j \leq \gamma + \psi + \zeta_0+ \eta$, 
$1\leq k \leq\zeta$, and $1\leq\ell\leq\psi$.
\item $\pi_{i, \ell} = \pi_{i, \ell}'$ for any $1 \leq i \leq \gamma + \psi + \zeta_0+  \eta$ and $1\leq \ell \leq\psi$. 
\item $\xi_{i, k} \equiv \xi_{i, k}' \pmod{\ord(z_k) / p}$ for any $\gamma <  i \leq \gamma + \psi + \zeta_0+ \eta$ and $1\leq k \leq\zeta$. 

\item $\rho(r_i) = \rho'(r_i')$ for any $1 \leq i \leq \psi + \zeta_0+ \eta$. 
\end{enumerate}

\end{definition}

If the third condition is strengthened to $\xi_{i, k} = \xi_{i, k}'$ for any $1 \leq i \leq \gamma + \psi + \zeta_0+\eta$, 
then Lemma~\ref{lem:restricted_group_iso_condition} implies that the two restrictions are group restriction isomorphic. 

For two representative vectors of two group restrictions,
the values in \eqref{equ:matching_pre_condition} 
\begin{equation}\label{equ:matching_pre_condition}\alpha_{i, j, k} = \alpha_{i, j, k}' \text{ and } \xi_{i, k} = \xi_{i, k}' \text{  for all the } 1 \leq i, j \leq \gamma, 1 \leq k \leq \zeta\end{equation} 
do not
depend on the choices of representative vectors.
Therefore, if these values differ, the two group restrictions
cannot be isomorphic.
Throughout the remainder of this section, we assume
that the input representative vectors satisfy
\eqref{equ:matching_pre_condition}.

\subsection{Extended representative vector}\label{sec:extended_vector}
We add the auxiliary elements of the group to a group representative vector, and call the resulting vector an extended representative vector.

The additional coordinates record generators of successive quotients defined by the order filtration and $\ker(\rho)$.  They are redundant as generators of the whole group, but useful as witnesses: they prevent a matrix change of basis from mixing coordinates in ways that violate the prescribed filtration.

Consider a representative vector $\vec r$ obtained from a group restriction $(z_1, \dots, z_\zeta, Y, g_1, \dots, g_\gamma, F, \rho)$ of group $G$. 
Let $K = \langle F, Z(G) \rangle / Z(G)$ ($K$ is exactly the subgroup \(K\) in condition (c) of Definition~\ref{def:restricted_group}) and $S = K^p$, i.e., the subgroup of $K$ generated by  all the elements that have $p$-th root. 
Without loss of generality, suppose $\rho$ function is also defined over $K$ such that $\rho({\widetilde{g}}) = \rho(g)$ for any $g \in F$. 
This function is well-defined as $\rho(g) = \rho(g')$ for any $g, g' \in F$ satisfying ${\widetilde{g}} = {\widetilde{g}'}$.

Let $K_0 = \{Z(G)\}$, and $K_i$ be the subgroup generated by elements of $K$ with order at most $p^i$ for $i \in [\tau]$. 
Let $S_1$ be $S \cap K_1 \cap \ker(\rho)$, where $\ker(\rho) = \{{\widetilde{g}} \in K:  \rho({\widetilde{g}}) = 0\}$ which is a subgroup of $K$. 
For $i > 1$, define $S_i := \langle K_{i - 1}, (S \cap K_i)\rangle \cap \ker(\rho)$. 
Let $\sigma_i$ be the rank of $S_i / (K_{i - 1} \cap S_i)$ and $\sigma = \sum_{i = 1}^{\tau} \sigma_i$. 

The extended representative vector is defined as follows.

\begin{definition}[Extended representative vector]\label{def:ext_rep_vec}
An extended (representative) vector $\vec{a}$ for a group restriction
is a vector $(a_1, \dots, a_{\phi})^T$ with $\phi = \psi + \zeta_0 + \eta + \sigma$ satisfying the following conditions:
\begin{enumerate}

\item $(a_1, \dots, a_{\psi + \zeta_0 + \eta})^T$ forms a representative vector for the group restriction. 

\item 
$(a_{\psi + \zeta_0 + \eta + 1}, \dots, a_{\psi + \zeta_0 + \eta + \sigma})^T = (w_{1, 1}, \dots, w_{1, \sigma_1}, w_{2, 1}, \dots, w_{2, \sigma_2}, \dots, w_{\tau, 1}, \dots, w_{\tau, \sigma_\tau})^T$ such that
for any $1 \leq i \leq \tau$, 
$w_{i, 1}, \dots, w_{i, \sigma_i} \in F$, and 
\[S_i / (K_{i - 1} \cap S_i) = \langle ({\widetilde{w}_{i, 1}}) (K_{i - 1} \cap S_i) \rangle \oplus \dots \oplus \langle ({\widetilde{w}_{i, \sigma_i}})(K_{i - 1} \cap S_i) \rangle.\]
\end{enumerate}
\end{definition}

For notational convenience, set $g_{\gamma+s}=a_s$ for every $1\leq s \leq \phi$ and define parameters $\alpha_{i, j, k}, \beta_{i, j, \ell}$, $\xi_{i, k}$, and $\pi_{i, \ell}$ for all $i, j \in [\gamma + \phi]$, $k \in [\zeta]$ and $\ell \in [\psi]$
as in Definition~\ref{def:parameters}
We say that two extended vectors match if their parameters $\alpha_{i,j,k}$,
$\beta_{i,j,\ell}$, and $\pi_{i,\ell}$ are equal for all admissible indices as in Definition~\ref{def:parameters}, 
$\xi_{i, k}$ are equal modulo $\ord(z_k) / p$,
and their corresponding $\rho$-values are equal.

Throughout the remainder of this section, we assume that the matching precondition holds. Otherwise, we conclude that the two group restrictions are not isomorphic.




We prove some useful properties of an extended vector. The following lemmas show that, to obtain a element of $F$, there is always a linear transform for the extended vector such that the element is obtained by multiplying a non-zero label element and several other zero label elements. 

\begin{fact}\label{fact:extended_representative_same_center}
Let $\vec{a} = (a_1, \dots, a_\phi)^T$  be an extended representative vector of a group restriction \[(z_1, \dots, z_\zeta, Y, g_1, \dots, g_\gamma, F, \rho)\] of group $G$
and $g \in G$ be an element equal to $\prod_{i = 1}^{\phi} a_i^{\alpha_i}$ with non-negative integers $\alpha_1, \dots, \alpha_{\phi}$. 
For another element $g' \in F$ such that ${\widetilde{g}'} = {\widetilde{g}}$, 
there exist integers $\alpha_1', \dots, \alpha_\phi'$ satisfying the following conditions:
\begin{enumerate}
\item $g' = z \cdot \prod_{i = 1}^{\phi} a_i^{\alpha_i'}$ for some $z \in \langle z_1\rangle \oplus \dots \oplus \langle z_{\zeta - \zeta_0}\rangle$. 
\item $\alpha_i' = \alpha_i$ for every $\psi + \zeta_0 + 1 \leq i \leq \phi$. 
\end{enumerate}
\end{fact}
\begin{proof}
Since $g$ is a product of elements in $F$, then $g$ is in $F$.
Hence, there exists $h \in Z(G) \cap F$ such that $g' = h \cdot g$.  
Since $F \cap Z(G) = Z_0 \oplus Y$, 
there are $z \in \langle z_1\rangle \oplus \dots \oplus \langle z_{\zeta-\zeta_0}\rangle$ and $\beta_1, \dots, \beta_\phi$ such that $h = z \cdot \prod_{i = 1}^{\phi} a_i^{\beta_i}$ with $\beta_i = 0$ for each $i > \psi + \zeta_0$. 
Let $\alpha_i' = \alpha_i + \beta_i$. The fact follows. 
\end{proof}

\begin{lemma}\label{lem:extended_representative_decompose_1}
Let $\vec{a} = (a_1, \dots, a_\phi)^T$  be an extended representative vector and $g \in F$ be an element such that $\ord({\widetilde{g}}) = p^m$ for some positive integer $m$ and $\rho(g) = 0$.
Then there exist $\alpha_1, \dots, \alpha_\phi$ such that the following conditions are satisfied:
\begin{enumerate}
\item  $g = z \cdot \prod_{i = 1}^\phi a_i^{\alpha_i}$  for some $z \in \langle z_1\rangle \oplus \dots \oplus \langle z_{\zeta - \zeta_0}\rangle$.
\item For any $\alpha_i \neq 0$, one of the following conditions holds:
\begin{enumerate}
\item $\ord({\widetilde{a}_i}) < p^m$.
\item $\ord({\widetilde{a}_i}) = p^m$ and $\alpha_i $ is a multiple of $p$. 
\item $\ord({\widetilde{a}_i}) = p^m$ and $\rho(a_i) = 0$. 
\end{enumerate}
\end{enumerate}
\end{lemma}
\begin{proof}
We prove by induction. 
For $m = 1$, we first consider the case of ${\widetilde{g}} \in S_1$, and thus ${\widetilde{g}}$ has a $p$-th root in $K$. 
By Definition~\ref{def:ext_rep_vec} and Fact~\ref{fact:extended_representative_same_center}, 
there exist required $\alpha_1, \dots, \alpha_\phi$ such that $\alpha_i \neq 0$ only for $i \leq \psi + \zeta_0$ or ${\widetilde{a}_i} \in S_1$. 

Now consider the remaining case $\ord({\widetilde{g}}) = p$,
by Definition~\ref{def:ext_rep_vec} and Fact~\ref{fact:extended_representative_same_center},
there are $g_1 = \prod_{i = 1}^\phi a_i^{\beta_i} $ and $g_2 = \prod_{i = 1}^\phi a_i^{\xi_i}$ satisfying the following conditions:
\begin{enumerate}
\item $g = g_1 \cdot g_2 \cdot x$ for some $x \in Z(G)$. 
\item $\beta_i \neq 0$ only for $1 \leq i \leq \psi + \zeta_0 + \eta$, $\ord({\widetilde{a}_i}) = p$, and $\rho(a_i) = 0$. 
\item $\xi_i \neq 0$ only for $i > \psi + \zeta_0 + \eta$, $\ord({\widetilde{a}_i}) = p$, and $\rho(a_i) = 0$. 
\end{enumerate}
Hence, $\prod_{i} a_i^{\beta_i + \xi_i}$ is an element in ${\widetilde{g}}$. By Fact~\ref{fact:extended_representative_same_center}, there exist required $\alpha_1, \dots, \alpha_\phi$.

Now suppose the lemma holds for every $g$ with $\ord({\widetilde{g}}) \leq p^m$ and prove the lemma for every $g$ with $\ord({\widetilde{g}}) \leq p^{m + 1}$. 
We first show that the lemma holds for every $g$ with ${\widetilde{g}} \in S_{m + 1}$.
Suppose ${\widetilde{g}}$ is in $S_{m + 1}$ but not in $K_m$ (the case of ${\widetilde{g}} \in K_m$ has been proved by induction). 
There exist $\beta_1, \dots, \beta_\phi$ such that 
\[({\widetilde{g}} )(K_{m } \cap S_{m+1}) =\left( \left(\prod_{i = 1}^\phi a_i^{\beta_i}\right) Z(G)\right) (K_{m} \cap S_{m+1})\] with $\beta_i \neq 0$ only for $i > \psi + \zeta_0 + \eta$ and $\ord({\widetilde{a}_i}) = p^{m+1}$ by Definition~\ref{def:ext_rep_vec}. 
Hence, $g = \prod_{i = 1}^{\phi} a_i^{\beta_i} \cdot g'$ for some ${\widetilde{g}'} \in K_{m} \cap S_{m + 1}$. 
By induction and Fact~\ref{fact:extended_representative_same_center}, the lemma holds for $g$ with ${\widetilde{g}} \in S_{m + 1}$.

Finally we show that the lemma holds for $g$ with $\ord({\widetilde{g}}) = p^{m+1}$. 
By Lemma~\ref{lem:representative_vector_decomposition}, 
there are $q_{\psi + \zeta_0 + 1}, \dots, q_{\psi + \zeta_0 + \eta}$ such that \[{\widetilde{g}} = \prod_{i = \psi + \zeta_0 + 1}^{\psi + \zeta_0 + \eta} a_i^{q_i}.\] 
Let \[Q_1 = \{ \psi + \zeta_0 + 1 \leq i \leq \psi + \zeta_0 + \eta : \ord({\widetilde{a}_i}) \neq \ord({\widetilde{g}}) \text{ or } q_i \text{ is a multiple of } p\},\] and \[Q_2 = \{\psi + \zeta_0 + 1, \dots, \psi + \zeta_0 + \eta\} \setminus Q_1.\] 
Let $g_1 = \prod_{i \in Q_1} a_i^{q_{i}}$ and $g_2 = \prod_{i \in Q_2} a_i^{q_i}$. 
Then $g_1$ is in $S_{m + 1}$ and $g_2$ has $\beta_1, \dots, \beta_{\phi}$ such that \[g_2  = \prod_{i = 1}^{\phi} a_i^{\beta_i}\] and $\beta_i \neq 0$ only for $\ord({\widetilde{a}_i}) = \ord({\widetilde{g}})$ with $\rho(a_i) = 0$.
Hence, $g = g_1 \cdot g_2 \cdot z$ for some $z \in Z(G)$. 
By the argument for $S_{m+1}$ and Fact~\ref{fact:extended_representative_same_center}, there exist required $\alpha_1, \dots, \alpha_\phi$. 
\end{proof}

\begin{lemma}\label{lem:extended_representative_decompose_2}
Let $\vec{a} = (a_1, \dots, a_\phi)^T$ be an extended representative vector. 
For any $\psi + \zeta_0 + 1 \leq s \leq \psi + \zeta_0 + \eta$ and any $g \in F$ such that $\rho(g) = \rho(a_s)$, $\ord({\widetilde{g}}) = \ord({\widetilde{a}_s})$, and 
${\widetilde{g}}$ has no $p$-th root, there exist $\alpha_1, \dots, \alpha_\phi$ satisfying the following conditions:
\begin{enumerate}
\item $g = z \cdot \prod_{i = 1}^\phi a_i^{\alpha_i}$ for some $z \in \langle z_1\rangle \oplus \dots \oplus \langle z_{\zeta - \zeta_0} \rangle$.
\item For any $i \neq s$ such that $\alpha_i \neq 0$, one of the following conditions holds:
\begin{enumerate}
\item $\ord({\widetilde{a}_i}) < \ord({\widetilde{a}_s})$.
\item $\ord({\widetilde{a}_i}) = \ord({\widetilde{a}_s})$ and $\alpha_i $ is a multiple of $p$. 
\item $\ord({\widetilde{a}_i}) = \ord({\widetilde{a}_s})$ and $\rho(a_i) = 0$. 
\end{enumerate}

\end{enumerate}
 \end{lemma}

\begin{proof}
By Fact~\ref{fact:representative_vector_to_denote_element} and the correspondence between a representative vector and an extended representative vector derived from the representative vector, there are $\beta_1, \dots, \beta_{\psi + \zeta_0 + \eta}$ such that \[{\widetilde{g}} = \prod_{i = 1}^{\psi + \zeta_0 + \eta} a_i^{\beta_i} \text{ and } 
\ord({\widetilde{a}_i^{\beta_i}}) \leq \ord({\widetilde{a}_s})\] for each $1 \leq i \leq \psi + \zeta_0 + \eta$.  

Let $\beta_i' = \beta_i$ for all $i \neq s$ and $\beta_s' = \beta_s - 1$. 
Let $h =\prod_i a_i^{\beta_i'}$. We have ${\widetilde{h}} = (g \cdot a_s^{-1}) Z(G)$ and $\rho(h) = 0$ because $\rho(z) = 0$ for any $z \in F \cap Z(G)$ and the additivity of $\rho$.
In addition, $\ord({\widetilde{h}}) \leq \ord({\widetilde{g}})$ because $\ord({\widetilde{g}})$ equals \[\max_{ \psi + \zeta_0 + 1 \leq i \leq \psi + \zeta_0 + \eta} \ord({\widetilde{a}_i^{\beta_i}})\] as ${\widetilde{a}_{\psi + \zeta_0 + 1}}, \dots, {\widetilde{a}_{\psi + \zeta_0 + \eta}}$ form a decomposition of $\langle F, Z(G) \rangle / Z(G)$.  
If $h \notin Z(G)$, then 
by Lemma~\ref{lem:extended_representative_decompose_1}, there exist $\alpha_1, \dots, \alpha_\phi$ such that the two conditions of Lemma~\ref{lem:extended_representative_decompose_1} hold for $h$ with $m$ being $\ord({\widetilde{h}}) = p^m$. 
If $h \in Z(G)$, then $h$ is in $F \cap Z(G)$, then there exists $\alpha_1, \dots, \alpha_\phi$ with $\alpha_i = 0$ for all the $i > \psi + \zeta_0$ such that $h = z \cdot \prod_{i = 1}^\phi a_i^{\alpha_i}$ for some $z \in \langle z_1\rangle \oplus \dots \oplus \langle z_{\zeta-\zeta_0}\rangle$. 

Now let $\alpha_s$ be replaced by $\alpha_s + 1$. 
Then  $\prod_{i = 1}^{\phi} {\widetilde{a}_i^{\alpha_i}} = {\widetilde{g}}$. Furthermore, $\alpha_1, \dots, \alpha_\phi$ satisfy the second condition of the lemma. 
By Fact~\ref{fact:extended_representative_same_center}, the lemma holds. 
\end{proof}

\subsection{Matrix tuple construction}\label{sec:encoding}

We define the matrix tuple construction based on a given extended vector.  
For an extended vector $\vec a = (a_1, \dots, a_\phi)^T$ for a group restriction $(z_1, \dots, z_\zeta, Y, g_1, \dots, g_\gamma, F, \rho)$ with parameter $\tau$ (and other parameters), 
we use an alternating matrix tuple over $R$  to encode the following parameters as defined in Definition~\ref{def:ext_rep_vec}: 
\begin{enumerate}
\item $\alpha_{i, j, k}$ and $\beta_{i, j, \ell}$ for any $1 \leq i, j \leq \gamma +  \phi$, $1\leq k \leq\zeta$, and $1\leq\ell\leq\psi$.
\item $\pi_{i, \ell}$ for any $1 \leq i \leq \gamma +\phi$ and $1\leq \ell \leq\psi$. 
\item $\xi_{i, k} $ for any $1 \leq i \leq \gamma +\phi$ and $1 \leq k \leq \zeta$. 
\item $\rho(a_i)$ for any $1 \leq i \leq \phi$. 
\end{enumerate}

Denote $\mu = \sum_{i = \psi + \zeta_0 + 1}^\phi \log_p \ord({\widetilde{a}_i})$.
Let $n_1=\phi$, $n_2=\psi+ \zeta_0 + 2\mu$, $n_3=\mu$,
$n_4=\eta$, and $n_5= 3 $.
For $1\leq i\leq 5$, let
$m_i=\sum_{j=1}^i n_j$, and set $m=m_5$.

Every row/column of the matrix tuple except the last three and $m_4 + 1$-th row/ column corresponds to an element of $F$. 
Let 
\[\vec{sp} = \begin{pmatrix} a_{\psi + \zeta_0 + 1} \\ a_{\psi +\zeta_0 +1}^p \\ \vdots \\  a_{\psi + \zeta_0 +1}^{\ord({\widetilde{a}_{\psi + \zeta_0 +1}}) / p} \\  \vdots
\\ a_{\phi} \\  \vdots \\ a_\phi^{\ord({\widetilde{a}_\phi})/ p}\end{pmatrix}\]
with length $\mu$ such that every $a_i$ for $\psi + \zeta_0 + 1\leq i \leq \phi$ contributes $\log_p \ord({\widetilde{a}_i})$ entries. 
The $i$-th row/column corresponds to 
\begin{enumerate}
\item $a_i$ (and thus $g_{\gamma + i}$)  for $1 \leq i \leq n_1$,
\item $a_{i - m_1}$ for $m_1 + 1 \leq i \leq m_1 + \psi + \zeta_0$,
\item $sp_{i - m_1 - \psi - \zeta_0}$ for $m_1 + \psi + \zeta_0 +  1 \leq i \leq m_2 - \mu$,
\item $sp_{i - m_1 - \psi - \zeta_0 - \mu}$ for $m_1 + \psi + \zeta_0 +  \mu + 1 \leq i \leq m_2$,
\item $sp_{i - m_2}$ for $m_2 + 1 \leq i \leq m_3$, and
\item $a_{(i - m_3 + \psi + \zeta_0)}$ for $m_3 + 1 \leq i \leq m_4$. 
\end{enumerate}

We construct the matrix tuple $\mathbf{A} \in \M(m, R)^s$ for some $s = \poly(m)$. 
For each matrix $A\in\mathbf A$, we specify only its strictly upper-triangular entries. All unspecified strictly upper-triangular entries are zero. We set $A[j,i]=-A[i,j]$ for $i < j$, and all diagonal entries are zero.
For convenience, let $\mathbf{A}_{i, j}$ denote $\mathbf{A}[m_{i - 1} + 1 : m_i, m_{j - 1} + 1: m_j]$, and 
$Q_{i, j}$ denote $Q[m_{i - 1} + 1 : m_i, m_{j - 1} + 1: m_j]$ for 
a matrix $Q \in \M(m, R)$. 
The construction detail is given in Figure~\ref{fig:construction}.

\begin{figure}
\begin{framed}\begin{flushleft}
\vspace{-.1cm}
\textbf{Matrix Tuple Construction} 

\begin{enumerate}
\item (commutator map for $F$) 
Construct $\zeta$ matrices (denoted as $A_1, \dots, A_\zeta$, below same). For each $k \in [\zeta], i, j\in [\phi]$ with $i < j$, set
$A_k[i, j] = \alpha_{\gamma + i, \gamma + j, k} \cdot p^{\tau} / \ord(z_k)$. 
Construct another $\eta$ matrices. 
Set $A_j[i, m_1 + \ell] = \beta_{\gamma + i, \gamma + \psi + \zeta_0 + j, \ell} \cdot p^\tau / \ord(a_\ell)$
for each $j \in [\eta]$, $i \in [\phi]$,  $\ell \in [\psi]$.

\item (power map for $F$ on $Y$) 
Construct $\tau$ matrices. For each $0 \leq j < \tau, i \in [\phi]$, 
if $a_i^{p^j} \notin Z(G)$, then set $A_{j+1} [i, \ell] = 1$ if the $\ell$-th column for $m_1 + \psi + \zeta_0 + 1 \leq \ell \leq m_2 - \mu$ corresponds to $a_i^{p^j}$, 
otherwise set $A_{j+1}[i, m_1 + \ell] = (\pi_{\gamma + i, \ell} \cdot p^j / \ord({\widetilde{a}_i})) \cdot p^{\tau} / \ord(a_\ell)$ for any $\ell \in [\psi]$. 
\item (power map for $F$ on $Z_0$) 
Construct  $\tau$ matrices. For each $0 \leq j < \tau, i \in [\phi]$, 
if $a_i^{p^j} \notin Z(G)$, then set $A_{j+1} [i, \ell] = 1$ if the $\ell$-th column for $m_2 - \mu + 1 \leq \ell \leq m_2$ corresponds to $a_i^{p^j}$, 
and set $A_{j+1}[i, m_1 + \psi + h] = (\xi_{\gamma + i, \zeta - \zeta_0 + h} \cdot p^j / \ord({\widetilde{a}_i})) \cdot p^{\tau + 1} / \ord(z_{\zeta - \zeta_0 + h})$ for each $h \in [\zeta_0]$.

\item (power map modulo $Z(G)$ for $F$) 
Construct $\tau$ matrices. For each $0 \leq j  < \tau, i \in [\phi]$, 
if $a_i^{p^j} \notin Z(G)$, then set $A_{j + 1}[i, \ell] = 1$ if the $\ell$-th column 
for $m_2 + 1 \leq \ell \leq m_3$ corresponds to $a_i^{p^j}$. 

\item (correspondence modulo $Z(G)$ for $F$) 
Construct one matrix $A$. 
For each $i \in [\phi]$, $j \in [\eta]$, 
set $A[i, m_3 + j] = x_{i, j} \cdot p^\tau / \ord({\widetilde{a}_{\psi + \zeta_0 +  j}})$
with $0 \leq x_{i, j} < \ord({\widetilde{a}_{\psi + \zeta_0 + j}})$ such that ${\widetilde{a}_i} = (\prod_{j = 1}^\eta a_{\psi + \zeta_0 + j}^{x_{i, j}})Z(G)$. 

\item ($\rho$ value for $F$) 
Construct $\kappa$ matrices. For each $i \in [\phi]$, $s \in [\kappa]$, 
set $A_{s}[i, m]$ is the $s$-th coordinate of $\rho(a_i)$. 

\item (auxiliary three rows) 
Construct three four with $A_1[m - 2, m -1] = 1$, $A_2[m - 2, m] = 1$,  and $A_3[m - 1, m] = 1$.

\item (correspondence for $Z_0$)
Construct $\zeta_0$ matrices such that for each $k \in [\zeta_0]$, $A_k[m_1 + \psi + k, m] = 1$. 

\item (power map for $g_1, \dots, g_\gamma$)
Construct $\gamma$ matrices. For each $i \in [\psi], \ell \in [\gamma]$, 
set $A_\ell[m_1 + i, m] = \pi_{\ell, i} \cdot p^\tau / \ord(a_i)$. 

\item (commutator map for $g_1, \dots, g_\gamma$) 
Construct another $\gamma^2$ matrices. 
For each $i, j \in [\gamma]$, construct one matrix $A$ with $A[m_1 + \ell, m] = \beta_{i, j, \ell} \cdot p^\tau / \ord(a_\ell)$ for every $\ell \in [\psi]$.

\item (commutator map for $g_1, \dots, g_\gamma$ and $F$)
Construct $\gamma \cdot \zeta$ matrices. 
For each $j \in [\gamma], k \in [\zeta]$, construct one matrix $A$ with $A[i, m] = \alpha_{\gamma + i,j, k} \cdot p^\tau / \ord(z_{k})$ for each $i \in [\phi]$. 
Construct another $\gamma$ matrices. 
For each $j \in [\gamma]$, construct one matrix $A$ with $A[i, m_1 + \ell] = \beta_{\gamma + i, j, \ell} \cdot p^\tau / \ord(a_\ell)$ for each $i \in [\phi], \ell \in [\psi]$. 
\end{enumerate}
\end{flushleft}
\end{framed}
\caption{Alternating matrix tuple construction}\label{fig:construction}
\end{figure}

The matrix tuple is well-defined as $\alpha_{i, j, k}$ is always a multiple of $\ord(z_k) / p^\delta$ as every element of $g \in F$ satisfies that $\ord({\widetilde{g}})$ is at most $p^\delta$ and $Z(G) = \oplus_{k = 1}^\zeta \langle z_k \rangle \oplus Y$. 
For $\beta_{i, j, \ell}$, and $\pi_{i, \ell}$, they are all at most $\tau$ as elements of $Y$ are of order at most $p^\tau$. 

In the rest of this subsection, we investigate the block diagonal solution for the matrix tuple. 
More specifically, we show that matrix tuples constructed from extended vectors of two group restrictions are isometric by some block diagonal matrix $Q = \diag(Q_1, \dots, Q_5)$ with $Q_i \in \M(n_i, R)$
if and only if the two group restrictions have matching extended vectors.

\begin{lemma}\label{lem:matching_congruence_correspondance}
Let $\vec a$ and  $\vec b$ be two extended vectors for two group restrictions, and $\mathbf{A}$, $\mathbf{B}$ be the matrix tuple constructed for $\vec a$ and $\vec b$ respectively. 
If $\rho(a_i) = \rho'(b_i)$ for each $1 \leq i \leq \psi + \zeta_0 + \eta$, 
then the  group restriction from which $\vec a$ is constructed has  an extended vector $\vec v$ that matches $\vec b$ if and only if there exists an invertible block diagonal matrix $Q$ with $Q_5$ being identity matrix such that $Q^T \mathbf{A} Q = \mathbf{B}$. 
\end{lemma}

While the proof of this lemma is complex, the intuition is clear. 
We prove the following properties:
\begin{enumerate}
\item If the two matrix tuples are isometric, then there is a block diagonal solution (Lemma~\ref{lem:block_invertible}).
\item If $\vec v$ exists, then we can construct a feasible block diagonal matrix $Q$ such that $Q^T \mathbf{A} Q$ equals the matrix tuple of $\vec v$ and thus equals $\mathbf{B}$ (Lemma~\ref{lem:corresponence_vec_to_tuple}). 
\item If there is a block-diagonal matrix $Q$ such that
$Q^T \mathbf{A} Q = \mathbf{B}$, then we can use $Q$ to construct an extended vector $\vec v$ for the group restriction which produces $\vec a$ such that $\vec v$ and $\vec b$ match (Lemma~\ref{lem:congruent_to_matching}).
\end{enumerate}

We start with some useful lemmas.
\begin{lemma}\label{lem:invertible_diagonal_matrix}
Let $W\in\M(r,R)$ be a diagonal matrix
whose diagonal entries are nonzero powers of $p$.
Let $A,B\in\M(n\times r,R)$ satisfy the following conditions:
\begin{enumerate}
\item every row of $A$ and $B$ is a linear combination of rows of $W$;
\item every row of $W$ is a linear combination of the rows of $B$. 
\end{enumerate}
If $Q\in\M(r,R)$ satisfies $AQ=B$, then $Q$ is invertible.
\end{lemma}
\begin{proof}
Let $M$ be the row module generated by $W$. By assumption, the rows
of $B$ generate $M$. 
Let $M'$ be the row module generated by $A$. 
$M'$ is a submodule of $M$. 
Since $AQ=B$, right multiplication
by $Q$ defines a homomorphism from $M'$ to $M$. 
Because rows of $B$ generate $M$, we have
$M'Q=M$.
And consequently,
\[
|M|=|M'Q|\leq |M'|\leq |M|.
\]
Therefore $M' = M$. 
Right multiplication by $Q$ is now a surjective endomorphism of the finite module $M$, so it is an automorphism.

Therefore, there exists a positive integer $k$ such that
$\vec xQ^k=\vec x$ for every row vector $\vec x\in M$. In particular, $WQ^k=W$.
Let  $W=\diag(p^{a_1},\ldots,p^{a_r})$ with $0\leq a_i<\tau$.
For each $i,j$, the equality $W(Q^k-I_r)=0$ implies
\[
p^{a_i}(Q^k-I_r)[i,j]=0\pmod{p^\tau}.
\]
Hence $p^{\tau-a_i}$ divides $(Q^k-I_r)[i,j]$, and since
$\tau-a_i\geq 1$, we obtain
\[
Q^k\equiv I_r\pmod p.
\]
Thus $(Q\bmod p)^k=I_r$, so $Q\bmod p$ is invertible over
$\mathbb F_p$. Consequently, $Q$ is invertible. 
\end{proof}

The following corollary is obtained from Lemma~\ref{lem:invertible_diagonal_matrix} with $A = C_\mathbf{A}$ and $B = C_\mathbf{B}$. 
\begin{corollary}\label{cor:inv_diagonal_matrix_tuple}
Let $W\in\M(r,R)$ be a diagonal matrix
whose diagonal entries are nonzero powers of $p$.
Let $\mathbf{A},\mathbf{B}\in\M(n\times r,R)^m$ satisfy the following conditions:
\begin{enumerate}
\item every row of every matrix in $\mathbf{A}$ and $\mathbf{B}$ is a linear combination of rows of $W$;
\item every row of $W$ is a linear combination of the rows of $C_\mathbf{B}$. 
\end{enumerate}
If $Q\in\M(r,R)$ satisfies $\mathbf{A}Q=\mathbf{B}$, then $Q$ is invertible.
\end{corollary}

The following lemma show that the block diagonal solution must be invertible. 
\begin{lemma}\label{lem:block_invertible}
Let $\mathbf{A}$ and $\mathbf{B}$ be matrix tuples constructed from extended representative vectors with the same parameters. 
Let $Q = \diag(Q_1, \dots, Q_5)$ with $Q_i$ of dimension $n_i \times n_i$ be a matrix such that $Q^T \mathbf{A} Q = \mathbf{B}$. 
Then $Q$ is invertible, and $Q_5 = q \cdot I$ for some $q \in R$ satisfying $q^2 = 1$. 
\end{lemma}

\begin{proof}
We use the standing preprocessing assumptions of this section: the corresponding cyclic orders in the center and in the quotient agree, and $\tau\geq2$.

The three matrices $J_{1,2}$, $J_{1,3}$,
and $J_{2,3}$ have trivial common kernel.
Since
$Q_5^T J_{i,j} Q_5=J_{i,j}$
for all $i<j$, the kernel of $Q_5$
is trivial. Thus $Q_5$ is invertible.
Moreover, $Q_5$ preserves the kernel
of each $J_{i,j}$. These kernels are
$\langle e_3\rangle$, $\langle e_2\rangle$,
and $\langle e_1\rangle$, respectively.
Hence $Q_5$ is diagonal.
Denote $Q_5$ by $\diag(q_1,q_2,q_3)$,
we obtain
$q_1q_2=q_1q_3=q_2q_3=1$.
Consequently, $q_1=q_2=q_3=q$
and $q^2=1$.


Let $t = \psi + \zeta_0$. Denote $Q_1$ as
\[\begin{pmatrix}U_{1, 1} & U_{1,2} \\ U_{2, 1} & U_{2,2} \end{pmatrix}\]
with $U_{1, 1}$ of dimension $t \times t$ and
$Q_2$ as
\[ \begin{pmatrix}W_{1, 1} & W_{1,2} \\ W_{2, 1} & W_{2,2} \end{pmatrix}\]
with $W_{1, 1}$ of dimension $t \times t$.

By the construction, we have the following observations for $\mathbf{A}_{1,3}$ on the matrices $A_1, \dots, A_\tau$ constructed in step 4:
\begin{enumerate}
\item the first $t $ rows of $\mathbf{A}_{1,3}$ are zero;
\item for each $1 \leq j \leq \mu$, there is a row of a matrix in $\mathbf{A}_{1,3}$ equal to $e_{j}^T$.
\item for each $t  + 1 \leq i \leq \phi$, there is a column of a matrix in $\mathbf{A}_{1,3}$ equal to $e_i$.
\end{enumerate}
The observations also hold for $\mathbf{B}_{1,3}$.
Hence, both the module spanned by rows of matrices in $\mathbf{A}_{1, 3}$
and the module spanned by rows of matrices in $\mathbf{B}_{1, 3}$ are $R^\mu$.
Since $Q_1^T \mathbf{A}_{1, 3} Q_3 = \mathbf{B}_{1, 3}$,
$Q_{3}$ is invertible by Corollary~\ref{cor:inv_diagonal_matrix_tuple} with $W$ being $I_{\mu}$.
Consequently, $U_{2, 1} = 0$ by the third property of $\mathbf{A}_{1,3}$.

Since $\mathbf{A}_{1, 2}[1: t, t + 1:t+ 2\mu] = \mathbf{B}_{1, 2}[1: t, t + 1: t + 2\mu] = 0$ by the construction,  we have
\[\begin{split}& Q_1^T \mathbf{A}_{1, 2} Q_2 \\ = &
\begin{pmatrix}U_{1, 1}^T &  \\ U_{1, 2}^T & U_{2,2}^T \end{pmatrix} \begin{pmatrix} \mathbf{A}_{1, 2}[1:t, 1:t]  & \\ \mathbf{A}_{1, 2}[t + 1 : \phi, 1:t]  & \mathbf{A}_{1, 2}[t + 1: \phi, t + 1:t + 2\mu]  \end{pmatrix}
\begin{pmatrix}W_{1, 1} & W_{1,2} \\ W_{2, 1} & W_{2,2} \end{pmatrix} \\
= & \begin{pmatrix} \mathbf{B}_{1, 2}[1:t, 1:t]  & \\ \mathbf{B}_{1, 2}[t + 1 : \phi, 1:t]  & \mathbf{B}_{1, 2}[t + 1: \phi, t + 1:t + 2\mu]  \end{pmatrix}.
\end{split}\]
Let $M$ be the sum of matrices in $\mathbf{A}_{1, 2}[1: t, 1 : t]$ corresponding to the first matrices constructed in step 2 and step 3 for $\mathbf{A}$,
and $N$ be the counterpart for $\mathbf{B}$.
Let $D_0=\diag(d_1,\ldots,d_t)$ with 
\[
 d_i=p^\tau / \ord(a_i) \text{ for } 1\leq i\leq\psi, \text{ and }
 d_{\psi+h}=p^{\tau+1} / \ord(a_{\psi+h}) \text{ for } 1\leq h\leq\zeta_0.\]
By the standing preprocessing assumptions, the corresponding central orders in the two restrictions agree. The first matrices in steps 2 and 3 give
\[
 M=\diag(d_1,\ldots,d_\psi,u_1d_{\psi+1},\ldots,u_{\zeta_0}d_t),\quad
 N=\diag(d_1,\ldots,d_\psi,v_1d_{\psi+1},\ldots,v_{\zeta_0}d_t),
\]
where each $u_h,v_h$ is a unit in $R$.
Indeed, the permitted order-$p$ central modifications contribute zero to these scaled entries modulo $p^\tau$, whereas the distinguished component of each $a_{\psi+h}$ contributes a unit multiple of $d_{\psi+h}$.
Thus the rows of $M$ and $N$ both generate the row module of $D_0$; in general $M$ and $N$ need not be equal.

Hence, $U_{1,1}^TMW_{1,1}=N$.
Every row of $U_{1,1}^TM$ is in the row module of $D_0$, and the rows of $N$ generate this module. Lemma~\ref{lem:invertible_diagonal_matrix}, with $W=D_0$, shows that $W_{1,1}$ is invertible.
Similarly, $W_{1,1}^TM^TU_{1,1}=N^T$ and the same lemma shows that $U_{1,1}$ is invertible.
We also have
\begin{equation}\label{equ:invertible_1}\mathbf{A}_{1, 2}[1: t, 1 : t]  W_{1, 2} = 0,\end{equation} and consequently by the construction, $\mathbf{A}_{1, 2}[t + 1: \phi, 1 : t]  W_{1, 2} = 0$ because the $(i, j)$-th entry of every matrix in $\mathbf{A}_{1, 2}[t + 1: \phi, 1 : t]$ is a multiple of the $j$-th diagonal entry of $M$. For the $Z_0$-columns in step 3, this uses the bound $\ord(g)\leq p^\tau$ for $g\in F$: whenever $\ord(z_k)>p^\tau$, the $z_k$-coefficient of any central element of $F$ is divisible by $\ord(z_k)/p^\tau$.
Hence,
\begin{equation}\label{equ:invertible_2}U_{2, 2}^T \mathbf{A}_{1, 2}[t + 1: \phi, t + 1:t + 2\mu] W_{2, 2} = \mathbf{B}_{1, 2}[t + 1: \phi, t + 1:t + 2\mu].\end{equation}
Since the last two properties of $\mathbf{A}_{1, 3}$ (and $\mathbf{B}_{1, 3}$) also hold for
$ \mathbf{A}_{1, 2}[1: \phi, t + 1:t + 2\mu]$ and ($ \mathbf{B}_{1, 2}[1: \phi, t + 1:t + 2\mu]$),
$W_{2, 2}$ is invertible by Corollary~\ref{cor:inv_diagonal_matrix_tuple}.
By taking the transpose of matrix tuples in Equation~\eqref{equ:invertible_2}, $U_{2, 2}$ is also invertible. Indeed, the transposed right-hand tuple has full row module, since for each $t<i\leq\phi$ one of the original matrices has a column equal to $e_i$.
Then $Q_1$ is invertible.

To show that $Q_2$ is invertible, observe that $U_{1,1}$ is invertible and the upper-right block of $Q_1^T\mathbf{A}_{1,2}Q_2=\mathbf{B}_{1,2}$ gives $MW_{1,2}=0$ as in Equation~\eqref{equ:invertible_1}.
Every diagonal entry of $M$ is a unit times a power $p^{a_i}$ with $0\leq a_i<\tau$. Therefore $MW_{1,2}=0$ implies $W_{1,2}\equiv0\pmod p$.
Since $W_{1,1}$ and $W_{2,2}$ are invertible, $Q_2\bmod p$ is block lower triangular with invertible diagonal blocks. Hence $Q_2$ is invertible over $R$.

Finally,
let $C$ and $D$ be the first matrix constructed in step 4 induced on $[t + 1:t + \eta, m_3 + 1:m_3 + \eta]$ for $\mathbf{A}$ and $\mathbf{B}$ respectively.
We have $C = D = \diag(p^\tau / \ord({\widetilde{a}_{t + 1}}), \dots, p^\tau / \ord({\widetilde{a}_{t + \eta}}))$. This equality uses the matching quotient invariant factors in the standing preprocessing assumptions.
Since every row of each matrix in $\mathbf{A}[1:\phi, m_3 + 1:m_3 + \eta]$ and $\mathbf{B}[1:\phi, m_3 + 1:m_3 + \eta]$ is a linear combination of rows of $C$,
by applying Corollary~\ref{cor:inv_diagonal_matrix_tuple} to $Q_1^T\mathbf{A}_{1,4}$ and $\mathbf{B}_{1,4}$, with the common diagonal matrix $C$,
$Q_4$ is also invertible.
\end{proof}

\begin{definition}
A matrix $Q \in \M(\phi, R)$ is compatible for an extended vector $\vec{a}$ if the following conditions are satisfied for each $i, j \in [\phi]$. 
\begin{enumerate}
\item If $\ord({\widetilde{a}_i}) > \ord({\widetilde{a}_j})$, then $Q[i, j]= 0$. 
\item If $\ord({\widetilde{a}_i}) = \ord({\widetilde{a}_j})$ and $i \neq j$, then either $Q[i, j]$ is a multiple of $p$ or $\rho(a_i) = 0$. 
\end{enumerate}
\end{definition}

\begin{lemma}\label{lem:power_pi}
Let $Q  \in \M(\phi, R)$ be a compatible matrix for an extended vector $\vec{a}$, and $\vec v := Q^T \vec a = (v_1, \dots, v_\phi)$. 
Then for each $i \in [\phi]$, 
$v_i^{\ord({\widetilde{a}_i})}$ is in $Z(G)$ 
and equals $w_i \cdot \prod_{j = 1}^{\phi} (a_j^{\ord({\widetilde{a}_i})})^{Q[j, i]}$ for some $w_i \in \oplus_{k = 1}^\zeta \langle z_k \rangle$ of order at most $p$. 

\end{lemma}

\begin{proof}
Let $s_i = \ord({\widetilde{a}_i})$. 
Since $G / Z(G)$ is an abelian group, 
$\ord({\widetilde{v}_i})$ is upper bounded by the maximum of the orders of $\ord({\widetilde{a}_j})$ with $Q[j, i] \neq 0$, and thus $v_i^{s_i}$ is in $Z(G)$ by the definition of compatible matrix.  

Now consider $v_i^{s_i}$. We have 
\[\begin{split}v_i^{s_i} = & \left(\prod_{j = 1}^{\phi} a_j^{Q[j, i]} \right)^{s_i} \\ = & \prod_{j = 1}^{\phi} \left(a_j^{Q[j, i]}\right)^{s_i} \cdot \prod_{1 \leq j <  k \leq \phi} \left[a_k^{Q[k, i]}, a_j^{Q[j, i]}\right]^{\binom{s_i}{2}} \\ = & \prod_{j = 1}^{\phi} a_j^{Q[j, i] \cdot s_i} \cdot \prod_{1 \leq j <  k \leq \phi} \left[a_k, a_j\right]^{\binom{s_i}{2}\cdot Q[j, i] \cdot Q[k, i]}.\end{split}\] 
For each $j < k$, if both $Q[j, i]$ and $Q[k, i]$ are non-zero, then we consider the following two cases. 
\begin{enumerate}
\item Case 1. At least one of $\ord({\widetilde{a}_j})$ and $\ord({\widetilde{a}_k})$ is strictly less than $s_i$, or at least one coefficient $Q[j,i], Q[k,i]$ is divisible by $p$. 
Since $\min\{\ord({\widetilde{a}_j}),\ord({\widetilde{a}_k})\}$ is always a multiple of $\ord([a_j,a_k])$ and smaller than or equal to $\ord({\widetilde{a}_i})$, 
we have \[\ord([a_j,a_k]^{Q[j, i] Q[k, i]}) \leq s_i / p.\]

On the other hand, if $p = 2$, then $s_i (s_i - 1) / 2 = s_i (s_i - 1) / p$. 
If $p > 2$, the $s_i (s_i - 1) / 2$ is a multiple of $s_i$. 
So, $s_i (s_i - 1) / 2$ is always a multiple of $s_i / p$. 
And thus 
\[\left[a_k, a_j\right]^{\binom{s_i}{2}\cdot Q[j, i] \cdot Q[k, i]} = 1.\]
\item Case 2. $\ord({\widetilde{a}_j}) = \ord({\widetilde{a}_k}) = s_i$ and neither $Q[j, i]$ nor $Q[k, i]$ is a multiple of $p$.
Then at least one of $j$ and $k$ have $\rho$ value being zero. 
Then $[a_j, a_k] \in  \oplus_{k = 1}^\zeta \langle z_k \rangle$ and is of order at most $s_i$.
Since $p\binom{s_i}{2}$ is a multiple of 
$s_i$, 
the order of \[\ord([a_j,a_k]^{
Q[j,i]Q[k,i]\binom{s_i}{2}}) \leq p. \]
\end{enumerate}
Then the lemma holds by Lemma~\ref{lem:high_bit}. 
\end{proof}

\begin{lemma}\label{lem:invertible_compatible}
For an extended vector $\vec a$ of a group restriction, 
let $Q_1 \in \M(\phi, R)$ be a compatible matrix for $\vec a$ such that $Q_1^T \vec a$ is another extended vector for the same group restriction. 
Then $Q_1$ is invertible. 
\end{lemma}

\begin{proof}
Since every row of $Q_1$ corresponds to an element in $\vec a$ and every column in $Q_1$ corresponds to an element of $Q_1^T\vec a$, 
we use $P$ to denote the set of all these elements. 
We partition $P$ into sets $U_0, W_1, U_1, \dots, W_\tau$, and $U_\tau$, where $W_i = \{g \in P: \ord({\widetilde{g}}) = p^i \text{ and } {\widetilde{g}} \in S_i\}$, and $U_i = \{g \in P: \ord({\widetilde{g}}) = p^i \text{ and } {\widetilde{g}} \notin S_i\}$. 
Without loss of generality, we reorder the rows and columns of $Q_1$ so that the rows and columns are ordered by $U_0, W_1, U_1, \dots, W_\tau, U_\tau$. 

By the definition of extended vector, for each $U_i$ or $W_i$, the number of rows and the number of columns corresponding to elements from the set are the same because the number of rows/columns from $W_i$ is always the rank of $S_i / (K_{i - 1} \cap S_i)$ and 
 the number of rows/columns from $U_i$ is $\psi+\zeta_0$ if $i = 0$, and the number of elements of order $p^i$ in an arbitrary direct sum decomposition of $\langle F, Z(G) \rangle / Z(G)$.
Then the induced submatrices by rows of $W_i \cup U_i$ and columns of $W_j \cup U_j$ for any $j > i$ are zero by the definition of compatible matrices.

For the induced submatrix by rows of $U_i$ and columns of $W_i$, every entry must be a multiple of $p$, because otherwise, the corresponding element in $Q_1^T \vec a$ is not in $S_i$ contradicting the definition of $W_i$.
Hence, for a column corresponding to an element of $W_i$, all the non-zero entries for rows not in $W_i$ contributes as an element of order less than $p^i$ in $\langle F, Z(G) \rangle / Z(G)$. 
Hence, to make the elements of $W_i$ corresponds to a generating set of $S_i$ modulo $K_{i-1} \cap S_i$, the induced submatrix on rows and columns of $W_i$ must be invertible. 

By the definition of direct sum decomposition of an abelian $p$-group, every element in the decomposition of order $p^i$ cannot be represented as an element of order less than $p^i$ and an element with $p$-th power. 
To make the elements of $U_i$ correspond to a generating set of $K_i$ when modulo $S_i$, the induced submatrix on rows and columns of $U_i$ must be invertible.

So, the induced submatrices by rows of $U_i$  and columns of $U_i$ are invertible by Theorem~\ref{thm:group_decomposition}. 
$Q_1$ module $p$ is a lower triangular block matrix with diagonal blocks being invertible. 
Hence, $Q_1$ is invertible. 
\end{proof}

The following lemma shows that a permitted change of extended representatives should induce a congruence between their matrix tuples.  

\begin{lemma}\label{lem:corresponence_vec_to_tuple}
For an extended vector $\vec a$ of a group restriction, 
let $Q_1 \in \M(\phi, R)$ be a compatible matrix for $\vec a$ such that $Q_1^T \vec a$ is another extended vector (for the same group restriction). There exists a block diagonal matrix $Q = \diag(Q_1, \dots, Q_5) \in \GL(m, R)$ with $Q_5$ being the identity matrix such that $Q^T \mathbf{A} Q = \mathbf{B}$, where $\mathbf{A}$ and $\mathbf{B}$ are the alternating matrix tuples constructed from $\vec a$ and $Q_1^T \vec a$ respectively.
\end{lemma}


\begin{proof}
Set $Q_5 = I$. 
By Lemma~\ref{lem:invertible_compatible}, $Q_1$ is invertible. 
We first show that there is an invertible matrix $Q_3$ such that $Q_1^T \mathbf{A}_{1,3} Q_3 = \mathbf{B}_{1,3}$. 
Since $Q_1$ is compatible with $\vec a$ and $Q_1$ is invertible, 
by the construction of $\mathbf{A}_{1, 3}$ and $\mathbf{B}_{1, 3}$, $\mathbf{A}_{1, 3} = \mathbf{B}_{1,3}$.
Hence, it is sufficient to show that there exists $Q_3$ such that $Q_1^T \mathbf{A}_{1,3} Q_3 = \mathbf{A}_{1,3}$. 

By the construction of the matrix tuple, we have the following properties for $\mathbf{A}_{1, 3}$:
\begin{enumerate}
\item for each matrix of $\mathbf{A}_{1,3}$, if a row is non-zero, then only one entry of the row is non-zero with value being 1. 
\item for each column of $\mathbf{A}_{1,3}$ as a column tuple, exactly one column in the column tuple is non-zero, and for the non-zero column, only one entry is non-zero with value being 1. 
\item For each $0 \leq j < \tau$, then the $i$-th row of the $(j+1)$-th matrix constructed in step 4 induced on $\mathbf{A}_{1,3}$ is zero if and only if $\ord({\widetilde{a}_i}) \leq p^j$. 
\end{enumerate}

Consider the $R$-module $M$ spanned by the columns of $C_{\mathbf{A}_{1,3}}$. 
By the properties of $\mathbf{A}_{1,3}$, every column of $C_{\mathbf{A}_{1,3}}$ has a unique coordinate with value being 1, and all the other coordinates are zero. Furthermore, different columns $C_{\mathbf{A}_{1,3}}$ have different value 1 coordinates. 
$M$ is a free $R$-module by the property of $\mathbf{A}_{1,3}$. 
Since 
both $\vec a$ and $Q_1^T \vec a$ are extended vectors for the same group restriction, 
$\ord({\widetilde{a}_i}) = \ord({\widetilde{v}_i})$ for each $i$, where $v_i$ denotes the $i$-th coordinate of $Q_1^T \vec a$. 
By compatibility, $Q_1[h,i]=0$ whenever $\ord({\widetilde{a}_h})>\ord({\widetilde{a}_i})$. Thus the rows of $Q_1^T A$ vanish wherever the corresponding rows of $A$ vanish, for every matrix $A$ constructed in step 4. 
Hence, the matrix $Q_1^T$ induces an endomorphism of $M$ by $P_1^T$ with \[
P_1= \operatorname{diag}(Q_1,\ldots,Q_1)
=I_s\otimes Q_1,
\]
where $s$ is the number of matrices in $\mathbf{A}$. 
Because $Q_1^T$ is invertible, the endomorphism is actually an automorphism. 
Hence, there exists an invertible $Q_3$ such that $Q_1^T \mathbf{A}_{1,3} Q_3  = \mathbf{A}_{1,3}$.


To construct $Q_4$, let $C\in \M(\eta, R)$ be the matrix such that $\prod_{j = 1}^\eta {\widetilde{a}_{\psi + \zeta_0 + j}^{C[i, j]}} = {\widetilde{v}_{\psi + \zeta_0 + i}}$, and 
\[D = \diag(p^\tau / \ord({\widetilde{a}_{\psi + \zeta_0 + 1}}), \dots, p^\tau / \ord({\widetilde{a}_{\psi + \zeta_0 + \eta}})).\] $Q_4$ is defined to be the matrix such that $C D Q_4 = D$. 
Since both ${\widetilde{a}_{\psi + \zeta_0 + 1}}, \dots, {\widetilde{a}_{\psi + \zeta_0 + \eta}}$ and ${\widetilde{v}_{\psi + \zeta_0 + 1}}, \dots, {\widetilde{v}_{\psi + \zeta_0 + \eta}}$ are direct sum decompositions of $K$, 
such a matrix exists and is invertible by Lemma~\ref{lem:inv_with_basis_new}.

To construct $Q_2$, set $t=\psi+\zeta_0$ and use $R$.
For $1\leq i\leq\psi$, let $s_i,s_i'\in Y$ denote the $Y$-components of $a_i,v_i$, respectively, as in Definition~\ref{def:parameters}.
Both $(s_1,\ldots,s_\psi)$ and $(s_1',\ldots,s_\psi')$
give direct sum decompositions of $Y$ with the same respective orders.
Let $C_Y\in\M(\psi,R)$ satisfy $C_Y(s_1,\ldots,s_\psi)^T=(s_1',\ldots,s_\psi')^T$,
and let
\[D_Y=\diag\left(p^\tau / \ord(s_1),\ldots,
                  p^\tau / \ord(s_\psi)\right).\]
By Lemma~\ref{lem:inv_with_basis_new}, there is an invertible
$P\in\GL(\psi,R)$ such that $C_YD_YP=D_Y$.
Thus right multiplication by $P$ changes the scaled $Y$-coefficients
from the basis $(s_i)$ to the basis $(s_i')$; see
Lemma~\ref{lem:inv_with_basis}.
\[ Q_2'=\diag(P,I_{\zeta_0},Q_3,Q_3)   \in\GL(t+2\mu,R).\]
In particular, $Q_2'$ fixes the rows indexed by the
$\zeta_0$ central coordinates in the second block.

Let
\[ U=\begin{pmatrix} A_{1,2,1}\\ \vdots\\ A_{1,2,s}\end{pmatrix},
W=\begin{pmatrix} B_{1,2,1}\\ \vdots\\ B_{1,2,s}\end{pmatrix}, P_1=I_s\otimes Q_1,\]
where $A_{1,2,j}$ and $B_{1,2,j}$ are the $j$-th matrices
of $\mathbf A_{1,2}$ and $\mathbf B_{1,2}$.
Denote
\[P_1^TUQ_2'=\begin{pmatrix} U_0 & T\end{pmatrix} \text{ and } W=\begin{pmatrix}W_0 &  T\end{pmatrix},\]
with the first $t$ and last $2\mu$ columns separated.
Indeed, the two copies of the $\mu$ power columns in steps 2 and 3 coincide with the columns used in step 4. Therefore
$Q_1^T\mathbf A_{1,3}Q_3=\mathbf B_{1,3}$ gives equality of
both copies of the last $\mu$ columns after multiplication by $Q_2'$.
By steps 2 and 3, each column of $T$ has precisely one nonzero
entry, equal to $1$, and these entries occur in distinct rows.

We claim that any row of $U_0$ and $W_0$ agrees if the corresponding row of $T$ is zero.
For rows belonging to step 2 or step 3, a zero row in $T$ means that $v_i^{p^j}\in Z(G)$, for the corresponding $i,j$.
By compatibility, all $a_h$ with $Q_1[h,i]\ne0$ have $\ord({\widetilde{a}_h})\leq\ord({\widetilde{a}_i})$.
Applying Lemma~\ref{lem:power_pi} to $v_i$ and then raising to the required $p$-power shows that its central power is the corresponding product of the central powers of the $a_h$, up to an element of $Z=\oplus_k\langle z_k\rangle$ of order at most $p$.
The $Y$-coefficients therefore agree after the change of basis $P$.
For step 3, the coefficients in the $Z_0$ columns also agree: those columns encode $\xi_{i,k}$ with the factor $p^{\tau+1}/\ord(z_k)$, which annihilates the possible order-$p$ discrepancy in $Z$.

For all remaining rows of $\mathbf A_{1,2}$, the power columns are zero. The rows in steps 9 and 10 agree after multiplication by $P$, since they are the coefficients of fixed central elements in the two bases of $Y$. The same change of basis works for the
commutator rows in step 11 by bilinearity of commutators.
For the commutator rows in step 1, observe that $\rho(v_{t+j})=\rho(a_{t+j})$ for $1\leq j\leq\eta$:
the first $t+\eta$ entries of both extended vectors are representative vectors, so their lexicographically minimal $\rho$-value sequences are identical.
Hence $\rho(v_{t+j}a_{t+j}^{-1})=0$; condition~(f) of Definition~\ref{def:restricted_group} implies that replacing $a_{t+j}$ by $v_{t+j}$ does not change the $Y$-component
of a commutator with an element of $F$.
Bilinearity in the other argument, followed by multiplication by $P$, therefore gives the required equality for these rows too.
This proves the claim.

Since $T$ consists of distinct unit columns, the claim implies
that there is a matrix $L\in\M(2\mu\times t,R)$ with
$W_0-U_0=TL$. Set
\[ Q_2=Q_2'\begin{pmatrix} I_t&0\\ L&I_{2\mu}\end{pmatrix}. \]
Then $Q_2\in\GL(t+2\mu,R)$, and
\[ P_1^TUQ_2=\begin{pmatrix} U_0 & T\end{pmatrix} \begin{pmatrix}I_t&0\\L&I_{2\mu}\end{pmatrix}  =\begin{pmatrix} W_0 & T\end{pmatrix}=W. \]
Thus $Q_1^T\mathbf A_{1,2}Q_2=\mathbf B_{1,2}$.
Moreover, the first $t$ rows of $Q_2$ equal those of
$\diag(P,I_{\zeta_0},Q_3,Q_3)$.
Consequently $Q_2^T\mathbf A_{2,5}=\mathbf B_{2,5}$:
step 8 is preserved because the $Z_0$ rows are fixed,
and steps 9 and 10 follow from the change of $Y$-basis by $P$.

With $Q_5=I$, by the commutator identity in a group of class~$2$, $Q_1^T\mathbf A_{1,1}Q_1=\mathbf B_{1,1}$ (step 1). The choice of $Q_4$ gives $Q_1^T\mathbf A_{1,4}Q_4=\mathbf B_{1,4}$ (step 5), while $Q_1^T\mathbf A_{1,3}Q_3=\mathbf B_{1,3}$ was shown above.
We have already verified the $(1,2)$- and $(2,5)$-blocks.
For the $(1,5)$-block, additivity of $\rho$ proves the identity for step 6, and bilinearity of commutators proves it for the $Z$-coefficients in step 11.
step 7 consists of fixed matrices on the last three coordinates, which are preserved by $Q_5=I$. All other upper-triangular blocks are zero by construction.
It follows that $Q^T\mathbf A Q=\mathbf B$.
\end{proof}

\begin{lemma}\label{lem:congruent_to_matching}
Let $\mathbf{A}$ and $\mathbf{B}$ be matrix tuples constructed from extended vectors $\vec a = (a_1, \dots, a_\phi)^T$ and $\vec  b = (b_1, \dots, b_{\phi})^T$ for $G$ and $H$ respectively. 
Suppose $\rho(a_i) = \rho'(b_i)$ for each $1 \leq i \leq \phi$. 
If there is a block diagonal $Q = \diag(Q_1, \dots, Q_5)$ with $Q_5$ being the identity matrix such that $Q^T \mathbf{A} Q = \mathbf{B}$, then $ Q_1^T \vec{a} $ is an extended vector such that $Q_1^T \vec a$ and $\vec b$ match. 
\end{lemma}

\begin{proof}
 Let $\vec v = (v_1, \dots, v_\phi)^T := Q_1^T \vec a$. We first prove that $Q_1^T \vec a$ is compatible with $\vec a$. 
 
 \begin{claim}\label{claim:compatible}
 $Q_1^T \vec a$ is compatible with $\vec a$.
 \end{claim}
 \begin{proof}
By $\mathbf{A}_{1,3}$ and $\mathbf{B}_{1,3}$,
$Q_1[i, j] = 0$ if $\ord({\widetilde{a}_i}) > \ord({\widetilde{a}_j})$.
Indeed, in the step 4 matrix indexed by
$k=\log_p\ord({\widetilde{a}_j})$, row $j$ of
$\mathbf B_{1,3}$ is zero, whereas row $i$ of
$\mathbf A_{1,3}$ is a standard unit row.  The
corresponding unit columns are distinct, and $Q_3$ is
invertible, forcing the stated vanishing.
For the second condition of compatible matrix, 
we observe that 
 $\rho(v_i) = \rho(a_i)$ for every $1 \leq i \leq \phi$, 
 which is by $\rho(v_i) = \rho'(b_i)$ using the fact $Q_1^T \mathbf{A}_{1, 5} Q_5 = Q_1^T \mathbf{A}_{1,5} = \mathbf{B}_{1, 5}$ and the encoding of $\rho$ by the matrix tuple construction. 
If $Q_1[i, j]$ is not a multiple of $p$ for some $i \neq j$ such that 
$\ord({\widetilde{a}_i}) = \ord({\widetilde{a}_j})$ and $\rho(a_i) \neq 0$, 
then 
$i$ must be at most $\eta + \psi + \zeta_0$ by the definition of extended vector. 
Let $\alpha_k = Q_1[k, j]$ for $1 \leq k \leq \psi + \zeta_0 + \eta$ with $k \neq j$, and if $j\leq\psi+\zeta_0+\eta$, let $\alpha_j = Q_1[j, j] - 1$.
Let $d = \prod_{k = 1}^{\psi + \zeta_0 + \eta} a_k^{\alpha_k}$.
All omitted entries of the extended vector have
zero $\rho$-value, because their images lie in some $S_i\leq\ker\rho$.
We have 
\[\rho(d) =\sum_{k=1}^{\psi + \zeta_0 + \eta} Q_1[k,j] \rho(a_k) - \rho(a_j) =\rho(v_j)-\rho(a_j)=0. \]
So
$\ord({\widetilde{d}}) \leq \ord({\widetilde{a}_j})$ because $Q[i, j]$ is non-zero only for $i$ with $\ord({\widetilde{a}_i}) \leq \ord({\widetilde{a}_j})$. 
Actually, 
$\ord({\widetilde{d}}) = \ord({\widetilde{a}_j})$ because $\alpha_i$ is not a multiple of $p$, $\ord({\widetilde{a}_j}) = \ord({\widetilde{a}_i})$, and ${\widetilde{a}_{\psi + \zeta_0 + 1}}, \dots, {\widetilde{a}_{\psi + \zeta_0 + \eta}}$ corresponds to a direct sum decomposition of $\langle F, Z(G) \rangle / Z(G)$. 
By Lemma~\ref{lem:representative_vector_decomposition}, 
for every $1 \leq k \leq \psi + \zeta_0 + \eta$ with $\rho(a_k) \neq 0$ and $\ord({\widetilde{a}_k}) = \ord({\widetilde{d}})$, 
$\alpha_k$ is a multiple of $p$, contradicting $\alpha_i$. 
Hence, the second condition of compatible matrix holds. 
And thus, $Q_1^T \vec a$ is compatible with $\vec a$.  
 \end{proof}
 
  \begin{claim}\label{claim:extended_vector_qa}
 $Q_1^T \vec a$ is an extended vector for $G$. 
  \end{claim}

\begin{proof}
Put $t=\psi+\zeta_0$ and use $R$.
By step 4, the first $t$ rows of $\mathbf B_{1,3}$ vanish,
whereas the stacked columns of $\mathbf A_{1,3}$ detect every
row with index greater than $t$. As $Q_3$ is invertible,
$Q_1^T\mathbf A_{1,3}Q_3=\mathbf B_{1,3}$ implies
$Q_1[i,j]=0$ for $i>t$ and $j\leq t$.
Thus $v_1,\ldots,v_t$ belong to $F\cap Z(G)=Y\oplus Z_0$.
Write the upper-left $t\times t$ block of $Q_1$ as
\[
U=\begin{pmatrix}U_{YY}&U_{YZ}\\U_{ZY}&U_{ZZ}\end{pmatrix},
\]
where the first $\psi$ coordinates correspond to $Y$ and
its last $\zeta_0$ coordinates correspond to $Z_0$.

Let $W$ be the upper-left $t\times t$ block of $Q_2$.
step 8, together with $Q_5=I$, says that the rows of $Q_2$
indexed by the $\zeta_0$ distinguished central coordinates are
standard unit rows. Consequently,
\[
W=\begin{pmatrix}P&C\\0&I_{\zeta_0}\end{pmatrix}.
\]
By Lemma~\ref{lem:block_invertible}, $W$ is invertible,
so $P\in\GL(\psi,R)$.

For $i\leq\psi$, write $a_i=s_i c_i$ with $s_i\in Y$ and
$c_i\in Z_0[p]$. Let
\[
D_Y=\diag\left(p^\tau / \ord(s_1),\ldots,
                    p^\tau / \ord(s_\psi)\right).
\]
The $(1,2)$-submatrix of the first matrix in step 2,
restricted to the first $t$ rows and columns, is
$\diag(D_Y,0)$. The corresponding matrix for $\mathbf B$
has the same $D_Y$, since the two $Y$-groups have the same
ordered invariant factors. Therefore its congruence gives
\begin{equation}\label{equ:compatible_2}
U_{YY}^T D_Y P=D_Y, U_{YZ}^T D_Y P=0.
\end{equation}
In particular $U_{YZ}^TD_Y=0$, so the $Y$-component of each
$v_{\psi+k}$ is trivial. Moreover, if $y_i$ denotes the
$Y$-component of $v_i$, the rows of $U_{YY}^TD_Y$
represent $y_1,\ldots,y_\psi$ under the faithful scaled-coordinate
embedding of $Y$ into $R^\psi$. Equation~\eqref{equ:compatible_2} and the
invertibility of $P$ show that these elements generate $Y$
and have the same respective orders as $s_1,\ldots,s_\psi$.
As the product of their orders is $|Y|$, they give
$Y=\oplus_{i=1}^\psi\langle y_i\rangle$.

For $1\leq k\leq\zeta_0$, put $d_k=\ord(a_{\psi+k})$.
By the representative-vector condition,
$a_{\psi+k}$ is a unit power of
$z_{\zeta-\zeta_0+k}^{\ord(z_{\zeta-\zeta_0+k})/d_k}$,
multiplied by an element of $Z_0[p]$.
Hence the central $(1,2)$-submatrix of the first matrix
in step 3 has, on the first $t$ rows and columns, the form
\[
\begin{pmatrix}0&0\\0&D_Z\end{pmatrix},
D_Z=\diag\left(u_k p^{\tau+1} / d_k:1\leq k\leq\zeta_0\right),
\]
where each $u_k\in R$ is a unit. The corresponding submatrix
for $\mathbf B$ has $D_Z'$ of the same form, with units $u_k'$.
(The possible order-$p$ factors are annihilated by the scaling
in step 3.) Since the last $\zeta_0$ rows of $W$ are standard
unit rows, congruence gives
\[U_{ZY}^T D_Z=0, U_{ZZ}^T D_Z=D_Z'.\]
Because $d_k\mid p^\tau$ and each diagonal entry of $D_Z$
is a unit times $p^{\tau+1}/d_k$, the first identity implies
that $U_{ZY}[k,i]$ is divisible by $d_k/p$ for each
$i\leq\psi$. Thus the $Z_0$-component of $v_i$ has order
at most $p$; this proves Property~1 in its corrected form.
The second identity implies that for $k\neq j$,
$U_{ZZ}[k,j]$ is divisible by $d_k/p$, and
$U_{ZZ}[j,j]u_j\equiv u_j'\pmod{d_j/p}$.
In particular, $U_{ZZ}[j,j]$ is a unit modulo $p$.
Since the $Y$-component of $v_{\psi+j}$ is trivial,
$v_{\psi+j}$ is a unit power of
$z_{\zeta-\zeta_0+j}^{\ord(z_{\zeta-\zeta_0+j})/d_j}$
up to multiplication by an element of $Z_0[p]$, and its
order is $d_j$. This is Property~3.

It remains to prove Properties~2 and 4. Let $K_G$ and
$K_H$ be the quotient groups $\langle F,Z(G)\rangle/Z(G)$
and $\langle E,Z(H)\rangle/Z(H)$, respectively.
For each of them, define the injective homomorphism
$\Phi_G:K_G\to R^\eta$ (respectively $\Phi_H:K_H\to R^\eta$)
by sending its $j$-th distinguished cyclic generator to
$\frac{p^\tau}{\ord({\widetilde{a}_{t+j}})}e_j$
(respectively the corresponding ${\widetilde{b}_{t+j}}$ to the same
vector). These two maps are well-defined because the ordered
quotient invariant factors agree.
The rows of the step 5 matrix $\mathbf A_{1,4}$ are precisely
$\Phi_G({\widetilde{a}_i})$, and similarly for $\mathbf B_{1,4}$.
In particular, the rows of each such matrix span the image
of the corresponding $\Phi$.

By $Q_1^T\mathbf A_{1,4}Q_4=\mathbf B_{1,4}$ and
Lemma~\ref{lem:block_invertible}, right multiplication by
$Q_4$ sends $\Phi_G(K_G)$ isomorphically onto $\Phi_H(K_H)$.
Consequently there is an isomorphism
\[
h:K_H\longrightarrow K_G,
\Phi_G(h(x))=\Phi_H(x)Q_4^{-1}.
\]
The same step 5 identity and the definition of $\vec v$
show that
\begin{equation}\label{equ:extended}
h({\widetilde{b}_i})={\widetilde{v}_i}(t<i\leq\phi).
\end{equation}
In particular, ${\widetilde{v}_{t+1}},\ldots,{\widetilde{v}_{t+\eta}}$ form
a direct sum decomposition of $K_G$ with the required
ordered cyclic orders. This proves Property~2.
Together with the prescribed quotient orders of the
additional $S_i$-generators, this also gives
$\ord({\widetilde{v}_i})=\ord({\widetilde{a}_i})$ for all $1\leq i\leq\phi$.

By step 6 and $Q_5=I$,
$\rho(v_i)=\rho'(b_i)$ for every $i$.
Since ${\widetilde{b}_{t+1}},\ldots,{\widetilde{b}_{t+\eta}}$ generate $K_H$,
Equation~\eqref{equ:extended} gives $\rho\circ h=\rho'$.
The isomorphism $h$ preserves element orders, $p$-th powers,
and $\ker\rho$, and therefore takes the $K_j$ and $S_j$
used in the definition of extended representative vectors
for $H$ onto the corresponding subgroups for $G$.
Applying $h$ to the direct-sum decompositions provided by
${\widetilde{b}_i}$, $i\in C_j$, proves Property~4.

Finally, the hypotheses of Lemma~\ref{lem:congruent_to_matching}
and step 6 imply
$\rho(v_i)=\rho'(b_i)=\rho(a_i)$ for all $i$.
Thus the entire lexicographic sequence in Condition~4 of
Definition~\ref{def:rep_vec} for $v_{t+1},\ldots,v_{t+\eta}$
is the same as that for $a_{t+1},\ldots,a_{t+\eta}$,
which is already minimal. Together with Properties~1--4,
this establishes that $\vec v$ is an extended representative
vector under Definition~\ref{def:ext_rep_vec}.
\end{proof}



It remains to prove that $\vec v$ and $\vec b$ match.
First consider the central coordinates.  Let $s_1,\ldots,s_\psi$
be the $Y$-components of $a_1,\ldots,a_\psi$,
let $y_1,\ldots,y_\psi$ be those of $v_1,\ldots,v_\psi$,
and let $s_1',\ldots,s_\psi'$ be the $X$-components of
$b_1,\ldots,b_\psi$.  For $u=\prod_\ell s_\ell^{c_\ell}\in Y$,
write its scaled $Y$-coordinate row as
\[
\lambda_Y(u)=\left(c_\ell p^\tau / \ord(s_\ell)\right)_{\ell=1}^\psi
\in R^\psi.
\]
Define $\lambda_X$ similarly using $s_1',\ldots,s_\psi'$.
Equation~(1) implies that the matrix $P$ converts
$\lambda_Y$ into $\lambda_X$ under the identification
$y_i\mapsto s_i'$: indeed,
$\lambda_Y(y_i)P=\frac{p^\tau}{\ord(s_i)}e_i^T$.
Since the $y_i$ generate $Y$ with the prescribed orders,
this determines the identification on the whole of $Y$.
Furthermore, the upper-right part of the same step 2
congruence gives $U_{YY}^TD_YC=0$; hence
$D_YC=0$, because $U_{YY}^TD_Y$ generates the same
row module as $D_Y$.  Thus $C$ acts trivially on every
scaled $Y$-coordinate row.
step 8 and the form
$W=\left(\begin{smallmatrix}P&C\\0&I_{\zeta_0}\end{smallmatrix}\right)$
show that the scaled $Z_0$-coordinates are left fixed.

For $1\leq i\leq\phi$, put
$d_i=\log_p\ord({\widetilde{v}_i})
=\log_p\ord({\widetilde{a}_i})
=\log_p\ord({\widetilde{b}_i})$.
We have $d_i<\tau$.  Consider row $i$ in the matrices
of steps 2 and 3 indexed by $j=d_i$.
By compatibility, every $a_h$ occurring in $v_i$ has
$\ord({\widetilde{a}_h})\leq p^{d_i}$.
Therefore the last $2\mu$ entries in row $i$ of the
corresponding $Q_1^T\mathbf A_{1,2}$ matrices are zero.
The lower-left block of $Q_2$ consequently has no effect
on these rows.  Lemma~\ref{lem:power_pi} shows that
$v_i^{p^{d_i}}$ and
$\prod_h a_h^{Q_1[h,i]p^{d_i}}$ differ by an element of
$Z$ of order at most $p$.  Thus their $Y$-components
are equal, and their $Z$-components are equal modulo
$Z[p]$.  The step 2 congruence, together with the
$Y$-coordinate conversion by $P$, now gives
\[ \pi_{\gamma+i,\ell}(\vec v)
 =\pi_{\gamma+i,\ell}'(\vec b)
(1\leq\ell\leq\psi). \]
The step 3 congruence gives
\[ \xi_{\gamma+i,k}(\vec v)\equiv
\xi_{\gamma+i,k}'(\vec b)
\pmod{\ord(z_k)/p}(1\leq k\leq\zeta). \]
Here step 3 records the $Z_0$-coordinates with the
scaling factor $p^{\tau+1}/\ord(z_k)$, whose kernel on
the allowed central coefficients is precisely the
order-$p$ ambiguity; for $\ord(z_k)=p$ the required
congruence is automatic.
steps 9 and 10, using the same conversion by $P$,
give the equality of the $\pi$- and $\beta$-coefficients
associated with the fixed generators $g_1,\ldots,g_\gamma$.
The fixed-generator $\xi$-coefficients agree exactly by
\eqref{equ:matching_pre_condition}.

We next compare commutators.  For $i,j\in[\phi]$,
the step 1 matrices supported on the $(1,1)$-block
and the identity
$Q_1^T\mathbf A_{1,1}Q_1=\mathbf B_{1,1}$
give equality of all the $\alpha$-coefficients of
$[v_i,v_j]$ and $[b_i,b_j]$.
The step 11 matrices supported on the $(1,5)$-block
similarly give the $\alpha$-coefficients of
$[v_i,g_j]$ and $[b_i,g_j']$ for every $j\leq\gamma$.
These scaled encodings are injective on the possible
commutators, whose orders divide $p^\delta$.
The fixed--fixed $\alpha$-coefficients agree by
\eqref{equ:matching_pre_condition}.

For the $Y$-components, steps 10 and 11, together
with the conversion by $P$, give equality of the
$\beta$-coefficients for fixed--fixed and
fixed--variable pairs, respectively.
For $1\leq q\leq\eta$, the $q$-th commutator matrix
in step 1 gives the equality of the $Y$-components of
$[v_i,a_{t+q}]$ and $[b_i,b_{t+q}]$ under the same
basis identification, for every $i\in[\phi]$.
Since $\rho(v_{t+q})=\rho(a_{t+q})$,
condition~(f) of Definition~\ref{def:restricted_group}
shows that
$[v_i,v_{t+q}]$ and $[v_i,a_{t+q}]$ have the same
$Y$-component.  Therefore
\[
\beta_{\gamma+i,\gamma+t+q,\ell}(\vec v)
=\beta_{\gamma+i,\gamma+t+q,\ell}'(\vec b)
\quad(1\leq i\leq\phi,\ 1\leq q\leq\eta,\ 1\leq\ell\leq\psi).
\]
For an arbitrary $j>t$, write
${\widetilde{b}_j}=\prod_{q=1}^\eta({\widetilde{b}_{t+q}})^{c_{j,q}}$.
Equation~(3) implies
${\widetilde{v}_j}=\prod_{q=1}^\eta({\widetilde{v}_{t+q}})^{c_{j,q}}$.
Because the groups have class~$2$, the $Y$-component
of the commutator is bilinear on the quotient.
Taking the same linear combinations of the preceding
equalities establishes
$\beta_{\gamma+i,\gamma+j,\ell}(\vec v)
=\beta_{\gamma+i,\gamma+j,\ell}'(\vec b)$
for all $i,j\in[\phi]$; the cases $j\leq t$ are trivial
because those coordinates are central.

Finally, step 6 and $Q_5=I$ yield
$\rho(v_i)=\rho'(b_i)$ for every $i$.
We have checked every coefficient required by
Definition~\ref{def:ext_rep_vec}, together with
\eqref{equ:matching_pre_condition}.  Therefore
$\vec v=Q_1^T\vec a$ is an extended representative
vector matching $\vec b$, as required.
\end{proof}

Together, the previous two directions turn an existential search over matching representatives into a block-diagonal matrix-tuple congruence problem. This equivalence is the bridge needed before invoking an unrestricted isometry algorithm.

\begin{proof}[Proof of Lemma~\ref{lem:matching_congruence_correspondance}]
We first show that if there is $\vec v$ which matches $\vec b$, then $Q$ exists.
Denote $\vec a$ as $(a_1, \dots, a_\phi)^T$, $\vec b$ as $(b_1, \dots, b_\phi)^T$, and $\vec v$ as $(v_1, \dots, v_\phi)^T$.
Let $t=\psi+\zeta_0$ and
$Z_{>}=\oplus_{k=\zeta-\zeta_0+1}^{\zeta}\langle z_k\rangle$.
The first $t$ entries of each representative vector generate the same
central subgroup $Y\oplus(F\cap Z_{>})$.
Indeed, their $Y$-components form a direct sum decomposition of $Y$;
for each $1\leq h\leq\zeta_0$, the prescribed component of the
$(\psi+h)$-th entry is a generator of
$\langle z_{\zeta-\zeta_0+h}\rangle\cap F$, while the other
components have order at most $p$.
As these cyclic factors have order at least $p^2$, the latter
modifications do not change the resulting basis modulo $p$.
Consequently, each $v_i$ with $i\leq t$ can be written \emph{exactly}
as a product of $a_1,\ldots,a_t$.

For $t<i\leq t+\eta$, apply
Lemma~\ref{lem:extended_representative_decompose_2} to $v_i$.
Its hypotheses hold because $\rho(v_i)=\rho(a_i)$ by the
lexicographic-minimality condition, and both vectors give
ordered direct sum decompositions of $K$.
For $t+\eta<i\leq\phi$, apply
Lemma~\ref{lem:extended_representative_decompose_1}:
these coordinates lie in $\ker\rho$ by the definition of $S_j$.
Together with the exact representations of the first $t$ entries,
these lemmas give a compatible matrix $Q_1$ such that, putting
$\vec s=Q_1^T\vec a$, we have
\[
s_i=v_i\quad(1\leq i\leq t),
s_i=v_i z_i\quad(t<i\leq\phi),
 z_i\in\oplus_{k=1}^{\zeta-\zeta_0}\langle z_k\rangle.
\]
The conditions on the coefficients in the two decomposition lemmas
are exactly the two compatibility conditions for $Q_1$.
Since the factors $z_i$ are central, the quotient coordinates,
quotient orders, and $\rho$-values of $\vec s$ and $\vec v$
are identical. Hence $\vec s$ is again an extended vector:
its first $t$ entries agree with those of $\vec v$, its next
$\eta$ entries give the same ordered decomposition and the same
lexicographically minimal $\rho$-sequence, and the remaining entries
represent the same prescribed generators of the quotients involving
$S_j$.
Moreover, $\vec s$ and $\vec v$ have identical matching parameters.
Their commutators, $Y$-power coefficients, and $\rho$-values are
unchanged; for $i>t$, the quotient order of $s_i$ is divisible by $p$,
so $s_i^{\ord({\widetilde{s}_i})}=v_i^{\ord({\widetilde{v}_i})}$.
Thus $\vec s$ also matches $\vec b$.

Then by lemma~\ref{lem:corresponence_vec_to_tuple}, there is a block diagonal matrix $Q$ with first block equal to $Q_1$ and last block equal to identity matrix such that $Q^T \mathbf{A} Q$ equal the matrix tuple constructed from $Q_1^T \vec a$.
It remains to verify that the tuples constructed from $\vec s$
and $\vec b$ are identical.  Matching gives equality of the entries
in steps 1, 6, 9, 10, and 11.  steps 2 and 4 are identical because
the corresponding quotient orders and $\pi$-coefficients agree;
step 3 is identical because the congruences
\[
\xi_{\gamma+i,k}(\vec s)\equiv
\xi_{\gamma+i,k}'(\vec b)\pmod{\ord(z_k)/p}
\]
become equalities after multiplication by the encoding factor
$p^{\tau+1}/\ord(z_k)$.  The auxiliary entries in steps 7 and 8
are fixed by the construction.

For completeness, we verify the remaining step 5 coordinates, which
are not explicitly part of the matching definition.  Write
\[
{\widetilde{s}_i}=\prod_{q=1}^{\eta}{\widetilde{s}_{t+q}^{x_{i,q}}},
{\widetilde{b}_i}=\prod_{q=1}^{\eta}{\widetilde{b}_{t+q}^{x'_{i,q}}}.
\]
Matching gives the same commutator coefficients for corresponding
pairs among $g_1,\ldots,g_\gamma,s_1,\ldots,s_\phi$
and $g'_1,\ldots,g'_\gamma,b_1,\ldots,b_\phi$.
Thus for every $j\leq\gamma$ and $r\leq\eta$, the element
\[
h_i=b_i\left(\prod_{q=1}^{\eta}b_{t+q}^{x_{i,q}}\right)^{-1}
\]
commutes with $g'_j$ and $b_{t+r}$: these commutator identities
hold for the corresponding word in $G$, because that word is central.
Since ${\widetilde{g}'_1},\ldots,{\widetilde{g}'_\gamma}$ and
${\widetilde{b}_{t+1}},\ldots,{\widetilde{b}_{t+\eta}}$ generate $H/Z(H)$,
we have $h_i\in Z(H)$.  Uniqueness of the coordinates in the
direct sum decomposition of $K_H$ now gives
\[
x_{i,q}\equiv x'_{i,q}
\pmod{\ord({\widetilde{b}_{t+q}})}
(1\leq q\leq\eta).
\]
Hence the scaled step 5 entries coincide.  Therefore the matrix
tuple constructed from $\vec s$ equals $\mathbf B$, and
$Q^T\mathbf A Q=\mathbf B$.

Now we show that if there is a block diagonal $Q$ with $Q_5$ being identity matrix such that $Q^T \mathbf{A} Q = \mathbf{B}$, then $Q_1^T \vec a$ is an extended vector that matches $\vec b$.
We observe that
the $\rho$ value for the $i$-th element of $Q_1^T \vec a$ equals the $\rho'(b_i)$ for any $1 \leq i \leq \phi$, and thus equal to $\rho(a_i)$.
For $\psi + \zeta_0 + \eta<i\leq\phi$, both
$\rho(a_i)$ and $\rho'(b_i)$ are zero
by the definition of an extended vector.
Hence $\rho(a_i)=\rho'(b_i)$ for all
$1\leq i\leq\phi$.
So by Lemma~\ref{lem:congruent_to_matching}, $Q_1^T \vec a$ is an extended vector that matches $\vec b$. The lemma is obtained by letting $\vec v = Q_1^T \vec a$.
\end{proof}

\subsection{Matching extended vectors}\label{sec:doubling}

We want to obtain block diagonal solutions for the matrix tuple isometry problem. However, the general solution might not be block diagonal. Hence, we construct another matrix tuple to enforce the block diagonal solutions. 


Let $n_{5 + i} = n_i$ for each $i \in [5]$, and $m_0 = 0$, $m_i = m_{i - 1} + n_i$ for each $i \in [10]$. 
The doubling matrix tuple for $\mathbf{A} \in \M(m, R)^s$ is another matrix tuple $\mathbf{D} \in \M(2m, R)^{s'}$ constructed from $\mathbf{A}$ as follows: 
For each matrix $A$ in $\mathbf{A}$, 
\begin{enumerate}
\item
if $A$ is non-zero on $A_{i, i}$ for some $i \in [5]$, then add one matrix $D$ in $\mathbf{D}$ with $D_{i, i} = A_{i, i}$, and another matrix $D'$ with $D'_{5 +i, 5+i} = A_{i, i}$;
\item if $A$ is non-zero on $A_{i, j}$ and $A_{j, i}$ for some $i < j$, then 
add the following matrices to $\mathbf{D}$: one matrix $D$ with $D_{i, j} = A_{i, j}$ and $D_{j, i} = A_{j, i}$, 
one matrix $D'$ with $D'_{i, j + 5} = A_{i, j}$ and $D'_{j + 5, i} = A_{j, i}$,
one matrix $D''$ with $D''_{i + 5, j} = A_{i, j}$ and $D''_{j, i+5} = A_{j, i}$,
and one matrix $D'''$ with $D'''_{i+5, j+5} = A_{i, j}$ and $D'''_{j+5, i+5} = A_{j, i}$. 
\end{enumerate}

We use the following lemma to show that 
doubling matrix tuple removes the need to solve a specially constrained block-diagonal isometry problem. The additional copies encode the block structure so that an ordinary isometry retains enough information to recover the desired correspondence.

\begin{lemma}\label{lem:doubling_matrix}
Let $\mathbf{A}$ and $\mathbf{B}$ be two matrix tuples constructed from extended vectors $\vec a$ and $\vec b$ for two groups. 
Let $\mathbf{D}$ and $\mathbf{E}$ be doubling matrix tuples for $\mathbf{A}$ and $\mathbf{B}$ respectively. 
The following properties hold:
\begin{enumerate}
\item For any $Q = \diag(Q_1, \dots, Q_5)$ such that $Q^T \mathbf{A}Q = \mathbf{B}$, letting $P = \diag(Q, Q)$, $P^T \mathbf{D} P = \mathbf{E}$. 
\item For any invertible $P$ such that $P^T \mathbf{D} P = \mathbf{E}$, 
letting $P' = P - \diag(P_{1, 1}, \dots, P_{10, 10})$ and $Q = \diag(P_{1, 1}, P_{2, 2}, P_{3, 3}, P_{4,4}, P_{5, 5})$, 
$\mathbf{D} P' = 0$ and 
 $Q^T \mathbf{A} Q = \mathbf{B}$ and $Q$ is invertible. 
 \item If $P[m, m] = 1$, then $P[1:\phi , 1:\phi]$ is a compatible matrix to $\vec a$ such that $(P[1:\phi, 1:\phi]^T) \vec a$ matches $\vec b$. 
\end{enumerate}
\end{lemma}

\begin{proof}
The first property holds by the construction of the doubling matrix tuple. Now we prove the second property. 

By the construction of the doubling matrix tuple,
each matrix $D\in\mathbf{D}$ is supported on
either one diagonal block or two symmetric
off-diagonal blocks $D_{j,k}$ and $D_{k,j}$.
In the latter case, if $v_j$ and $v_k$
are supported on the $j$-th and $k$-th
coordinate blocks, respectively, then
$Dv_j$ and $Dv_k$ are supported on
different coordinate blocks. Hence,
$Dv_j\neq 0$ implies
$D(v_j+v_k)\neq 0$.

It is sufficient to show that $\mathbf{D} P' = 0$. If this is the case, then 
\[(P - P')^T \mathbf{D} (P - P') = P^T \mathbf{D} P = \mathbf{E}.\] And consequently, $Q^T \mathbf{A} Q = \mathbf{B}$ by the correspondence between $\mathbf{A}$ and $\mathbf{D}$.  And $Q$ is invertible by Lemma~\ref{lem:block_invertible}. 

For an arbitrary integer $1 \leq \alpha \leq 2m$, 
Let $w$ be the $\alpha$-th column of $P$, i.e., $P e_\alpha$, and $v$ be the $\alpha$-th column of $P'$, i.e., $P' e_\alpha$. 
Let $i$ be the integer such that $m_{i - 1} + 1 \leq \alpha \leq m_i$. 
We use $v_j$ to denote the vector in $\oplus_{s = m_{j - 1} + 1}^{m_j} \langle e_s \rangle$ such that $v = \sum_j v_j$. 
We have $v_i = 0$. 

We show that $\mathbf{D} v_j = 0$ for every $j \neq i$ by contradiction.
Suppose there is a $j$ not equal to $i$ such that $\mathbf{D} v_j \neq 0$. 
Then there is a matrix $D$  in the tuple $\mathbf{D}$ such that $D v_j \neq 0$. Let $E$ be the counterpart in the tuple $\mathbf{E}$. 
Suppose $D_{j, k}$ and $D_{k, j}$ are non-zero for some $k, \ell \in [10]$. 
(If $D_{k, \ell}$ and $D_{\ell, k}$ for some $k \neq j$ and $\ell \neq j$ are non-zero, then $D v_j = 0$, contradicting to $D v_j \neq 0$.)

If $k = j$, then $D (v - v_j) = 0$, and thus $D v = D v_j \neq 0$. But by the construction of $\mathbf{D}$ and $\mathbf{E}$, the $\alpha$-th column of $E$ is zero because $j \neq i$. 
Since $P^TDP=E$ and $Ee_\alpha=0$,
we have $P^TDw=0$.
As $P$ is invertible, $Dw=0$.
But it should be $D w = D v \neq 0$ for this case. Contradiction. 

Now consider the case of $k \neq j$. 
If $k \neq i$, then $D v = D(v_j + v_k)$. 
By the construction of $D$, if $D v_j \neq 0$, then $D (v_j + v_k) \neq 0$. 
Hence, $D w = D v \neq 0$.
However, the $\alpha$-th column of $E$ is zero because $k \neq i$ and $j \neq i$. 
This contradicts the condition that $P^T \mathbf{D} P = \mathbf{E}$, which implies that $P^T D w = E e_\alpha$.

If $k = i$, then let $k' = k + 5$ if $k \leq 5$, or $k' = k - 5$ if $k > 5$. 
By the definition of the doubling construction, the tuple contains the corresponding matrix \(D'\), obtained by replacing the \(k\)-th block with its paired block \(k'\), with all other blocks zero.
Then $D'$ is zero on all the other induced submatrices. 
Let $E'$ be the counterpart of $D'$ in $\mathbf{E}$. 

Since $D v_j \neq 0$, by the correspondence between $D$ and $D'$, $D' v_j \neq 0$. Hence, $D' v = D' (v_j + v_{k'}) \neq 0$. 
Thus $D' w = D' v \neq 0$
However, the $\alpha$-th column of $E'$ is zero because $k' \neq i$ and $j \neq i$. 
This contradicts the condition that $P^T \mathbf{D} P = \mathbf{E}$, which implies that $P^T D' w = E' e_\alpha$. 

For the last property, by the second property, we have $Q^T \mathbf{A} Q = \mathbf{B}$. 
By Lemma~\ref{lem:block_invertible}, $Q_5$ is the identity matrix. 
The property is obtained by Lemma~\ref{lem:congruent_to_matching}
\end{proof}




Now we are ready to present our algorithm to determine if there are extended vector that match. 

\begin{framed}
\noindent \textbf{Extended Vector Matching algorithm}

\noindent \textbf{Input}: Group restrictions $(z_1, \dots, z_\zeta, Y, g_1, \dots, g_\gamma, F, \rho)$ and $(z_1', \dots, z_\zeta', X, g_1', \dots, g_\gamma', E, \rho')$

\noindent \textbf{Output}: Yes, or no. Representative vectors $\vec b$ and $\vec{a'}$, and a  set  $\Gamma_1$ of matrices.  
\begin{enumerate}
\item Compute representative vectors
$\vec s=(s_1,\ldots,
s_{\psi+\zeta_0+\eta})^T$
and
$\vec s'=(s_1',\ldots,
s_{\psi+\zeta_0+\eta}')^T$
for the two group restrictions.
Output no if
$\rho(s_i)\ne\rho'(s_i')$
for some
$1\leq i\leq\psi+\zeta_0+\eta$.
\item Compute arbitrary extended representative vectors $\vec{a} = (a_1, \dots, a_{\phi})^T$ for $\vec{s}$ and $\vec{a'} = (a_1', \dots, a_{\phi}')^T$ for $\vec{s'}$. 
\item Construct alternating matrix tuples $\mathbf{A}$ from $\vec{a}$ and $\mathbf{A}'$ from $\vec{a'}$ over $R$.
\item Construct doubling matrix tuple $\mathbf{D}$ and $\mathbf{D}'$ for $\mathbf{A}$ and $\mathbf{A}'$ respectively. 
\item Run the alternating matrix tuple isometry algorithm for $\mathbf{D}$ and $\mathbf{D}'$. Return no if the output is no, otherwise, let $P'$ be the output matrix, and $P = P'[m, m] \cdot P'$. 
\item 
 Compute a generating set $\Gamma_0$ of the group generated by the invertible $P'$ s.t. $P'^T \mathbf{D}' P' = \mathbf{D}'$. Let $\Gamma_1 = \{(P'[m, m]  \cdot P') : P' \in \Gamma_0\}$.  
\item Let $\vec b = (P[1:\phi, 1:\phi])^T \vec a$. 
\item Output yes, and Return $\vec b$, $\vec a'$,  and $\Gamma_1$.
\end{enumerate}
\end{framed}

\begin{lemma}\label{lem:matching_algo}
Given two group restrictions of two finite $p$-groups $G$ and $H$ of class two with order $p^n$, 
there is an algorithm to determine if the two group restrictions have extended representative vectors that match in time 
\[p^{O(\tau \cdot \log(\tau \cdot n) \cdot \max\{( n^{7 / 4} \cdot \delta^{7/ 2},  n^{3/2}\delta^7)\})},
\]
Furthermore, if the matching extended vectors exist, then the algorithm outputs two extended vectors $\vec a'$ and $\vec b$ for the two restrictions that match, and 
$\Gamma_1$ with size 
\[p^{\tau \cdot n^{7 / 4} \cdot \delta^{7/ 2} \log(\tau \cdot n)},\] a generating set of the invertible automorphisms $Q$ of $\mathbf{D}'$ with $Q[m, m] = 1$. 


\end{lemma}

\begin{proof}
Suppose the two group restrictions have two extended vectors that match. 
Then by Lemma~\ref{lem:matching_congruence_correspondance}, there exists an invertible block diagonal $Q$ such that $Q^T \mathbf{A} Q = \mathbf{A}'$. 
Hence, 
by Lemma~\ref{lem:doubling_matrix}, 
there exists a matrix $P'$ such that $P'^T \mathbf{D} P' = \mathbf{D}'$.
By Lemma~\ref{lem:doubling_matrix} and Lemma~\ref{lem:congruent_to_matching},  $\vec b$ is an extended vector that matches $\vec a'$.

If the two group restrictions does not have two extended vectors that match, 
then by Lemma~\ref{lem:matching_congruence_correspondance}, there does not exist an invertible block diagonal $Q$ such that $Q^T \mathbf{A} Q = \mathbf{A}'$. 
Then by Lemma~\ref{lem:doubling_matrix}, the algorithm outputs no. 

We show that $\Gamma_1$ generates the group of
isometries of $\mathbf D'$ with $(m,m)$-entry one.
Let
\[ S=\{P\in\GL(2m,R): P^T\mathbf D'P=\mathbf D'\}.
\]
By the construction, every $P\in S$
satisfies $P[m,j]=0$ for $j\ne m$ and
$P[m,m]^2=1$. Thus the map
$\chi:S\to R^\times$ defined by
$\chi(P)=P[m,m]$ is a homomorphism.

Define $\nu(P)=\chi(P)P$. Since
$\chi(P)^2=1$, we have
$\nu(P)\in S$ and
$\chi(\nu(P))=1$.
Moreover, for $P,Q\in S$,
\[
\nu(PQ)=\chi(PQ)PQ
       =\nu(P)\nu(Q).
\]
Hence $\nu$ is a homomorphism onto
$\ker\chi$: indeed, every
$Q\in\ker\chi$ satisfies $\nu(Q)=Q$.

Since $\Gamma_0$ generates $S$,
its image
\[
\Gamma_1=\{\nu(P):P\in\Gamma_0\}
\]
generates $\ker\chi$, as required.


To bound the running time, 
by the construction of the matrix tuple and doubling matrix tuple, 
 $m$, the dimension of the matrices in the matrix tuple, is $O(n \cdot \delta^2)$. 
 And the number of matrices in the matrix tuple is $\poly(n, \tau)$.
 By Lemma~\ref{lem:representative_vec_construction} and Theorem~\ref{thm:matrix_tuple_isometry_overall}, 
 the running time is obtained. 
\end{proof}

\subsection{Group restriction isomorphism}\label{sec:group_restriction_iso}

We give the final group restriction isomorphism algorithm based on the result given by the matrix tuple isometry solution. 
We solve the following problem: 
given two extended vectors $\vec a$ and $\vec b$ for two groups restrictions that match and a generating set of the automorphism group of the doubling matrix tuple constructed for $\vec a$, 
determine whether there is an extended vector $\vec c$ for the group restriction of $\vec a$ such that $\vec c $ and $\vec b$ implies restricted group isomorphism.





Given two matching extended vectors
$\vec a=(a_1,\ldots,a_\phi)^T$ and
$\vec b=(b_1,\ldots,b_\phi)^T$, let $\mathbf A$
be the matrix tuple constructed from $\vec a$.
Put $L=2m+\zeta$ and $t=\psi+\zeta_0$.
We use the following notation and generators.
\begin{enumerate}
\item For $1\leq i\leq\phi$, let
$d_i=\ord({\widetilde{a}_i})$, and let $d_i=1$ for
$\phi<i\leq L$. For $k\in[\zeta]$, let
\[q_k=\ord(z_k), L_k=q_k/p, 
 s_k=\max\{1,q_k/p^{\delta+1}\},
 h_k=z_k^{s_k}, \text{ and } e_k=\ord(h_k)=q_k/s_k.
\]
In particular, $e_k\leq p^{\delta+1}\leq p^\tau$, so $h_k\in F\cap Z(G)$.

\item Augment $\vec a$ and $\vec b$ to
$\vec x=(x_1,\ldots,x_L)^T\in G^L$ and
$\vec y=(y_1,\ldots,y_L)^T\in H^L$ by setting
\[
\begin{array}{ll}
 x_i=a_i,\ y_i=b_i & (1\leq i\leq\phi),\\
 x_i=1_G,\ y_i=1_H & (\phi<i\leq2m),\\
 x_{2m+k}=h_k,\ y_{2m+k}=(z_k')^{s_k}& (k\in[\zeta]).
\end{array}
\]
The orders $q_k$, and hence $s_k$, agree for the two restrictions.

\item For an augmented vector $\vec v$ of this form, denote
 \[v_i^{d_i}=\left(\prod_{k=1}^\zeta
 z_k^{\alpha_{i,k}}\right)y_i, \]
for  $y_i\in Y$ and  $0\leq\alpha_{i,k}<q_k$.
Define $\ell_{i,k}=\alpha_{i,k}\bmod L_k$ and
\[
 \theta_{\vec v}[i,k]=
 \frac{\alpha_{i,k}-\ell_{i,k}}{L_k}\in\mathbb F_p.
\]
For every augmented vector whose first $\phi$ entries match
$\vec a$, the numbers $\ell_{i,k}$ are independent of the
vector: this follows from the matching condition for $i\leq\phi$,
and from the fixed remaining coordinates for $i>\phi$.
We denote these fixed residues by $\ell_{i,k}$ throughout.

\item For each normalized generator $T\in\Gamma_1$ of the
doubled-tuple isometry group, let
\[
 Q_T=\diag(T_{1,1},\ldots,T_{5,5})
\]
be its first-copy block-diagonal part, using the five blocks
of the Matrix Tuple Construction. By
Lemma~\ref{lem:doubling_matrix}, $Q_T^T\mathbf A Q_T=\mathbf A$;
by normalization, its fifth block is $I_3$.
Add $\diag(Q_T,Q_T,I_\zeta)$ to $\Gamma$.

For each $t<i\leq\phi$ and $k\in[\zeta]$ with $q_k>d_i$, add
\[
 I_L+\frac{e_k}{p d_i}E_{2m+k,i}
\]
to $\Gamma$. The displayed coefficient is an integer because
$d_i\leq p^\delta$ and $q_k>d_i$.
For each $j\in[\zeta_0]$, putting $i=\psi+j$ and
$k=\zeta-\zeta_0+j$, also add
\[
 I_L+\frac{e_k}{p}E_{2m+k,i}
\]
to $\Gamma$. These are the only elementary central transformations.

\item For a generator $P\in\Gamma$, choose its entries
$\widehat P[j,i]$ as integers in $[0,p^\tau)$ and let $c_{ij}(P)= d_i\widehat P[j,i] / d_j$.

These are integers for the generators considered here.
Let $\beta_{\ell,j,k}$ be the coefficient of $z_k$ in
$[a_\ell,a_j]$ for $j,\ell\leq\phi$, and set
$\beta_{\ell,j,k}=0$ whenever $j>\phi$ or $\ell>\phi$.
For a matrix $\theta\in\mathbb F_p^{L\times\zeta}$, define
\begin{equation}\label{equ:final_encoding}
\begin{split}
 & P(\theta)[i,k] \\
 \equiv {}  & \sum_{j=1}^L c_{ij}(P)\theta[j,k]
 +\left\lfloor\frac{
 \displaystyle\sum_{j=1}^L c_{ij}(P)\ell_{j,k}
 +\binom{d_i}{2}\displaystyle\sum_{1\leq j<r\leq L}
 \widehat P[j,i]\widehat P[r,i]\beta_{r,j,k}}
 {L_k}\right\rfloor\pmod p.
\end{split}
\end{equation}
The second term records the carry from the lower coefficients.
All arithmetic inside the floor is over the integers, using
representatives $0,\ldots,p-1$ for the entries of $\theta$,
before reduction modulo $p$.
\end{enumerate}

\begin{lemma}\label{lem:group_action}
Let $\vec v\in G^L$ satisfy:
\begin{enumerate}
\item its first $\phi$ entries form an extended vector matching $\vec a$;
\item $v_i=1_G$ for $\phi<i\leq2m$;
\item $v_{2m+k}=h_k$ for $k\in[\zeta]$.
\end{enumerate}
For every generator $P\in\Gamma$, the ordered-product vector
$\vec v'=P^T\vec v$ satisfies these three conditions, and
\[
 P(\theta_{\vec v})=\theta_{\vec v'}.
\]
\end{lemma}

\begin{proof}
First consider a generator
$P=\diag(Q_T,Q_T,I_\zeta)$ obtained from $T\in\Gamma_1$.
Since $Q_T$ is block diagonal, all dummy identity coordinates
and the appended central coordinates remain fixed.
By Lemma~\ref{lem:congruent_to_matching}, the first $\phi$
coordinates of $\vec v'$ form an extended vector matching
$\vec v$ and hence $\vec a$.
For these generators, $d_j\mid d_i\widehat P[j,i]$ for
$j,i\leq\phi$ by compatibility; the other nonzero entries
lie among coordinates with $d_j=d_i=1$.

Next consider an elementary generator. For $t<i\leq\phi$ and
$q_k>d_i$, it multiplies $v_i$ by
\[
 h_k^{e_k/(p d_i)}=z_k^{q_k/(p d_i)}.
\]
This central element lies in $F$, has zero $\rho$-value,
and does not change quotient images, commutators, or the
$Y$-parts of powers. Its $d_i$-th power is $z_k^{L_k}$,
so the $Z$-power coefficients change only modulo $q_k$
by a multiple of $L_k$, as allowed by matching.
For $i=\psi+j$ and $k=\zeta-\zeta_0+j$, the elementary
generator instead multiplies $v_i$ by
$h_k^{e_k/p}=z_k^{L_k}$. This preserves the requirement
that $v_i$ generate its prescribed central cyclic subgroup:
the corresponding exponent changes by a multiple of $p$
after dividing by $\max\{1,q_k/p^\tau\}$.
It also preserves matching, and leaves the other entries fixed.
For elementary generators the divisibility
$d_j\mid d_i\widehat P[j,i]$ is immediate.
Thus all generators preserve the three conditions.

To verify the coefficient formula, for the given admissible
$\vec v$ write
\[
 v_j^{d_j}=\left(\prod_{k=1}^\zeta
 z_k^{\ell_{j,k}+L_k\theta_{\vec v}[j,k]}\right)y_j,
  y_j\in Y,
\]
where exponent representatives are taken modulo $q_k$.
By matching, the coefficients of $z_k$ in
$[v_r,v_j]$ equal $\beta_{r,j,k}$ when $j,r\leq\phi$;
for the other indices the commutators are trivial.
The class-$2$ power formula applied to
$v_i'=\prod_jv_j^{\widehat P[j,i]}$ gives
\[
\begin{split}
\alpha'_{i,k}\equiv{}&
 \sum_{j=1}^L c_{ij}(P)
 (\ell_{j,k}+L_k\theta_{\vec v}[j,k])\\
&+\binom{d_i}{2}
 \sum_{1\leq j<r\leq L}
 \widehat P[j,i]\widehat P[r,i]\beta_{r,j,k}
 \pmod{q_k},
\end{split}
\]
where $\alpha'_{i,k}$ is the coefficient of $z_k$ in
$(v_i')^{d_i}$. Since $q_k=pL_k$, taking the quotient by
$L_k$ and reducing modulo $p$ gives exactly
Equation~\eqref{equ:final_encoding}.
Therefore $P(\theta_{\vec v})=\theta_{\vec v'}$.
The lemma applies successively to every generator in an
orbit computation.
\end{proof}

\begin{framed}

\noindent \textbf{Group Restriction Isomorphism Algorithm}

\noindent \textbf{Input}: Two group restrictions $(z_1, \dots, z_\zeta, Y, g_1, \dots, g_\gamma, F, \rho)$ and $(z_1', \dots, z_\zeta', X, g_1', \dots, g_\gamma', E, \rho')$. 
\noindent \textbf{Output}: Yes or no.
\begin{enumerate}
\item Output no if the projection of $g_i^{\ord({\widetilde{g}_i})}$ onto $z_1, \dots, z_\zeta$ has different coefficients to the projection of $(g_i')^{\ord({\widetilde{g}_i'})}$ onto $z_1', \dots, z_\zeta'$ for some $1 \leq i \leq \gamma$. 

\item Run the Extended Vector Matching Algorithm for the two group restrictions. If the output is no, then return no. Otherwise, let $\vec a, \vec b$ and $\Gamma_1$ be the output. 


\item Construct the augmented vectors
$\vec x$ and $\vec y$ and the group
generators $\Gamma$.

\item Let $\Psi=\{\theta_{\vec x}\}$.

\item Repeat until $\Psi$ does not change:
for each $\theta\in\Psi$ and $P\in\Gamma$,
add $P(\theta)$ to $\Psi$.

\item Output yes if
$\theta_{\vec y}\in\Psi$,
and no otherwise.

\end{enumerate}
\end{framed}

\begin{lemma}\label{lem:vec_in_psi}
For every extended vector $\vec c=(c_1,\ldots,c_\phi)^T$
matching $\vec a$ for the same group restriction, let
$\vec w\in G^L$ be its augmented vector as defined above.
Then $\theta_{\vec w}\in\Psi$.
\end{lemma}

\begin{proof}
Let $t=\psi+\zeta_0$, and $\mathbf A(\vec u)$ denote the tuple constructed from an extended
vector $\vec u$ of the given restriction.  As $\vec c$ matches $\vec a$,
the tuples $\mathbf A(\vec c)$ and $\mathbf A(\vec a)$ are identical.
Indeed, matching gives equality in every step of the construction
except possibly step 5; equality of the step 5 coordinates follows
from the commutator-pairing argument in the proof of
Lemma~\ref{lem:matching_congruence_correspondance}.

Apply 
Lemma~\ref{lem:matching_congruence_correspondance}, with
$\vec b=\vec c$ and $\vec v=\vec c$.  It supplies a compatible
$Q_1\in\GL(\phi,R)$ and an invertible block diagonal matrix 
$Q=\diag(Q_1,Q_2,Q_3,Q_4,I_3), Q^T\mathbf A Q=\mathbf A$,
such that, writing $\vec s=Q_1^T\vec a$ using the ordered-product
convention,
\begin{equation}\label{eq:vecpsi-close}
 s_i=c_i\quad(1\leq i\leq t),
 s_i=c_i z_i\quad(t<i\leq\phi),
 z_i\in\oplus_{k=1}^{\zeta-\zeta_0}\langle z_k\rangle.
\end{equation}
Let $\vec s'$ be the augmented vector obtained from $\vec s$
by adjoining the prescribed dummy and central coordinates.
In particular, $z_i^p=1$.  Since $d_i=\ord({\widetilde{a}_i})$ is divisible by
$p$ for $i>t$, Equation~\eqref{eq:vecpsi-close} implies
\begin{equation}\label{eq:vecpsi-target}
 s_i^{d_i}=c_i^{d_i}\quad(1\leq i\leq\phi),
 \theta_{\vec s'}=\theta_{\vec w}
 \quad\text{on the first $\phi$ rows}.
\end{equation}
(The equality of the $\theta$-matrices also holds on the fixed
auxiliary rows.)

Let $\mathbf D$ be the doubling of $\mathbf A$ and put
$P=\diag(Q,Q)$.  By Lemma~\ref{lem:doubling_matrix}, $P$ is a
normalized isometry of $\mathbf D$.  We next justify that a word in
the normalized generators, after replacing each factor by its
block-diagonal part, produces exactly the power coefficients in
\eqref{eq:vecpsi-target}.  This point cannot be obtained merely from
agreement modulo $Z(G)$.

Let
\[
 N=\bigcap_{A\in\mathbf A}\ker A\subseteq R^m,
  N_D=\bigcap_{D\in\mathbf D}\ker D
       =N\oplus N\subseteq R^{2m}.
\]
The construction gives the following two properties of
$u\in N$:
\begin{equation}\label{eq:vecpsi-kernel}
 \begin{aligned}
 &u_i=0 &&(t<i\leq\phi),\\
 &u_i\in\ord(a_i)R
       &&(1\leq i\leq\psi),\\
 &u_i\in(\ord(a_i)/p)R
       &&(\psi<i\leq t),\\
 &u\in pR^m.&&
 \end{aligned}
\end{equation}
Here the first line follows from the unit entries in step 4;
the second from the diagonal $Y$-power entries in step 2;
and the third from the distinguished $Z_0$-power entries in
step 3.  To see the last line, apply steps 2 and 3 to the
remaining columns of block~2, step 8 to its distinguished
$Z_0$ columns, step 4 to block~3, step 5 to block~4, and
step 7 to block~5.  Their unit or nonzero scaled entries force
every remaining coordinate of a common-kernel vector to be a
multiple of $p$.  In the first block this already follows from
the preceding three lines, since the distinguished $Z_0$
coordinates have order at least $p^2$.  In particular,  if $u\in pN$,
\begin{equation}\label{eq:vecpsi-pkernel}
 u_i=0\ (i>t), a_i^{u_i}=1\ (i\leq t).
\end{equation}

For an isometry $T$ of $\mathbf D$, let $H(T)$ be its ten-block
diagonal part and put $E(T)=T-H(T)$.  By
Lemma~\ref{lem:doubling_matrix}, every column of $E(T)$ lies in
$N_D$.  By \eqref{eq:vecpsi-kernel}, every entry of $E(T)$
is divisible by $p$.  For any two isometries $T,U$,
\[
 H(TU)-H(T)H(U)=H\bigl(E(T)E(U)\bigr).
\]
Every column of $E(T)E(U)$ lies in $pN_D$: the columns
of $E(T)$ lie in $N_D$ and the entries of $E(U)$ are multiples
of $p$.  The common kernel $N_D$ is a direct sum of its
intersections with the ten coordinate blocks: for each tuple matrix,
its two nonzero off-diagonal blocks have disjoint image supports, so
the separate coordinate-block components of a common-kernel vector
again belong to the common kernel.  Thus the block-diagonal projection
$H$ preserves $pN_D$.  Moreover, $H(T)$ is an invertible isometry by
Lemma~\ref{lem:doubling_matrix}, so it preserves
$N_D$.  Induction therefore shows that, for any word
$P=T_r\cdots T_1$ in normalized isometries,
\begin{equation}\label{eq:vecpsi-diagonal-word}
 H(T_r)\cdots H(T_1)-H(P)
 \quad\text{has every column in }pN_D.
\end{equation}

Since $\Gamma_1$ generates all normalized isometries, write
$P=\diag(Q,Q)$ as a word in $\Gamma_1$.
For each factor $T$ of that word, use the corresponding
$\diag(Q_T,Q_T,I_\zeta)\in\Gamma$.
If inverses occur in the word, replace them by positive powers,
since the group is finite.  Let $M$ be the product, in the same
order, of the first $\phi\times\phi$ blocks of these projected
generators.  Since $H(P)=\diag(Q,Q)$,
\eqref{eq:vecpsi-diagonal-word} and \eqref{eq:vecpsi-pkernel}
show that the ordered-product vector $M^T\vec a$ is \emph{exactly}
$Q_1^T\vec a=\vec s$: the only possible discrepancies in the
first block occur in central coordinates, with exponents
annihilating their corresponding elements.

Now apply the projected generators successively to the augmented
vector $\vec x$, and call the result $\vec v$.  Every intermediate
vector is admissible by Lemma~\ref{lem:group_action}.
For $i\leq t$, all factors contributing to the $i$-th coordinate
are central, by compatibility, so successive ordered products
agree exactly with $M^T\vec a$ in that coordinate.  For $t<i\leq\phi$,
compatibility ensures that every original element $a_j$ contributing
to that coordinate has $\ord({\widetilde{a}_j})\leq d_i$.  Reassociating and
reordering the successive products therefore changes the resulting
element only by a product of commutators $[a_j,a_k]$ of order
dividing $d_i$, because $G$ has class~$2$.  These commutators are
central and vanish upon taking the $d_i$-th power.  Consequently,
\[
 v_i^{d_i}=(M^T\vec a)_i^{d_i}=s_i^{d_i}
 (1\leq i\leq\phi).
\]
The dummy and appended coordinates remain fixed.  Hence
\[
 \theta_{\vec v}=\theta_{\vec s'}=\theta_{\vec w}.
\]
By Lemma~\ref{lem:group_action}, $\theta_{\vec v}\in\Psi$.
Therefore $\theta_{\vec w}\in\Psi$, as required.
\end{proof}

\begin{proof}[Proof of Theorem~\ref{thm:restricted_group_iso_algo}]
Suppose that the two group restrictions are isomorphic. 
By Lemma~\ref{lem:matching_algo}, the extended vector matching algorithm must output two extended vectors $\vec a$ and $\vec b$ that match and a generating set which generates all the matrices $P'^T \mathbf{D} P' = \mathbf{D}$, where $\mathbf{D}$ is the doubling matrix tuple for $\vec a$. 

Since they are group restriction isomorphic, let $f$ be a group restriction isomorphism. 
Let $\vec c=(c_1,\dots,c_\phi)^T$,
where $c_i=f^{-1}(b_i)$ for $1\leq i\leq \phi$.
We have $\theta_{\vec x_c} = \theta_{\vec x_b}$, where $\vec x_c$ and $\vec x_b$ are the vectors of length $L$ constructed for $\vec c$ and $\vec b$ respectively. 
Then $\vec c$ is a valid extended vector that matches $\vec a$.  
By Lemma~\ref{lem:vec_in_psi}, $\theta_{\vec x_c}$ is in $\Psi$. So 
$\theta_{\vec x_b}$ is in $\Psi$, and thus
the algorithm outputs yes.

Now we show that if the algorithm outputs yes, then there is a restriction isomorphism.
Since the algorithm outputs yes, we have
$\theta_{\vec y}\in\Psi$.
By Lemma~\ref{lem:group_action}, there is
an admissible vector $\vec v$ such that
$\theta_{\vec v}=\theta_{\vec y}$.
The lower $Z$-power coefficients agree by matching, and
the highest digits agree by the equality of $\theta$.
Hence the full projected power coefficients
of $v_i^{d_i}$ and $b_i^{d_i}$ agree for
every $1\leq i\leq \phi$.
Together with the matching conditions,
Lemma~\ref{lem:restricted_group_iso_condition}
implies that the two group restrictions
are isomorphic.

Now we bound the running time.
Let $n = \log_p |G|$. 
By Lemma~\ref{lem:matching_algo},  the running time of extended vector matching algorithm is 
\[p^{O(\tau \cdot \log(\tau \cdot n) \cdot \max\{n^{7 / 4} \cdot \delta^{7/ 2}, n^{3/2}\delta^7\})},
\]
with a generating set of size 
\[p^{O\left(\tau \cdot  n^{7 / 4} \cdot \delta^{7/2} \log(\tau \cdot n)\right)}.\]  
By Lemma~\ref{lem:group_action} and Lemma~\ref{lem:vec_in_psi}, only the first $\phi$ rows of $\theta$ vary; hence $|\Psi|\leq p^{\phi\zeta}=p^{O(n\zeta)}$. 
So the running time of the Step 4 of the algorithm is 
\[p^{O\left(\tau \cdot  n^{7 / 4} \cdot \delta^{7/2} \log(\tau \cdot n)\right)}\cdot p^{O( n \cdot \zeta)} = p^{O\left( \max\{ \tau n^{7/4} \delta^{7/2} \log(\tau \cdot n), n \zeta \}\right)}.\]  
So the overall running time is 
\[\begin{split} 
& p^{O(\tau \cdot \log(\tau \cdot n) \cdot \max\{n^{7 / 4} \cdot \delta^{7/ 2}, n^{3/2}\delta^7\})}
+ p^{O\left( \max\{\tau n^{7/4} \delta^{7/2} \log(\tau \cdot n), n \zeta \}\right)} \\ = &  
p^{O\left(    \max\left\{ \tau n^{7 / 4} \cdot \delta^{7/ 2} \log(\tau \cdot n) ,  \tau \cdot n^{3/2} \delta^7 \log(\tau \cdot n),  n\cdot \zeta \right\}\right)}.\qedhere
\end{split}\]  
\end{proof}

\bibliography{references}
\bibliographystyle{alpha}

\appendix

\section{Alternating matrix tuple isometry}\label{sec:matrix_tuple_isometry}

We prove Theorem~\ref{thm:matrix_tuple_isometry_overall} by proving the two theorems below in this section. 
The theorem mostly follows the framework of~\cite{grochow2026beyond} for alternating matrix tuple isometry over $\F_2$. 
Here, we generalize the result to matrix tuples over the finite ring $\Z / 2^\tau \Z$.
Throughout this section, we suppose $p = 2$, but the result holds for any prime.

\begin{theorem}\label{thm:matrix_tuple_isometry}
Let $R = \Z / p^\tau \Z$ for some prime $p$ and positive integer $\tau$. 
Let $\mathbf{A}$ and $\mathbf{B}$ be two alternating matrix tuples in $\M(n, R)^m$ for some positive integers $n, m, \tau$ and a prime number $p$. 
There is an algorithm to determine if there is a matrix $P \in \GL(n, R)$ such that $P^T \mathbf{A} P = \mathbf{B}$ in time $p^{O(\tau n^{7/4} \log (\tau n))} \cdot \poly(m)$. Furthermore, if the output is yes, the algorithm also outputs an invertible matrix $P$ such that $P^T \mathbf{A} P = \mathbf{B}$. 
\end{theorem}

In addition, we also prove the following result on computing a generating set of the automorphism group of a matrix tuple. 

\begin{theorem}\label{thm:matrix_tuple_automorphism}
Let $R = \Z / p^\tau \Z$ for some prime $p$ and positive integer $\tau$. 
Let $\mathbf{A}$  be an alternating matrix tuple in $\M(n, R)^m$ for some positive integers $n$ and  $m$. 
There is an algorithm to output a generating set of the invertible automorphism group of size $p^{\tau\cdot n^{1.5}} \cdot \poly(m)$ in time $p^{O(\tau n^{7/4} \log (\tau n))}$.
\end{theorem}



Throughout this section, unless specified, we assume that all matrix tuples are non-degenerate, i.e., 
there are no invertible matrices $X$ and $Y$ such that some row of $X^T \mathbf{A}Y$ is zero in every matrix of the tuple
or some column of $X^T \mathbf{A}Y$ is zero in every matrix of the tuple. 
This assumption is without loss of generality because we can preprocess the common zero direct summands and reduce to non-degenerate case. 

We use $R$ to denote $\Z / p^\tau \Z$ in this section. 
Throughout the paper, when we compute a new matrix tuple using an invertible congruence operation, we keep the matrix used for backtracking.

\subsection{Left-right block diagonalization of matrix tuples}

We show some results for left-right block diagonalization of matrix tuples here. In this subsection, the matrix tuples are not necessarily square matrix tuples.


\begin{lemma}\label{lem:non-degenerate_matrix_tuple_invertible}
Let $\mathbf{A}$ be an $n \times r$ non-degenerate matrix tuple, and $Y$ be a matrix in $\M(r, R)$. 
If, for every $v \in R^r$, there exists $v' \in R^r$ such that 
\begin{equation}\label{equ:non-degenerate_matrix_tuple}
\mathbf{A} v = \mathbf{A} Y v',
\end{equation} then $Y$ is invertible. 
\end{lemma}

\begin{proof}

Let $\Gamma = \{v \in R^r : \mathbf{A} v = 0 \}$. 
For every $v \in \Gamma$, every coordinate of $v$ lies in $pR$. Indeed, if some coordinate of $v$ were a unit, then $v$ could be extended to a basis of $R^r$. Hence there would exist $Z\in \GL(r,R)$ having $v$ as one of its columns, so $\mathbf{A}Z$ would contain a zero column, contradicting the non-degeneracy of $\mathbf{A}$.

By Equation~(\ref{equ:non-degenerate_matrix_tuple}), for every $1 \leq i \leq r$,  there exists a vector $v_i'$ such that $\mathbf{A} e_i = \mathbf{A} Y v_i'$, where $e_i$ is the vector whose $i$-th coordinate is one and whose other coordinates are zero. 
Hence, $Y v_i' = e_i + z_i$ for some $z_i \in \Gamma$.   
Since every coordinate of $z_i$ is a multiple of $p$,  for each $i$, there is a linear combination of $v_1', \dots, v_r'$, denoted by $x_i$, such that $Y x_i = e_i + y_i$ for some $y_i \in \langle e_{i+1}, \dots, e_r\rangle$. 
Then there exists $x_1', \dots, x_r'$ such that $Y x_i' = e_i$, and thus \[Y \begin{pmatrix}x_1' & \dots & x_r'\end{pmatrix} = I.\]
Consequently, $Y$ is invertible. 
\end{proof}

\begin{definition}
    Two matrix tuples $\mathbf{A}, \mathbf{A}' \in \M(n \times r, R)^m$ are \emph{left-right equivalent} if there are invertible matrices $U \in \GL(n, R)$ and $W \in \GL(r, R)$ such that $U^T \mathbf{A} W = \mathbf{A}'$. 

    Two matrix tuples $\mathbf{A}, \mathbf{A}' \in \M(n \times n, R)^m$ are \emph{congruent} 
    if there is an invertible matrix $U \in \GL(n, R)$ such that $U^T \mathbf{A} U = \mathbf{A}'$. 
\end{definition}

\begin{definition}[Left-right block and minimal left-right block]\label{def:block}
For a matrix tuple $\mathbf{A} \in \M(n \times r, R)^m$, let $X \in \M(n \times n', R)$, $Y \in \M(r \times r' , R)$, where $n'$ and $r'$ are positive integers.
We say that 
$X^T \mathbf{A} Y$ is a \emph{left-right block} of $\mathbf{A}$ if there exist matrices $X' \in \M(n \times (n - n'), R )$ and $Y' \in \M(r\times (r - r'), R)$ such that invertible matrices 
\[
\begin{pmatrix}
    X & X'
\end{pmatrix} \in \GL(n, R) \text{ and }\begin{pmatrix}
    Y & Y'
\end{pmatrix} \in \GL(r, R)
\] satisfy
\[
\begin{pmatrix}
    X & X'
\end{pmatrix}^T \mathbf{A} \begin{pmatrix}
    Y & Y'
\end{pmatrix}
=\begin{pmatrix}
    X^T \mathbf{A} Y & \\ & X'^T \mathbf{A} Y'  
\end{pmatrix}. 
\]
$X^T \mathbf{A} Y$ is a \emph{minimal left-right block} of $\mathbf{A}$ if there does not exist $U \in \M(n' \times n'', R)$ and $W \in \M(r' \times r'', R)$ for some $1 \leq n'' < n'$ and $1 \leq r'' < r'$ such that $U^T X^T \mathbf{A} Y W$ is a left-right block of $\mathbf{A}$.

Moreover, we say $\diag(X_1^T\mathbf{A}Y_1, \dots, X_k^T \mathbf{A}Y_k)$ is a minimal left-right block diagonalization of $\mathbf{A}$ if every $X_i^T \mathbf{A} Y_i$ is a minimal left-right block of $\mathbf{A}$ and 
both \[\begin{pmatrix}
    X_1 & \dots & X_k
\end{pmatrix} \text{ and } \begin{pmatrix}
    Y_1 & \dots & Y_k
\end{pmatrix}\]
are invertible. 

\end{definition}

Since we assume that the matrix tuples considered in this section are non-degenerate,
every minimal left-right block is also non-degenerate. 

\begin{lemma}
Every minimal left-right block diagonalization of a matrix tuple over $R$ is unique up to permutation of the blocks and left-right equivalence of corresponding blocks.
\end{lemma}
\begin{proof}
Let $U = R^r$, and $V = R^n$. Each $A_i$ defines an $R$-linear map $A_i: U \rightarrow V$ with $A_i(u)$ being $A_i u$ for any $u \in U$. 
Consider the quiver
\begin{center}
    $ \begin{tikzcd}
            U \arrow[r, "A_1{,}\dots{,} A_m"] & V 
        \end{tikzcd} $
 \end{center}
The representation $\mathcal{A} = (U, V; A_1, \dots, A_m)$ is a representation of the quiver. 
Hence, for any matrices $P \in M(n, R)$ and $Q \in M(r, R)$ such that $P A_i = A_i Q$ for any $1 \leq i \leq m$, $(P, Q)$ is an endomorphism of the representation $\mathcal{A}$.

A direct-sum decomposition of this representation into $k$ components is exactly a simultaneous block diagonalization of the matrix tuple into $k$ blocks. 
Furthermore, if each block of the block diagonalization is minimal, then each component of the direct sum decomposition is an indecomposable representation. 

Since $R$ is Artinian and $U,V$ are finitely generated $R$-modules, the representation has finite length. Hence the Krull–Schmidt theorem applies.

So, if $M_1 \oplus \dots \oplus M_\alpha$ and $M_1' \oplus \dots \oplus M_\beta'$ are two decompositions into indecomposable representations, then $\alpha = \beta$, and after a suitable permutation, $M_i \cong M_i'$. 
\end{proof}

\begin{lemma}\label{lem:fitting}
Let $X^T \mathbf{A} Y$ be a non-degenerate $s \times u$ minimal left-right block of $\mathbf{A}$. 
For any $S \in \M(s, R)$ such that there exists $U \in M(u, R)$ satisfying $S X^T \mathbf{A} Y = X^T \mathbf{A} Y U$, 
if $S$ is not an invertible matrix, then $(I + S) X^T \mathbf{A} Y (I + U)^{-1}= X^T \mathbf{A} Y$. 
\end{lemma}
\begin{proof}
Let $\mathbf{B} = (B_1, \dots, B_m)$ be the matrix tuple $X^T \mathbf{A} Y$. 
Since $\mathbf{B}$ contains only a single minimal left-right block, 
$\mathcal{B} = (R^u, R^s; B_1, \dots, B_m)$ is indecomposable and has finite length. 
By the Fitting lemma, the endomorphism ring of $\mathcal{B}$ is local. Hence every non-invertible endomorphism of $B$ is nilpotent. Since $S$ is not invertible, the endomorphism $(S,U)$ is nilpotent. In particular, both $S$ and $U$ are nilpotent. Therefore $I+S$ and $I+U$ are invertible.
Then we have
$(I + S) X^T \mathbf{A} Y (I + U)^{-1} = X^T \mathbf{A} Y$.
\end{proof}

\begin{definition}\label{def:block_space}
For a minimal left-right block $\mathbf{B} = X^T \mathbf{A} Y$ of $\mathbf{A}$ consisting of $n'$ rows and $r'$ columns, let $\Gamma_\mathbf{B}$ be the set 
\[\begin{split}\Gamma_\mathbf{B} := \{X' \in \M(n \times n', R): & \exists Y' \in \M(r \times r', R)\text{ s.t. }  \\ & X'^T \mathbf{A} Y' \text{ is a left-right block and } X'^T \mathbf{A} Y' =\mathbf{B}\}. \end{split}\] 
Let   
$V_{\mathbf{B}} := \langle \Gamma_\mathbf{B} \rangle$, i.e., the linear span of the matrices in $\Gamma_{\mathbf{B}}$ as elements of an $R$-module, and $K_\mathbf{B}$ be $V_\mathbf{B} \setminus \Gamma_\mathbf{B}$. 

\end{definition}

For invertible matrices $U$ and $W$, we say that $U^T \mathbf{A} W$ is a left-right block diagonalization of $\mathbf{A}$ if $U^T \mathbf{A} W = \diag(\mathbf{A}_1,\dots, \mathbf{A}_k)$ such that each of the $\mathbf{A}_i$ is a left-right block of $\mathbf{A}$, and the left-right block diagonalization is minimal if each of $\mathbf{A}_i$ is a minimal left-right block of $\mathbf{A}$.


Let $V_1, \dots, V_k$ be submodules of $R^n$ such that $R^n = V_1 \oplus \dots \oplus V_k$. 
Then any vector $v\in R^n$ has a unique decomposition $v = v_1 + \dots + v_k$, where $v_i \in V_i$.  
Let $W$ be a matrix with $n$ rows over $R$. The projection of $W$ onto $V_i$ with respect to $V_1 \oplus \dots \oplus V_k$ is another matrix $W'$ of the same dimensions as $W$ such that the $j$-th column of $W'$ is the projection of the $j$-th column of $W$ onto $V_i$ with respect to $V_1 \oplus \dots \oplus V_k$.

The following theorem was proved in Theorem 1.3 and Lemma 7.2 of \cite{qiao2024canonical} for $\F_p$. 
We show that the same result holds over $R$, including the case $p=2$.

\newcommand{\columnspan}{\mathrm{columnspan}}

\begin{theorem}\label{thm:left_right_decomposition}
Let $\mathbf{A} \in \M(n \times r, R)^m$ be a matrix tuple over $R$ for some prime $p$. 
Let $\mathbf{B} = X^T \mathbf{A} Y$ be a minimal left-right block of $\mathbf{A}$ with $s$ rows and $t$ columns. 
Then we have
\begin{enumerate}
\item

$\langle K_\mathbf{B} \rangle  = K_\mathbf{B}$. Thus $K_B$ is a submodule of $V_B$.


\item 
Let $U = \begin{pmatrix}U_1 \cdots U_k \end{pmatrix}$ and $W =\begin{pmatrix}W_1  \cdots  W_k \end{pmatrix}$ be two invertible matrices such that 
$U^T \mathbf{A} W = \diag(U_1^T \mathbf{A} W_1, \dots, U_k ^T\mathbf{A} W_k)$ is a left-right block diagonalization of $\mathbf{A}$. 
The projection of $X$ on any $U_i$ with respect to $\columnspan(U_1) \oplus \dotsb \oplus \columnspan(U_k)$ is in $V_\mathbf{B}$, where $\columnspan(U_i)$ denotes the linear span of the columns of $U_i$.  
\item Let $X_1, \dots, X_k$ be the projection of $X$ onto $U_1, \dots, U_k$. 
Let $\Gamma$ be a subset of $[k]$.
If there exists $i \in \Gamma$ such that $X_i \in V_\mathbf{B} \setminus K_\mathbf{B}$, then $\sum_{i \in \Gamma} X_i$ is in $V_\mathbf{B} \setminus K_\mathbf{B}$.  
\end{enumerate}
\end{theorem}
\begin{proof}
We regard the transposed tuples as representations
\[
\mathcal M=(R^s,R^t;B_1^T,\ldots,B_m^T),
\mathcal N=(R^n,R^r;A_1^T,\ldots,A_m^T),
\]
where $\mathbf B=(B_1,\ldots,B_m)$ and
$\mathbf A=(A_1,\ldots,A_m)$.
Since the ambient tuple is non-degenerate, every direct summand
with zero source or zero target is excluded. Thus minimality of
$\mathbf B$ implies that $\mathcal M$ is an indecomposable
direct summand of $\mathcal N$.
Thus $E=\operatorname{End}(\mathcal M)$ is a local ring
by the Fitting lemma. Write $\mathfrak m=\operatorname{Rad}(E)$.

Let $H=\operatorname{Hom}(\mathcal M,\mathcal N)$.
For $f\in H$, write $f=(f_0,f_1)$ for its maps on
$R^s$ and $R^t$, and let
\[
\alpha:H\longrightarrow\M(n\times s,R),
\alpha(f)=f_0.
\]
The set $\Gamma_{\mathbf B}$ is precisely the set of
$\alpha(f)$ for split monomorphisms $f:\mathcal M\to\mathcal N$:
a split monomorphism can be completed to a direct-sum
decomposition, giving a left-right block equal to $\mathbf B$,
and conversely each such block gives a split monomorphism
of the transposed representations.

Define
\[
H_0=\{f\in H:\ gf\in\mathfrak m
\text{ for every }g\in\operatorname{Hom}(\mathcal N,\mathcal M)\}.
\]
This is an $R$-submodule of $H$. Moreover, $f\notin H_0$
if and only if $f$ is a split monomorphism: indeed,
$gf\notin\mathfrak m$ means that $gf$ is invertible in $E$,
and then $(gf)^{-1}g$ is a left inverse of $f$.

Choose a fixed split monomorphism $e:\mathcal M\to\mathcal N$
and a retraction $q:\mathcal N\to\mathcal M$ with $qe=I$.
If $f\in H_0$, then $qf\in\mathfrak m$, so $I+qf$
is invertible. Consequently $e+f$ is a split monomorphism,
since $(I+qf)^{-1}q$ is a left inverse of $e+f$.
Thus $f=(e+f)-e$ is a difference of two split
monomorphisms. Every element outside $H_0$ is itself a split
monomorphism. It follows that
\[
V_{\mathbf B}=\alpha(H).
\]

Furthermore, $\ker\alpha\subseteq H_0$.
Indeed, if $f_0=0$, the first component of $gf$ is zero
for every $g$, so $gf$ cannot be invertible in $E$.
Therefore $\alpha(H_0)$ and $\alpha(H\setminus H_0)$
are disjoint: otherwise the difference of two preimages
would lie in $\ker\alpha\subseteq H_0$.
Since $\Gamma_{\mathbf B}=\alpha(H\setminus H_0)$,
we obtain
\[
K_{\mathbf B}=V_{\mathbf B}\setminus\Gamma_{\mathbf B}
=\alpha(H_0).
\]
Hence $K_{\mathbf B}$ is an $R$-submodule of $V_{\mathbf B}$,
proving the first property.

For the second property, let $e:\mathcal M\to\mathcal N$
be the split monomorphism corresponding to the given block
$\mathbf B=X^T\mathbf A Y$, so $e_0=X$.
For the left-right block diagonalization in the theorem,
let $\pi_i:\mathcal N\to\mathcal N$ be the idempotent
projection onto its $i$-th direct summand.
Then $\pi_i e\in H$, and the first component of
$\pi_i e$ is exactly the projection of $X$ onto
$\columnspan(U_i)$ with respect to
$\bigoplus_j\columnspan(U_j)$.
It therefore belongs to $\alpha(H)=V_{\mathbf B}$,
as required.

For the third property, put $\pi_\Gamma=\sum_{i\in\Gamma}\pi_i$, an
idempotent endomorphism of $\mathcal N$. Since the first component of
$\pi_i e$ is $X_i$, we have
\[
  \alpha(\pi_i e)=X_i, \alpha(\pi_\Gamma e)=\sum_{i\in\Gamma}X_i .
\]
The submodule $H_0$ is closed under left composition with
$\operatorname{End}(\mathcal N)$: if $f\in H_0$ and
$\varphi\in\operatorname{End}(\mathcal N)$, then for every
$g\in\operatorname{Hom}(\mathcal N,\mathcal M)$ we have
$g(\varphi f)=(g\varphi)f\in\mathfrak m$, because
$g\varphi\in\operatorname{Hom}(\mathcal N,\mathcal M)$.

Let $i\in\Gamma$ satisfy $X_i\in V_{\mathbf B}\setminus K_{\mathbf B}
=\Gamma_{\mathbf B}$. Since $\alpha(H_0)=K_{\mathbf B}$, this gives
$\pi_i e\notin H_0$. Suppose, for contradiction, that
$\pi_\Gamma e\in H_0$. The $\pi_j$ are orthogonal idempotents, so
$\pi_i\pi_\Gamma=\pi_i$, and by the closure property
$\pi_i e=\pi_i(\pi_\Gamma e)\in H_0$, a contradiction. Hence
$\pi_\Gamma e\in H\setminus H_0$, and therefore
\[
  \sum_{i\in\Gamma}X_i=\alpha(\pi_\Gamma e)\in
  \alpha(H\setminus H_0)=\Gamma_{\mathbf B}
  =V_{\mathbf B}\setminus K_{\mathbf B}.\qedhere
\]
\end{proof}

\subsection{Congruent block diagonalization of alternating matrix tuples}

For an alternating matrix tuple $\mathbf{A} \in \M(n, R)^m$ and a matrix $X \in \M(n \times n',  R)$ for some $0 < n' < n$, we say that $X^T \mathbf{A} X$ is a congruent block of $\mathbf{A}$ if there exists another matrix $X' \in \M( n \times (n - n'),  R )$ such that the invertible matrix
\[
\begin{pmatrix}
    X & X'
\end{pmatrix} \in \GL(n,  R)
\] satisfies
\[
\begin{pmatrix}
    X & X'
\end{pmatrix}^T \mathbf{A} \begin{pmatrix}
    X & X'
\end{pmatrix}
=\begin{pmatrix}
    X^T \mathbf{A} X & \\ & X'^T \mathbf{A} X'
\end{pmatrix}. 
\]

The following lemma was proved in~\cite{grochow2026beyond} for $\F_2$. 
We show that the same result holds over $R = \Z / p^\tau \Z$, including the case $p=2$.

\begin{theorem}\label{thm:congruence_block_diagonalization}
Let $X^T \mathbf{A}Y$ be a minimal left-right block for an alternating matrix tuple $\mathbf{A}$ over  $R$ such that the number of columns of $X$ is less than or equal to the number of columns of $Y$. Then one of the following two conditions holds:
\begin{enumerate}
\item $X^T \mathbf{A}X$ is a congruent block of $\mathbf{A}$, and $X$ has the same number of columns as $Y$. 
\item $\begin{pmatrix}X & Y \end{pmatrix}^T  \mathbf{A}\begin{pmatrix}X & Y \end{pmatrix}$ is a congruent block of $\mathbf{A}$. 
\end{enumerate}
\end{theorem}

We first prove the following lemma. 

\begin{lemma}\label{lem:block_diagonal_basic}
Suppose $X^T \mathbf{A} Y$ is a minimal left-right block of an alternating matrix tuple $\mathbf{A} \in \M(n,  R)^m$, and the number of columns of $X$ is less than or equal to that of $Y$.
If there exists an invertible matrix Z such that $X^T \mathbf{A} X Z = X^T \mathbf{A} Y$, then $X^T \mathbf{A} X$ is both a left-right block and a congruent block of $\mathbf{A}$. 
\end{lemma}

\begin{proof}
Let $d_x$ and $d_y$ denote the number of columns of $X$ and $Y$ respectively. 
We have $d_x = d_y$ because $X^T \mathbf{A} X Z = X^T \mathbf{A} Y$ for some invertible $Z$. Hence, $X$ has the same number of columns as $Y$. 

Since $X^T \mathbf{A} Y$ is a left-right block of $\mathbf{A}$, there exist $X'$ and $Y'$ such that 
\[\begin{pmatrix}X & X'\end{pmatrix}^T \mathbf{A} \begin{pmatrix}Y & Y'\end{pmatrix} = \diag(X^T \mathbf{A} Y, X'^T \mathbf{A} Y')\]
is a block diagonalization of $\mathbf{A}$. Thus, both $\begin{pmatrix}X & X'\end{pmatrix}$ and $ \begin{pmatrix}Y & Y'\end{pmatrix}$ are invertible matrices. 

On the other hand, 
\[\begin{pmatrix}X & X'\end{pmatrix}^T \mathbf{A} \begin{pmatrix}XZ & Y'\end{pmatrix} = 
\begin{pmatrix}X^T \mathbf{A} Y & \\  \mathbf{B} & X'^T \mathbf{A} Y'  \\ \end{pmatrix}\]
for some matrix tuple $\mathbf{B}$ whose columns are linear combinations of columns of $ X'^T \mathbf{A} Y' $. 
Thus, there exists an invertible matrix $Q$ such that 
\[\begin{pmatrix}X & X'\end{pmatrix}^T \mathbf{A} \begin{pmatrix}XZ & Y'\end{pmatrix}Q = \begin{pmatrix}X & X'\end{pmatrix}^T \mathbf{A} \begin{pmatrix}Y & Y'\end{pmatrix},\]
and consequently $\begin{pmatrix}XZ & Y'\end{pmatrix}$ is an invertible matrix, and also $\begin{pmatrix}X & Y'\end{pmatrix}$ is an invertible matrix.

Hence, 
\[\begin{pmatrix}X & Y'\end{pmatrix}^T \mathbf{A} \begin{pmatrix}X & Y'\end{pmatrix} = 
\begin{pmatrix}X^T \mathbf{A} X & \\  & Y'^T \mathbf{A} Y'  \\ \end{pmatrix}.\]\
Hence, $X^T \mathbf{A} X$ is both a left-right block and a congruent block of $\mathbf{A}$. 
 \end{proof}

\begin{proof}[Proof of Theorem~\ref{thm:congruence_block_diagonalization}]
Let $r$ and $s$ be the numbers of columns of $X$ and $Y$,
respectively, where $r\leq s$, and write
$\mathbf{B}=X^T\mathbf{A}Y$ and
$\mathbf{C}=Y^T\mathbf{A}X=-\mathbf{B}^T$.
Since $\mathbf{B}$ is a minimal left-right block, there are
invertible matrices
\[
U=\begin{pmatrix}X & X'\end{pmatrix},
W=\begin{pmatrix}Y & Y'\end{pmatrix}
\]
such that
\[
U^T\mathbf{A}W
=\diag(\mathbf{B},X'^T\mathbf{A}Y').
\]
Write
\[
X=YP+Y'P', Y=XQ+X'Q'
\]
for some $P\in\M(s\times r,R)$ and
$Q\in\M(r\times s,R)$.
Since $X^T\mathbf{A}Y'=0$ and $Y^T\mathbf{A}X'=0$,
we have
\[
X^T\mathbf{A}X=\mathbf{B}P,
Y^T\mathbf{A}Y=\mathbf{C}Q.
\]
Because $\mathbf{A}$ is alternating, both $\mathbf{B}P$ and
$\mathbf{C}Q$ are alternating. Thus
\[
P^T\mathbf{C}=\mathbf{B}P,
 Q^T\mathbf{B}=\mathbf{C}Q.
\]
Consequently,
\[
(QP)^T\mathbf{B}=P^TQ^T\mathbf{B}
=P^T\mathbf{C}Q=\mathbf{B}PQ.
\]
Therefore $((QP)^T,PQ)$ is an endomorphism of the minimal
left-right block $\mathbf{B}$. Since $\mathbf{B}$ is indecomposable,
its endomorphism ring is local by the Fitting lemma.

Suppose first that $((QP)^T,PQ)$ is invertible.
Then both $QP$ and $PQ$ are invertible. Reducing modulo $p$
and using $r\leq s$, we obtain $r=s$ and $P$ invertible.
In this case,
\[
\begin{pmatrix}X & Y'\end{pmatrix}
=\begin{pmatrix}Y & Y'\end{pmatrix}
\begin{pmatrix}P&0\\P'&I\end{pmatrix}
\]
is invertible. Moreover, $X^T\mathbf{A}Y'=0$, and by
alternation $Y'^T\mathbf{A}X=0$. Hence
\[
\begin{pmatrix}X & Y'\end{pmatrix}^T\mathbf{A}
\begin{pmatrix}X & Y'\end{pmatrix}
=\diag(X^T\mathbf{A}X,Y'^T\mathbf{A}Y').
\]
Thus the first condition holds.

Otherwise, $((QP)^T,PQ)$ is not invertible.
Since its endomorphism ring is local, the endomorphism
\[
\bigl(I_r-(QP)^T,\,I_s-PQ\bigr)
\]
is invertible. In particular, $I_r-QP$ is invertible, and so is
\[
T=\begin{pmatrix}I_r&Q\\P&I_s\end{pmatrix}.
\]
Let $\pi_U:R^n\to R^r$ and $\pi_W:R^n\to R^s$ denote
the projections onto the first block coordinates with respect
to the bases $U$ and $W$, respectively. Define
\[
F:R^n\longrightarrow R^r\oplus R^s,
 F(v)=(\pi_U(v),\pi_W(v)).
\]
By the definitions of $P$ and $Q$,
\[
F\begin{pmatrix}X&Y\end{pmatrix}
=\begin{pmatrix}I_r&Q\\P&I_s\end{pmatrix}=T.
\]
Since $T$ is invertible, $F$ is split surjective and
\[
R^n=\columnspan\begin{pmatrix}X&Y\end{pmatrix}
\oplus\ker F.
\]
The module $\ker F$ is a direct summand of $R^n$, hence is
free over the local ring $R$. Choose a matrix $U''$ whose
columns form a basis of $\ker F$. Then
$\begin{pmatrix}X&Y&U''\end{pmatrix}$ is invertible.

Since $\pi_W(U'')=0$, the columns of $U''$ belong to
$\columnspan(Y')$. Therefore $X^T\mathbf{A}U''=0$.
Similarly, $\pi_U(U'')=0$ implies that the columns of $U''$
belong to $\columnspan(X')$, and hence
$Y^T\mathbf{A}U''=0$.
By alternation, the corresponding lower-left blocks also vanish.
Thus
\[
\begin{pmatrix}X&Y&U''\end{pmatrix}^T\mathbf{A}
\begin{pmatrix}X&Y&U''\end{pmatrix}
=\diag\left(
\begin{pmatrix}X&Y\end{pmatrix}^T\mathbf{A}
\begin{pmatrix}X&Y\end{pmatrix},
U''^T\mathbf{A}U''\right).
\]
This proves the second condition.
\end{proof}

Theorem~\ref{thm:congruence_block_diagonalization} implies the following corollary. 
 \begin{corollary}\label{cor:standard_block_diagonalization}
Any alternating matrix tuple $\mathbf{A}$ has a block diagonalization under the congruence action such that every block satisfies one of the following two conditions:
\begin{enumerate}
\item The block is $X^T \mathbf{A}X$ for some $X$ which is also a left-right block.
\item The block is $\begin{pmatrix} X &  Y\end{pmatrix}^T \mathbf{A} \begin{pmatrix} X & Y\end{pmatrix} $, where $X ^T \mathbf{A}Y$ is a minimal left-right block of $\mathbf{A}$. 
\end{enumerate}
Furthermore, one can construct a block diagonalization of $\mathbf{A}$ under the congruence action as follows: 
Arbitrarily select a minimal left-right block  $X^T \mathbf{A} Y$ of $\mathbf{A}$ such that the number of columns of $X$ is less than or equal to the number of columns of $Y$,
if $X$ satisfies the first condition of Theorem~\ref{thm:congruence_block_diagonalization}, then let $U_0 = X$; otherwise, let $U_0 = \begin{pmatrix}X & Y\end{pmatrix}$. 
Let $U = \begin{pmatrix} U_0 & U_1\end{pmatrix}$ be an arbitrary invertible matrix such that $U^T \mathbf{A} U = \diag(U_0^T \mathbf{A} U_0, U_1^T \mathbf{A} U_1)$.
Let $U_0^T \mathbf{A} U_0$ be a congruent block and recurse on $U_1^T \mathbf{A} U_1$.
\end{corollary}

\begin{lemma}\label{lem:two_blocks_right_multi}
Let $\mathbf{A}$ be an alternating matrix tuple and $X^T \mathbf{A} Y$ be a minimal left-right block such that $\begin{pmatrix}X & Y \end{pmatrix}^T \mathbf{A} \begin{pmatrix}X & Y \end{pmatrix}$ is a congruent block of $\mathbf{A}$. 
Then the following properties hold:
\begin{enumerate}
\item There is an invertible matrix $U$ and a matrix $Z$ such that 
\[\begin{pmatrix}X & Y \end{pmatrix}^T \mathbf{A} \begin{pmatrix}X & Y \end{pmatrix} U = \begin{pmatrix} X^T \mathbf{A} Y & \\ & Y^T \mathbf{A} X\end{pmatrix} Z.\]
\item There is an invertible matrix $U'$ and a matrix $Z'$ such that 
\[U'\begin{pmatrix}X & Y \end{pmatrix}^T \mathbf{A} \begin{pmatrix}X & Y \end{pmatrix}  = Z'\begin{pmatrix} Y^T \mathbf{A} X & \\ & X^T \mathbf{A} Y\end{pmatrix} .\]
\end{enumerate}
\end{lemma}
\begin{proof}

Since $X^T\mathbf{A} Y$ is a minimal left-right block of $\mathbf{A}$, there exists an invertible matrix $U_0$ such that the following two conditions hold:
\begin{enumerate}
\item $ \begin{pmatrix}X & Y \end{pmatrix} U_0 = \begin{pmatrix} Y & U_0'\end{pmatrix}$ for some $U_0'$. 
\item \[\begin{pmatrix}X & Y \end{pmatrix}^T \mathbf{A} \begin{pmatrix}X & Y \end{pmatrix} U_0 = \begin{pmatrix} X^T \mathbf{A} Y & \\ \mathbf{B} & \mathbf{C}\end{pmatrix}\]
for some matrix tuples $\mathbf{B}$ and $\mathbf{C}$. 
\end{enumerate}
Since $Y^T \mathbf{A} X$ is a block, every column of $\mathbf{B}$ and $\mathbf{C}$ is a linear combination of the columns of $Y^T \mathbf{A} X$.
Then the first property holds. 

The second property is obtained by applying the transpose of the first property. 
\end{proof}

We define the standard congruent block diagonalization of an alternating matrix tuple as follows. 

\begin{definition}\label{def:standard_congruent_block}
A congruent block $U^T\mathbf{A}U$ of an alternating matrix tuple $\mathbf{A}$ for some matrix $U$ is a \emph{standard congruent block} if one of the following two conditions holds:
\begin{enumerate}
    \item $U^T \mathbf{A} U$ is a minimal left-right block of $\mathbf{A}$.
    \item $U = \begin{pmatrix}X & Y \end{pmatrix}$ such that $X^T \mathbf{A} X$ is not a minimal left-right block of $\mathbf{A}$, and $X^T \mathbf{A} Y$ is a minimal left-right block of $\mathbf{A}$.
\end{enumerate}

We say that a standard congruent block $U^T\mathbf{A}U$ is obtained from a minimal left-right block $X^T \mathbf{A} Y$ if one of the following two cases is satisfied:
\begin{enumerate}
    \item $X = U$ and $Y$ is a matrix such that $X^T \mathbf{A} Y$ is a minimal left-right block of $\mathbf{A}$. 
    \item $U = \begin{pmatrix}X & Y \end{pmatrix}$. 
\end{enumerate}

We say that an alternating matrix tuple $\mathbf{A}$ is \emph{standard congruent block diagonalized} if $\mathbf{A} = \diag(\mathbf{S}_1, \dots, \mathbf{S}_k)$ such that every $\mathbf{S}_i$ is a standard congruent block. 
\end{definition}

\begin{lemma}\label{lem:standard_second_type_lr}
Let $\mathbf{A}$ be an alternating matrix tuple over $R$, and let
\[
\mathbf{H} = \begin{pmatrix}X & Y\end{pmatrix}^T
\mathbf{A}\begin{pmatrix}X & Y\end{pmatrix}
\]
be a standard congruent block of the second type. Let
$\mathbf{B} = X^T\mathbf{A}Y$. Then $\mathbf{H}$ is left-right
equivalent to $\diag(\mathbf{B},-\mathbf{B}^T)$.
Consequently, every minimal left-right block of $\mathbf{H}$ is
left-right equivalent to either $\mathbf{B}$ or $\mathbf{B}^T$.
\end{lemma}
\begin{proof}
Let $r$ and $s$ be the numbers of columns of $X$ and $Y$,
respectively, and write $\mathbf{C}=Y^T\mathbf{A}X=-\mathbf{B}^T$.
Since $\mathbf{B}$ is a minimal left-right block of $\mathbf{A}$,
there exist matrices $X'$ and $Y'$ such that
$\begin{pmatrix}X & X'\end{pmatrix}$ and
$\begin{pmatrix}Y & Y'\end{pmatrix}$ are invertible and
\[
\begin{pmatrix}X & X'\end{pmatrix}^T\mathbf{A}
\begin{pmatrix}Y & Y'\end{pmatrix}
=\diag(\mathbf{B},X'^T\mathbf{A}Y').
\]
Since $\begin{pmatrix}Y & Y'\end{pmatrix}$ is invertible,
we may write $X=YP+Y'P'$ for some matrices $P,P'$.
The equality $X^T\mathbf{A}Y'=0$ implies
$X^T\mathbf{A}X=\mathbf{B}P$.
Similarly, writing $Y=XQ+X'Q'$ and using
$Y^T\mathbf{A}X'=0$, we have
$Y^T\mathbf{A}Y=\mathbf{C}Q$.
Hence
\[
\mathbf{H}=\begin{pmatrix}
\mathbf{B}P & \mathbf{B}\\
\mathbf{C} & \mathbf{C}Q
\end{pmatrix}.
\]

Since $\mathbf{A}$ is alternating, both $\mathbf{B}P$ and
$\mathbf{C}Q$ are alternating. Thus
\[
P^T\mathbf{C}=\mathbf{B}P,
 Q^T\mathbf{B}=\mathbf{C}Q.
\]
Consequently,
\[
(QP)^T\mathbf{B}
=P^TQ^T\mathbf{B}
=P^T\mathbf{C}Q
=\mathbf{B}PQ.
\]
Therefore $((QP)^T,PQ)$ is an endomorphism of the minimal
left-right block $\mathbf{B}$.

Suppose that this endomorphism is invertible. Then both
$QP\in\M(r,R)$ and $PQ\in\M(s,R)$ are invertible.
Reducing modulo $p$, we obtain $r=s$, and then $P$ is
invertible. Thus
$X^T\mathbf{A}X=\mathbf{B}P$ is right equivalent to
$\mathbf{B}$. By Lemma~\ref{lem:block_diagonal_basic},
$X^T\mathbf{A}X$ is a minimal left-right block of
$\mathbf{A}$, contradicting the assumption that
$\mathbf{H}$ is a standard congruent block of the second type.
Hence $((QP)^T,PQ)$ is not invertible.

Since $\mathbf{B}$ is indecomposable, its endomorphism ring is
local by the Fitting lemma. Therefore $I_r-QP$ is invertible.
It follows that
\[
Z=\begin{pmatrix}I_s&P\\Q&I_r\end{pmatrix}
=\begin{pmatrix}I_s&0\\Q&I_r\end{pmatrix}
\begin{pmatrix}I_s&P\\0&I_r-QP\end{pmatrix}
\]
is invertible.
Let $U$ denote the permutation matrix that exchanges the
$X$-columns and the $Y$-columns. Then
\[
\mathbf{H}U
=\begin{pmatrix}\mathbf{B}&\mathbf{B}P\\
\mathbf{C}Q&\mathbf{C}\end{pmatrix}
=\diag(\mathbf{B},\mathbf{C})Z.
\]
Thus $\mathbf{H}UZ^{-1}=\diag(\mathbf{B},-\mathbf{B}^T)$,
so $\mathbf{H}$ is left-right equivalent to
$\diag(\mathbf{B},-\mathbf{B}^T)$.
Since $\mathbf{B}$ and $\mathbf{B}^T$ are minimal
left-right blocks, the last property follows from the
uniqueness of minimal left-right block diagonalization.
\end{proof}

\subsection{Frequent blocks alternating matrix tuple isometry}

We remark that the parameter $r$ in this section will correspond to $n^\Delta$. 
In this section, we prove the following theorem. 

\begin{theorem}\label{thm:small_block}
Let $\mathbf{A}_1, \mathbf{A}_2 \in \M(n, R)$ be two non-degenerate alternating matrix tuples such that every minimal left-right block of $\mathbf{A}_1$ and $\mathbf{A}_2$ has at most $r$ rows and columns for some positive integer $r$. 
There is an algorithm to determine if $\mathbf{A}_1$ and $\mathbf{A}_2$ are congruent in $p^{O(\tau \cdot n \cdot r+ n \log n)}\cdot \mathrm{poly}(m)$ time. 
\end{theorem}

We first bound the number of different standard congruent block diagonalized alternating matrix tuples congruent to a given alternating matrix tuple.

\begin{lemma}\label{lem:small_block_diag_number}
Let $\mathbf{A}$ be a matrix tuple over $n$ variables over $R$ such that every minimal left-right block has at most $r$ rows and $r$ columns. 
There are at most $p^{O(\tau \cdot n \cdot r + n \log n)}$ different possible standard congruent block diagonalized alternating matrix tuples congruent to $\mathbf{A}$.

\end{lemma}

\begin{proof}
Let $\mathbf{B}_1, \dots, \mathbf{B}_\ell$ be arbitrary minimal left-right blocks of $\mathbf{A}$ such that every minimal left-right block is left-right equivalent to exactly one of $\mathbf{B}_1, \dots, \mathbf{B}_\ell, \mathbf{B}_1^T, \dots, \mathbf{B}_\ell^T$.
We say that a standard congruent block diagonalization $\diag(\mathbf{S}_1, \dots, \mathbf{S}_k)$ is ordered with respect to $\mathbf{B}_1, \dots, \mathbf{B}_\ell$ if  $\beta_i \leq \beta_j$ for any $1 \leq i < j \leq k$, where $\beta_i$ is the integer such that the congruent block $\mathbf{S}_i$ is obtained from a left-right block left-right equivalent to $\mathbf{B}_{\beta_i}$ or $\mathbf{B}_{\beta_i}^T$ under left-right action.

We first bound the number of different possible standard congruent block diagonalized alternating matrix tuples congruent to $\mathbf{A}$ with respect to $\mathbf{B}_1, \dots, \mathbf{B}_\ell$.

For a standard congruent block diagonalization of $\mathbf{A}$, let $k$ denote the number of standard congruent blocks, and let the $k$ numbers $\alpha_1, \dots, \alpha_k$ each be either $1$ or $2$, indicating the number of minimal left-right blocks in the $i$-th standard congruent block of the standard congruent block diagonalization.

We bound the number of possible different standard congruent block diagonalized matrix tuples for fixed $k$ and $\alpha_1, \dots, \alpha_k$. 
Let $\mathbf{A}'$ be an arbitrary such matrix tuple. 
For any $1 \leq i \leq k$, let $m_i$ denote the number of rows and columns of the $i$-th standard congruent block and $g_i$ be the integer such that the $i$-th standard congruent block is obtained from a minimal left-right block equivalent to $\mathbf{B}_{g_i}$. 
By the definition of standard congruent block diagonalization, $m_1, \dots, m_k$ and $g_1, \dots, g_k$ are determined once $k$ and $\alpha_1, \dots, \alpha_k$ are fixed. 

We show that there are at most $p^{O( \tau m_i^2)}$ different possible $\mathbf{S}_i$ for fixed $m_i$ and $g_i$. 
By the definition of standard congruent block, there exist $S$ and $W$ such that $\mathbf{S}_i$ is obtained from the block $X^T \mathbf{A} Y$, which equals $S^T \mathbf{B}_{g_i} W$. 
By Lemma~\ref{lem:two_blocks_right_multi}, there exists a matrix $U$ such that $\mathbf{S}_i$ equals either $S^T \mathbf{B}_{g_i} W U$ (if $S_i$ corresponds to a left-right block of $\mathbf{A}$) or $\diag(S^T \mathbf{B}_{g_i} W, W^T \mathbf{B}_{g_i} S) U$ (if $S_i$ corresponds to two left-right blocks of $\mathbf{A}$). 
Since there are at most $(p^\tau)^{3 m_i^2}$ possible $S, W$ and $U$, there are at most $p^{3 \tau m_i^2}$ different possible $\mathbf{S}_i$. 
Hence, there are at most \[\prod_{i = 1}^k p^{O(\tau m_i^2)} = p^{O(\tau \cdot \sum_{i} m_i^2)} = p^{O(\tau \cdot n \cdot r)}\] different possible  standard congruent block diagonalized matrix tuples for fixed $k$ and  $\alpha_1, \dots, \alpha_k$.

Since there are at most $n \cdot 2^n$ different possible $k$ and $(\alpha_1, \dots, \alpha_k)$, the total number of possible standard congruent block diagonalizations for $\mathbf{A}$ with respect to $\mathbf{B}_1, \dots, \mathbf{B}_\ell$ is upper bounded by 
\[ n \cdot 2^n \cdot p^{O(\tau \cdot n \cdot r)}= p^{O(\tau \cdot n \cdot r)}.\]

Finally, since every standard congruent block diagonalization can be obtained by permuting the blocks for a standard congruent block diagonalization with respect to $\mathbf{B}_1, \dots, \mathbf{B}_\ell$, the total number of standard block diagonalizations is upper bounded by 
\[ n! \cdot p^{O(\tau \cdot n \cdot r)}= p^{O(\tau \cdot n \cdot r + n \log n)}.\]
\end{proof}

Now for a given alternating matrix tuple $\mathbf{A}$, we give an algorithm to enumerate all the possible standard block diagonalizations of $\mathbf{A}$.

In our algorithm, we obtain the diagonalizations by iteratively performing swap and local update operations, defined as follows: For a standard congruent block diagonalization $\mathbf{A} = \diag(\mathbf{S}_1, \dots, \mathbf{S}_k)$, let $m_i$ denote the number of rows and columns of $\mathbf{S}_i$, and $r_i = \sum_{j = 1}^{i} m_j$ for any $1 \leq i \leq k$.

\begin{enumerate}
    \item For $1 \leq i < j \leq k$, the swap operation on the $i$-th and $j$-th blocks produces 
    \[\diag(\mathbf{S}_1, \dots, \mathbf{S}_{i - 1}, \mathbf{S}_j, \mathbf{S}_{i + 1}, \dots, \mathbf{S}_{j - 1}, \mathbf{S}_i, \mathbf{S}_{j + 1}, \dots, \mathbf{S}_k).\] 
    
    \item For any $1 \leq i \leq j \leq k$ with $j$ being either $i$ or $i + 1$, 
    the local update operation produces 
     \[\diag(\mathbf{S}_1, \dots, \mathbf{S}_{i - 1}, \mathbf{L}, \mathbf{S}_{i + 1}, \dots, \mathbf{S}_k),\]
     where $\mathbf{L}$ is a standard congruent block diagonalization of $\mathbf{S}_i$ if $i = j$ or $\diag(\mathbf{S}_i, \mathbf{S}_{i+1})$ if $j = i + 1$.

\end{enumerate}

\begin{lemma}\label{lem:correctness_small_block}
Let $\mathbf{A}_1$ and $\mathbf{A}_2$ be two standard congruent block diagonalizations of $\mathbf{A}$. 
Then there exists a sequence of swap and local update operations that transforms $\mathbf{A}_1$ into $\mathbf{A}_2$. 
\end{lemma}
\begin{proof}
Let $U$ be an arbitrary invertible matrix such that $U^T \mathbf{A}_1 U = \mathbf{A}_2$. 
Let $n_1$ denote the number of rows/columns of the first congruent block of $\mathbf{A}_2$. 
Let $U_1$ be the submatrix of $U$ consisting of its first $n_1$ columns. 
Then the first congruent block of $U^T \mathbf{A}_1 U$ is constructed from $U_1^T \mathbf{A}_1 U_1$. 
Let $X$ and $Y$ be matrices such that the congruent block $U^T \mathbf{A}_1 U$ is obtained from $X^T \mathbf{A}_1 Y$. 
Then the columns of both $X$ and $Y$ are linear combinations of columns of $U$. 

Let $m_i$ be the number of rows/columns of the $i$-th congruent block of $\mathbf{A}_1$. 
Let $E_i$ be the induced submatrix of $I_n$ on the $(1 + \sum_{j = 1}^{i - 1}m_j)$-th column to the $(\sum_{j = 1}^i m_j)$-th column. 
Let $X_i$ be the projection of $X$ onto $E_i$ with respect to $E_1\oplus \dots  \oplus E_\ell$, where $\ell$ denotes the number of blocks of $\mathbf{A}_1$.  
Let $\alpha$ denote the number of nonzero $X_i$.

We show that after a sequence of operations, $\alpha$ decreases to one, and $X_i$ is nonzero only for $i = 1$. 
Without loss of generality, we can assume $X_1$ is in $V_{X^T \mathbf{A}Y} \setminus K_{X^T \mathbf{A}Y}$ because if not, we can use one swap operation to enforce this condition by Theorem~\ref{thm:left_right_decomposition}. 

Suppose $\alpha$ is greater than one. 
Assume that $X_2$ is nonzero; otherwise, we can swap the second block with the $i$-th block for some $i > 2$ to make $X_2$ nonzero. 
Since $X_1$ is in $V_{X^T \mathbf{A}_1Y} \setminus K_{X^T \mathbf{A}Y}$, 
by Theorem~\ref{thm:left_right_decomposition}, 
$X_1 + X_2$ is also in $V_{X^T \mathbf{A}_1Y} \setminus K_{X^T \mathbf{A}_1Y}$. 
Hence, since the matrix tuple is congruent block diagonal, 
we can find a matrix $Y$ using only columns corresponding to the first two blocks such that 
$(X_1 + X_2)^T \mathbf{A}_1 Y$ is a block. 
Such a $Y$ must exist because of the congruent block diagonalization. 
We construct a new standard congruent block diagonalization for the first two congruent blocks of $\mathbf{A}_1$
such that the first standard congruent block is constructed from $(X_1 + X_2)^T \mathbf{A}_1 Y$ by Theorem~\ref{thm:congruence_block_diagonalization}. 
Then we use this new standard congruent block diagonalization for the first two congruent blocks of $\mathbf{A}_1$ to replace the original first two blocks in $
\mathbf{A}_1$. 
After this process, we update $U$ accordingly, and $\alpha$ decreases by one. 
If we repeat the above process, after a finite number of operations $\alpha$ becomes one.

For $Y$, let $Y_i$ be the projection of $Y$ onto $E_i$ with respect to $E_1\oplus \dots  \oplus E_\ell$, and 
$\beta$ denote the number of nonzero $Y_i$. 
Using reasoning similar to that for $\alpha$, we can reduce $\beta$ to one through a finite sequence of operations while maintaining $\alpha$ equal to one, by constructing the first block from $X^T \mathbf{A}_1 Y'$ for a suitable $Y'$ so that $\beta$ decreases by one at each step. 
Then $X = X_1$, $Y= Y_1$, and $X_i$, $Y_i$ are all zero matrices for all $i > 1$. 
Then we can use a local update operation on the first block to make it equal to the first block of $\mathbf{A}_2$. 

By recursing on the remaining blocks, the lemma follows. 
\end{proof}
We have the following algorithm for Theorem~\ref{thm:small_block}.

\begin{framed}\vspace{-.3cm}\begin{flushleft}
\textbf{Frequent Block Matrix Tuple Isometry Algorithm}

\textbf{Input:} Two alternating matrix tuples $\mathbf{A}$ and $\mathbf{B}$. 


\textbf{Output:}
Yes or no. If yes, a matrix $S$.

\begin{enumerate}
    \item Compute arbitrary standard congruent block diagonalizations of $\mathbf{A}$ and $\mathbf{B}$, denoted by $\mathbf{A}'$ and $\mathbf{B}'$. 
    Let $\Gamma = \{\mathbf{A}'\}$. 
    \item Iteratively perform the following process until $\Gamma$ does not change: For any $\mathbf{A}'' \in \Gamma$,
    enumerate all the possible swap and local operations for $\mathbf{A}''$, and add the results to $\Gamma$.  
    \item Output yes if $\mathbf{B'} \in \Gamma$ and the matrix $S$ such that $S^T \mathbf{A} S = \mathbf{B}$. Otherwise output no. 
\end{enumerate}
\end{flushleft}
\vspace{-.3cm}
\end{framed}

\begin{proof}[Proof of Theorem~\ref{thm:small_block}]
The correctness of the algorithm follows from Lemma~\ref{lem:correctness_small_block}. 
So we only bound the running time here. 
The first step of the algorithm can be done by enumerating $X$ and $Y$ such that $X^T \mathbf{A} Y$ is a minimal left-right block of at most $r$ rows and columns by Corollary~\ref{cor:standard_block_diagonalization}.
Hence, the first step of the algorithm takes $p^{\tau \cdot r \cdot n} \cdot \poly(n, m)$ time.

For the second step, every $\mathbf{A}''$ in $\Gamma$ is a standard congruent block diagonalization of $\mathbf{A}$. 
By Lemma~\ref{lem:small_block_diag_number}, the set $\Gamma$ is of size at most $p^{O(\tau \cdot n \cdot r + n \log n)}$. For each standard block diagonalization, there are at most $n^2 + n \cdot p^{O(\tau \cdot r^2)}$ possible swapping and local update operations. Hence, the overall running time is $p^{O(\tau \cdot n \cdot r  + n \log n)}\cdot \mathrm{poly}(m)$.  
\end{proof}

\subsection{Infrequent block alternating matrix tuple isometry}

We prove the following theorem in this section. 

\begin{theorem}\label{thm:large_block}
Let $R = \Z / p^\tau \Z$, $D$ be a positive integer and $\mathbf{A}_1, \mathbf{A}_2 \in \M(n, R)^m$ be two non-degenerate alternating matrix tuples such that 
for every minimal left-right block $\mathbf{B}$, there are at most $D$ minimal left-right blocks in an arbitrary left-right block diagonalization of $\mathbf{A}_i$ left-right equivalent to $\mathbf{B}$ for any $i \in \{1, 2\}$. 
There is an algorithm to determine if $\mathbf{A}_1$ and $\mathbf{A}_2$ are congruent in $p^{O( n \cdot D)}\cdot \mathrm{poly}(p, \tau, n, m)$ time. 
If the answer is yes, then the algorithm outputs a matrix $S$ such that $S^T \mathbf{A}_1 S = \mathbf{A}_2$. 

Furthermore, the algorithm outputs a generating set of size $p^{n D} \cdot \poly(n, m, p, \tau)$ for the automorphism group of $\mathbf{A}_2$, which has size $O(p^{nD}\cdot \tau \cdot n^2)$. This is a generating set for the group of matrices $S$ satisfying $S^T \mathbf{A}_2 S = \mathbf{A}_2$.   
\end{theorem}

We start by looking for an invertible matrix $S$ such that 
$S^T\mathbf{A}_1 S =  \mathbf{A}_2$. We can limit the search space
based on the following observation.

\begin{lemma}
\label{lem:in-adjoint}
Let $S,S_0,S_0'\in \GL(n,R)$ such that
$\mathbf{A}_2 = S^T \mathbf{A}_1 S$ and 
$\mathbf{A}_2 ={S_0}^T\mathbf{A}_1S_0'$.
Then the matrix $S_1:=S_0^{-1}S$ satisfies the
following properties:
\begin{itemize}
\item $\mathbf{A}_2=(S_0S_1)^T\mathbf{A}_1S_0S_1$;
\item 
there exists a matrix $S_1'$ 
such that 
$S_1^T\mathbf{A}_2=\mathbf{A}_2S_1'$.
\end{itemize}
\end{lemma}

\begin{proof}
The first property of $S_1$ is obvious. To see the second, 
put $S_1'=S^{-1}S_0'$. 
We have 
$S^T\mathbf{A}_1=\mathbf{A}_2S^{-1}$ which can be used to show
\[S_1^T\mathbf{A}_2=S^T{S_0^T}^{-1}\mathbf{A}_2=
S^T{S_0^T}^{-1}S_0^T\mathbf{A}_1S_0'=S^T\mathbf{A}_1S_0'=
\mathbf{A}_2S^{-1}S_0'=\mathbf{A}_2S_1'.\]
\end{proof}

We shall begin our algorithm with the relatively easy task of
finding $S_0$ and $S_0'$ as in the lemma. Then the congruence of
$\mathbf{A}_1$ and $\mathbf{A}_2$ is equivalent to the existence
of $S_1$ in the set 
$$G=\{S_1\in\GL(n, R):\exists S_1'\in\GL(n,R)\mbox{~such~that~} 
S_1^T\mathbf{A}_2=\mathbf{A}_2S_1'\}$$
such that $\mathbf{A}_2=(S_0 S_1)^T \mathbf{A}_1 S_0S_1$.
It turns out that $G$ is a matrix group having an efficiently computable
 unipotent normal subgroup $J$ of index at most $p^{nD}$, 
and for
any invertible matrix $Y$, the problem of deciding the existence of (and finding) $X\in J$
such that $\mathbf{A}_2=X^T Y^T\mathbf{A}_1 Y X$ can be solved in time
$\poly(n, m, p, \tau)$. We take a system $\Gamma$ of coset representatives of $J$ in $G$
and for all $Z\in \Gamma$ attempt to solve the above search problem for
$Y=S_0Z$. 

Before describing the formal algorithm, we introduce some
notation and terminology. Also, we present some basic facts
and some tools that can be used in the algorithm.
We shall consider the set 
\begin{equation}\label{eq:calA}
{\cal A}=\{S_1\in \M(n,R):\exists S_1'\mbox{~such~that~} 
S_1^T\mathbf{A}_2=\mathbf{A}_2S_1'\}
\end{equation}
closely related to $G$. 
Since the condition
is linear, ${\cal A}$ is a linear submodule of $\M(n\times n, R)$.
Also, if $S_1,S_2\in{\cal A}$, that is,
$S_1^T\mathbf{A}_2=\mathbf{A}_2S_1'$ and $S_2^T\mathbf{A}_2=\mathbf{A}_2S_2'$ for
some matrices $S_1'$ and $S_2'$, then
\[(S_1S_2)^T\mathbf{A}_2=S_2^TS_1^T\mathbf{A}_2=S_2^T\mathbf{A}_2S_1'=\mathbf{A}_2S_2'S_1',\]
whence $S_1S_2\in {\cal A}$. 
Thus ${\cal A}$ is in fact a subalgebra of $\M(n,R)$ and
$G$ is the group of invertible elements of ${\cal A}$.
Indeed, every $S\in\mathcal A\cap\GL(n,R)$ has its inverse in
$\mathcal A$ by Cayley--Hamilton.  The projection
$\mathcal E\to\mathcal A^T$ in the proof of
Lemma~\ref{lem:index_bound} has nilpotent kernel, so the unit
$S^T\in\mathcal A^T$ lifts to a unit $(S^T,T)\in\mathcal E$.
Thus $T$ is invertible and $S\in G$.
The Jacobson radical $\Rad({\cal A})$ is the largest nilpotent
ideal of ${\cal A}$.  We define $J$
as the set 
$$J=I+\Rad({\cal A})=\{I+S_1:S_1\in \Rad({\cal A})\}.$$ 
Using the fact
that $\Rad({\cal A})$ is an ideal of ${\cal A}$, it is straightforward
to check that $J$ is a normal subgroup of $G$.

\begin{lemma}\label{lem:upper_tri_generating_set}
Let $R = \Z / p^\tau \Z$ and $N \subseteq M(n, R)$ be a nilpotent 
associative $R$-algebra, given by $R$-module generators.   Given generators $\Gamma$ for any subgroup
$H\leq I+N$, one can compute a generating set for $H$
of size at most $\tau n^2$ and perform membership tests
in $H$ with running time  $\poly(n,p,\tau, 
|\Gamma|)$.
\end{lemma}
\begin{proof}
Write $N^k=0$ for a sufficiently large $k$.
Refine the chains of additive $p$-groups
\[
  N^i/N^{i+1}, 1\leq i<k,
\]
into composition series with factors of order $p$.  This produces
a chain of additive subgroups
\[
 N=T_0\supset T_1\supset\cdots\supset T_t=0,
  |T_j/T_{j+1}|=p,
  t=\log_p|N|\leq\tau n^2,
\]
where each $T_j$ lies between two consecutive powers
$N^i$ and $N^{i+1}$.
Set $U_j=I+T_j$.  Every $U_j$ is a group: if
$x,y\in T_j\subseteq N^i$, then
$xy\in N^{i+1}\subseteq T_j$, and inverses are given by
finite geometric series.  Moreover,
\[
 U_0\trianglerighteq U_1\trianglerighteq\cdots
 \trianglerighteq U_t=1,
 \quad |U_j/U_{j+1}|=p,
 \quad [U_0,U_j]\subseteq U_{j+1}.
\]
Thus this is a computable central series with cyclic factors of
order $p$.  Membership in $U_j$ and the coordinate of an element
of $U_j/U_{j+1}\cong\mathbb F_p$ are obtained by additive
submodule computations.

Here is a collection procedure, included to make the
complexity explicit.  Store at most one \emph{pivot} $a_j\in
H\cap(U_j\setminus U_{j+1})$ for each level $j$.
To \emph{sift} an element $h\in H$, visit the levels in increasing
order.  If its coordinate at level $j$ is nonzero and a pivot
$a_j$ is present, multiply by a suitable power of $a_j^{-1}$
to cancel that coordinate.  The resulting element is in $U_{j+1}$.
If no pivot exists at the first nonzero level, store the current
element as the new pivot.  Enqueue, whenever a pivot is added,
its $p$-th power and its commutators with all previously stored
pivots, and sift these elements as well.  Begin with the given
generators $\Gamma$ in the queue.

There can be at most $t$ pivot insertions; each insertion adds
at most $t+1$ elements to the queue.  Hence the procedure uses
polynomially many group operations.  Upon termination, every
original generator, every pivot $p$-th power, and every pairwise
pivot commutator sifts to the identity.  Because the above series
is central, the usual collection argument now rewrites every
product of pivots as an ordered product
$\prod_j a_j^{c_j}$ with $0\leq c_j<p$.
These normal forms are distinct (compare the first level at
which their exponents differ).  Sifting is therefore a
constructive membership test in the subgroup generated by the
pivots.  The original generators all sift to the identity,
so this subgroup is exactly $H$.
There are at most $t\leq\tau n^2$ pivots, as required.
\end{proof}

\begin{lemma}\label{lem:upper_triangularize}
Let $R=\mathbb Z/p^\tau\mathbb Z$, $V=R^n$, and let $F\leq\GL(n, R)$
be a finite $p$-group given by generators $\Gamma$.  There is an algorithm with $\poly(n, p, \tau, |\Gamma|)$ running time that constructs a chain of $F$-invariant additive subgroups
\[
 0=V_0\subset V_1\subset\cdots\subset V_{\tau n}=V,
  |V_i/V_{i-1}|=p,
\]
such that
\[
   (Q-I)V_i\subseteq V_{i-1}
   \text{for all $Q\in F$ and all $i$.}
\]
\end{lemma}
\begin{proof}
Let $B$ be the nonunital associative $R$-algebra
generated by $Q-I$ as $Q$ ranges through the supplied generators
of $F$.  A finite $p$-subgroup of $\GL(n, \mathbb F_p)$ is
conjugate to a subgroup of the upper unitriangular group.
Therefore the image of $B$ modulo $p$ is a nilpotent
algebra with $B^n\subseteq pM(n, R)$, and hence
\[
 B^{n\tau}=0.
\]
Compute $B$ by closing the $R$-span of the generators
under multiplication, and compute the sequence of submodules
\[
 V\supseteq BV\supseteq
 B^2V\supseteq\cdots\supseteq0.
\]
The linear part of every element of $F$ lies in
$I+B$, so every term is $F$-invariant.  Furthermore,
$B(B^iV)\subseteq B^{i+1}V$.
Refine each additive quotient
$B^iV/B^{i+1}V$ into a composition series
with factors of order $p$, and concatenate these refinements
from $0$ to $V$.  Every intermediate subgroup $W$ between
$B^{i+1}V$ and $B^iV$ satisfies
\[
 (Q-I)W\subseteq B^{i+1}V
\]
for every $Q\in F$.  Thus the asserted property holds at each
successive step.  All module generators, products, quotient
calculations and composition refinements are computable by
standard algorithms for modules over $\mathbb Z/p^\tau\mathbb Z$.
Since $|V|=p^{\tau n}$, the refined chain has exactly $\tau n$
steps.  This also proves the running-time bound.
\end{proof}

\begin{lemma}\label{lem:radical-generators}
There is an algorithm, with running time $\poly(n,m,p,\tau)$,
that computes a multiplicative generating set $\Gamma$ of $J = I + \Rad(\mathcal{A})$ containing at most $O(\tau n^2)$ invertible matrices. 
\end{lemma}
\begin{proof}
Let $Q=\Rad(\mathcal A)$ and $J=I+Q$.
First compute an $R$-module generating set for $\mathcal A$ by solving
the linear equations of~\eqref{eq:calA} over $R$.
Standard algorithms for linear systems and submodules over
$\mathbb Z/p^\tau\mathbb Z$ give a polynomial-time procedure.
These generators also permit effective computations with submodules,
quotients, and products of submodules.

Since $(p\mathcal A)^\tau=0$, the ideal $p\mathcal A$ is contained in
$Q $. For $\overline{\mathcal A}=\mathcal A/p\mathcal A$,
the radical-quotient identity gives
\begin{equation}\label{eq:rad-quotient}
\Rad(\overline{\mathcal A})=Q /p\mathcal A.
\end{equation}
Represent $\overline{\mathcal A}$ as an abstract finite-dimensional
$\mathbb F_p$-algebra by computing an $\mathbb F_p$-basis of the quotient
module $\mathcal A/p\mathcal A$ and its multiplication table. Importantly,
this quotient should not be identified without justification with the
algebra obtained by reducing the ambient matrices modulo $p$:
in general, $\mathcal A\cap p\M_n(R)$ can be larger than $p\mathcal A$.
Compute an $\mathbb F_p$-basis of
$\Rad(\overline{\mathcal A})$ using the polynomial-time algorithm for
finite-dimensional algebras over finite fields (e.g., R\'onyai's
algorithm). Lift these basis elements to $\mathcal A$ and adjoin
an $R$-module generating set of $p\mathcal A$. By
Equation~\eqref{eq:rad-quotient}, the resulting matrices generate
$Q $ as an $R$-module. If desired, this set can be reduced
to at most $\tau n^2$ generators, since every submodule of $\M_n(R)$
can be generated by at most $n^2$ elements over $R$.

Write $Q ^i$ for the $i$th ideal power. Starting from
$Q $, compute successively
$Q ^{i+1}=Q ^iQ $ using pairwise products
of module generators and submodule reduction. Let $k\geq1$ be the
smallest integer such that $Q ^{k+1}=0$; if $Q =0$,
output $\Gamma=\varnothing$ and the conclusion is immediate. Otherwise
$Q \supsetneq Q ^2\supsetneq\cdots
\supsetneq Q ^k\supsetneq0$: equality at two consecutive
powers would force all subsequent powers to be equal, contradicting
nilpotence. As $|Q |\leq p^{\tau n^2}$, we have
$k\leq\tau n^2$.

We now construct a generating set for $J$ under \emph{matrix
multiplication}.  For each $i=k,k-1,\ldots,1$, compute a set
$T_i\subseteq Q ^i$ whose cosets generate
$Q ^i/Q ^{i+1}$ as an additive group.  Such a set
can be found by module computations over $R$: start with the zero
submodule of the quotient, and repeatedly adjoin an element of
$Q ^i$ whose coset is not yet generated, until the quotient
is generated.  Write
\[
H_i=\{I+X:X\in T_i\},
H=\bigcup_{i=1}^k H_i,
 C=\bigcup_{i=1}^k T_i.
\]
We claim that $H$ generates $J$ \emph{multiplicatively}:
\begin{equation}\label{eq:mult-gen}
  J=\langle H\rangle_{\mathrm{mult}}
   =\left\langle I+X:X\in C\right\rangle_{\mathrm{mult}}.
\end{equation}

Indeed, set $J_i=I+Q ^i$ for $1\leq i\leq k+1$.
Then $J_1=J$, $J_{k+1}=\{I\}$, and $J_{i+1}$ is a normal
subgroup of $J_i$.  For each $i\geq1$, the map
\[
\begin{aligned}
Q ^i/Q ^{i+1}&\longrightarrow J_i/J_{i+1},\\
X+Q ^{i+1}&\longmapsto (I+X)J_{i+1}
\end{aligned}
\]
is an isomorphism from the \emph{additive} group on the left to the
\emph{multiplicative} quotient group on the right.  It is
well-defined, bijective, and a homomorphism because
\[
(I+X)(I+Y)=I+X+Y+XY,
 XY\in Q ^{2i}\subseteq Q ^{i+1}.
\]
Thus $H_i$ maps onto a generating set of $J_i/J_{i+1}$.
Descending induction proves
\[
J_i=\left\langle H_i\cup H_{i+1}\cup\cdots\cup H_k
\right\rangle_{\mathrm{mult}}.
\]
The base case $J_{k+1}=\{I\}$ is immediate.  For the induction
step, any $g\in J_i$ is congruent modulo $J_{i+1}$ to a product of
matrices from $H_i$ and their inverses; the remaining factor lies
in $J_{i+1}$ and is generated by $H_{i+1},\ldots,H_k$ by induction.
Taking $i=1$ proves Equation~\eqref{eq:mult-gen}.

Finally, choose each $T_i$ by adding only cosets that enlarge the
additive subgroup already generated in
$Q ^i/Q ^{i+1}$.  Every addition increases that
subgroup's order by at least $p$.  Consequently,
\[
|H|\leq\sum_{i=1}^k|T_i|
\leq\sum_{i=1}^k
 \log_p\big|Q ^i/Q ^{i+1}\big|
=\log_p|Q |\leq\tau n^2.
\]

\end{proof}

A square matrix $M$ is called 
unipotent if $M-I$ is nilpotent. A matrix group is unipotent if all its
elements are unipotent.  

Let $L$ be a $d$-dimensional matrix
group over $R$ given by generators.
The {\em vector transporter problem} for $L$ is the following. Given
$d$-dimensional vectors $v_1,v_2$, compute the set
$L_{v_1\rightarrow v_2}=\{M \in L:v_2=Mv_1\}$. It is either
the empty set or a right coset of the stabilizer $L_{v_2}=L_{v_2\rightarrow
v_2}$. By computing $L_{v_1\rightarrow v_2}$ (if it is non-empty) we mean
giving a list of generators for the stabilizer $L_{v_2}$ and a 
single representative of the coset of $L_{v_1\rightarrow v_2}$.
We shall refer to $v_1$ as the {\em initial vector} and $v_2$ as the {\em target vector}. In fact, looking for
$S$ such that $\mathbf{A}_2= S^T \mathbf{A}_1 S$ can be cast as 
(a weaker version of) an instance of the vector transporter problem for
$\GL(d,R)$. 
However, we shall consider instances of the vector transporter problem
for the unipotent group $J$. 

To apply the transporter lemma to a tuple $\mathbf{A}\in M(n, R)^m$,
we use the
linear left action
\[
  f(X)(\mathbf{A})=X^{-T} \mathbf{A} X^{-1}
\]
for $X\in J$.
Thus $X^T\mathbf A_1 X=\mathbf A_2$
holds precisely when
$f(X)(\mathbf{A}_2)=\mathbf A_1$.
So, we consider the vector transporter problem with $d=mn^2$.  Group elements can be maintained
as $n\times n$ matrices throughout the stabilizer computation;
the action matrices on $V$ need not be explicitly enumerated.
Since this action need not be faithful, all Schreier computations
are carried out on the original matrices in $J$, using $f$ only
to evaluate their action.  We compress subgroup generators using
Lemma~\ref{lem:upper_tri_generating_set} with
$N=\Rad(\mathcal A)$; this retains the kernel of the action.

Let $V=R^d$.  An affine transformation of $V$ is a map
$\Phi:V\to V$ of the form $\Phi(v)=Qv+b$, where
$Q\in\M(d,R)$ and $b\in V$.  Its linear part is
$\operatorname{lin}_\Phi(v)=\Phi(v)-\Phi(0)=Qv$.
Compositions of affine transformations are affine.  The invertible
affine transformations (those with $Q\in\GL(d,R)$) form a group,
and $\Phi\mapsto\operatorname{lin}_\Phi$ is a homomorphism from
this group to $\GL(d,R)$. 
 We call a group of affine transformations unipotent if their linear parts form
a unipotent linear group.

\begin{lemma}\label{lem:unipotent-vector-transporter}
Let $R=\mathbb Z/p^\tau\mathbb Z$ and $V=R^d$.
Suppose that $G$ is a finite group of affine transformations of $V$,
given by a generating set $\Gamma$, and that the group is unipotent.  Given $v_1,v_2\in V$, there is a
deterministic algorithm, polynomial in $d,p,\tau,$ and $|\Gamma|$,
that decides whether
\[
  G_{v_1\to v_2}=\{g\in G:g(v_1)=v_2\}
\]
is empty.  If it is nonempty, the algorithm outputs one element
$g_0\in G_{v_1\to v_2}$ and a multiplicative generating set for
$G_{v_2\to v_2}$, so that
\[
   G_{v_1\to v_2}=G_{v_2\to v_2}\,g_0.
\]
\end{lemma}

\begin{proof}
We first explain why all intermediate group computations have
polynomial complexity.  Represent an affine transformation
$g(v)=Q_gv+b_g$ by the invertible $(d+1)\times(d+1)$ matrix
\[
 \widehat g=\begin{pmatrix}Q_g&b_g\\0&1\end{pmatrix}.
\]
The kernel of the map from $G$ to its group of linear parts is a
subgroup of the additive $p$-group $V$.  Every element of the finite
group of linear parts is unipotent over $R$, hence has $p$-power
order; by Cauchy's theorem this image is a $p$-group.  Therefore
$G$ itself is a $p$-group.  Let $N $ be the (nonunital) $R$-algebra generated
by $\widehat g-I$ for $g\in\Gamma$.  As the reduction of $G$
modulo $p$ is a $p$-subgroup of $\GL(d+1,\mathbb F_p)$, it is
simultaneously conjugate over $\mathbb F_p$ to a group of upper
\emph{unitriangular} matrices.  Thus every product of $d+1$
elements of $N $ vanishes modulo $p$, and
\[
    N ^{(d+1)\tau}=0.
\]
Consequently $G\leq I+N $.  The algebra $N $ and
its powers can be computed by $R$-module operations.  Applying
Lemma~\ref{lem:upper_tri_generating_set} to this algebra gives a
polynomial-time routine that replaces any generating set of any
subgroup of $G$ by one having at most $\tau(d+1)^2$ generators.
We use this routine after each stabilizer computation below.

By Lemma~\ref{lem:upper_triangularize}, we can compute a chain of
$G$-invariant additive subgroups
\[
  0=V_0\subset V_1\subset\cdots\subset V_L=V,
   |V_i/V_{i-1}|=p,
   L=\tau d,
\]
with $(Q_g-I)V_i\subseteq V_{i-1}$ for every $g\in G$.
Notice that the $V_i$ need not be free $R$-submodules.

We maintain a group $H\leq G$ by generators and an element $a\in G$
with the following invariant at stage $i$:
\[
  \{g\in G:g(v_1)-v_2\in V_i\}=Ha.
\]
In particular, $a(v_1)-v_2\in V_i$ and
$h(v_2)-v_2\in V_i$ for every $h\in H$.
Initially $i=L$, $H=G$, and $a=1$, so the invariant holds.
Assume it holds for some $i\geq1$.  For $h\in H$ define
\[
 f_i(h)=h(v_2)-v_2+V_{i-1}\ \in V_i/V_{i-1}.
\]
This is a group homomorphism to the additive group of order $p$.
Indeed, for $h,k\in H$,
\[
 h(k(v_2))-v_2=
 \bigl(h(v_2)-v_2\bigr)
       +Q_h\bigl(k(v_2)-v_2\bigr),
\]
and $Q_h$ acts trivially on $V_i/V_{i-1}$.
Put $\delta=a(v_1)-v_2+V_{i-1}$.
The same calculation shows
\[
 h(a(v_1))-v_2+V_{i-1}=\delta+f_i(h).
\]
Hence a correction exists at this stage precisely when
$-\delta\in f_i(H)$.  We can decide this by evaluating $f_i$
on the generators of $H$, since $V_i/V_{i-1}\cong\mathbb F_p$.
If no correction exists, the transporter is empty.
Otherwise choose $h_0\in H$ with $f_i(h_0)=-\delta$ and replace
$a$ by $h_0a$.  Such $h_0$ can be constructed as a power of a
generator having nonzero $f_i$-value, or as the identity when
$\delta=0$.

Next replace $H$ by $\ker f_i$.  When $f_i(H)=0$ nothing changes.
Otherwise choose a generator $s\in H$ with $f_i(s)\ne0$,
and set $t_j=s^j$ for $0\leq j<p$.  If $\eta(g)\in\mathbb F_p$
is defined by $f_i(g)=\eta(g)f_i(s)$, Schreier's lemma gives
the following generators of $\ker f_i$:
\[
  t_j g\,t_{\,j+\eta(g)\bmod p}^{-1},
   g\in\Gamma_H,\quad j=0,\ldots,p-1,
\]
where $\Gamma_H$ is the current generating set of $H$.
Apply Lemma~\ref{lem:upper_tri_generating_set} to compress the
resulting list to at most $\tau(d+1)^2$ generators.
The invariant holds at level $i-1$, because the elements
$g$ satisfying $g(v_1)-v_2\in V_{i-1}$ are exactly
\[
 \{ha:h\in H,\ f_i(h)=-\delta\}
   =(\ker f_i)h_0a.
\]

After $L=\tau d$ stages, the invariant gives
$G_{v_1\to v_2}=Ha$.  In particular, $a(v_1)=v_2$ and
$H=G_{v_2\to v_2}$, proving correctness and completeness.  The number of stages and the generator
set sizes are polynomial, and every Schreier step considers at
most $p$ cosets.  The claimed running time follows.
\end{proof}

\begin{lemma}\label{lem:index_bound}
Under the hypotheses of Theorem~\ref{thm:large_block},
$|{\cal A} / \Rad({\cal A})| \leq p^{n \cdot D} $.  
\end{lemma}
\begin{proof}
Let $\mathbf A=\mathbf A_2$.  Regard $\mathbf A$ as the
representation $(R^n,R^n;A_1,\ldots,A_m)$ of the $m$-arrow
quiver, and write
\[
\mathcal E=\{(P,T)\in\M(n,R)^2:
  PA_j=A_jT\text{ for every }j\}
\]
for its endomorphism ring (componentwise multiplication).
Projection onto the first coordinate gives a surjective ring
homomorphism $\mathcal E\to\mathcal A^T=\{S^T:S\in\mathcal A\}$.
Its kernel is
\[
 \mathcal K=\{(0,T):A_jT=0\text{ for every }j\}.
\]
Every column of such a $T$ belongs to the common right kernel of
the tuple.  By non-degeneracy, that kernel contains no primitive
vector, and hence is contained in $pR^n$.  It follows that
$T\in p\M(n,R)$ and $\mathcal K^\tau=0$.
Because $\mathcal K$ is a nilpotent two-sided ideal,
\[
 \mathcal E/\Rad(\mathcal E)
 \cong\mathcal A^T/\Rad(\mathcal A^T)
 \cong(\mathcal A/\Rad(\mathcal A))^{\mathrm{op}}.
\]
(The last identification uses matrix transposition.)

By Krull--Schmidt, write the representation as
$\bigoplus_{j=1}^t M_j^{\oplus a_j}$, with the $M_j$
pairwise nonisomorphic and indecomposable.  The ring
$\operatorname{End}(M_j)$ is finite and local, so its residue
division ring is a finite field, say $\mathbb F_{p^{e_j}}$.
The standard description of the semisimple quotient of the
endomorphism ring of a Krull--Schmidt decomposition gives
\[
 \mathcal E/\Rad(\mathcal E)
 \cong\bigoplus_{j=1}^t
    \M(a_j,\mathbb F_{p^{e_j}}).
\]
Here $a_j\leq D$ by hypothesis.  Matrix algebras over finite fields
are isomorphic to their opposite rings, so
\[
 \mathcal A/Q\cong\bigoplus_{i=1}^s
           \M(d_i,\mathbb F_{p^{m_i}}),
  Q=\Rad(\mathcal A),\quad d_i\leq D.
\]

It remains to show $\sum_i d_i m_i\leq n$.  Reduce the matrices of
$\mathcal A$ entrywise modulo $p$, writing
$\rho:\mathcal A\to\M(n,\mathbb F_p)$ and
$\mathcal B=\rho(\mathcal A)$.  Its kernel is
$\mathcal A\cap p\M(n,R)$, a nilpotent ideal because its
$\tau$-th power is zero.  Hence
\[
 \mathcal B/\Rad(\mathcal B)
 \cong\mathcal A/Q
 \cong\bigoplus_i\M(d_i,\mathbb F_{p^{m_i}}).
\]
Lift the mutually orthogonal central idempotents of this
semisimple quotient to a complete set of orthogonal idempotents
$E_1,\ldots,E_s$ in $\mathcal B$ with $\sum_iE_i=I$.
Put $V=\mathbb F_p^n$ and $V_i=E_iV$.
Since $\mathcal B\subseteq\M(n,\mathbb F_p)$ acts faithfully on
$V$, each $V_i$ is nonzero, and
$V=\bigoplus_i V_i$.
The corner algebra $\mathcal B_i=E_i\mathcal B E_i$ has
semisimple quotient
$\M(d_i,\mathbb F_{p^{m_i}})$ and radical
$E_i\Rad(\mathcal B)E_i$.
Because the radical is nilpotent and $V_i\ne0$,
$V_i/\Rad(\mathcal B_i)V_i$ is a nonzero module over this
matrix algebra.  Its unique simple module has
$\mathbb F_p$-dimension $d_i m_i$.  Therefore
$\dim_{\mathbb F_p}V_i\geq d_i m_i$, and summing gives
$\sum_i d_i m_i\leq n$.
Finally,
\[
 \log_p|\mathcal A/Q|=\sum_i d_i^2m_i
 \leq D\sum_i d_i m_i\leq nD.\qedhere
\]
\end{proof}

\begin{framed}\vspace{-.3cm}\begin{flushleft}

\textbf{Infrequent Block Alternating Matrix Tuple Isometry Algorithm}

\textbf{Input:} Two alternating matrix tuples $\mathbf{A}_1$, $\mathbf{A}_2$.
\noindent 

\textbf{Output:}
A set of matrices $\Gamma$. 
Yes or no. If yes, also output a matrix. 

\begin{enumerate}

\item Decide whether there is a pair $(S_0,S_0')$ of invertible matrices
such that $\mathbf{A}_2=S_0^T\mathbf{A}_1S_0'$; if so, compute it.
Otherwise, record that the tuples are not congruent and skip
Step~5(a), but continue computing the automorphism group. 
\item Compute a generating set of ${\cal A}$, a generating set of $\Rad(\cal{A})$ and a system of generators for the group 
$J=I+\Rad(\cal{A})$
\item Compute a complete set $\mathcal C\subseteq\mathcal A$
of representatives of the additive quotient
$\mathcal A/\Rad(\mathcal A)$.

\item Run the vector transporter algorithm for $J$ with both initial and
 target vectors $\mathbf{A}_2$, retaining subgroup generators as matrices
 in $J$.  Let $\Gamma$ be the returned generators of
 $K_J=\{Y\in J:Y^T\mathbf{A}_2Y=\mathbf{A}_2\}$.
 Compress these generators using Lemma~\ref{lem:upper_tri_generating_set}
 with $N=\Rad(\mathcal A)$.
\item For each invertible matrix $Z$ in ${\cal C}$, 
\begin{enumerate}
\item If $S_0$ exists, run the vector transporter algorithm for $J$ with
initial vector $\mathbf A_2$ and target vector
$Z^T S_0^T\mathbf A_1 S_0Z$.
If it returns $X\in J$, then
$
\mathbf A_2=(S_0ZX)^T\mathbf A_1(S_0ZX)$.
\item Run the vector transporter algorithm for $J$ with initial vector
$\mathbf{A}_2$ and target $Z^T\mathbf{A}_2Z$.  If it returns
$X_Z\in J$, add $ZX_Z$ to $\Gamma$.

\end{enumerate}
\item Output $\Gamma$. Output $S_0 ZX$ if there is $\mathbf{A}_2= (S_0 Z X)^T  \mathbf{A_1} S_0ZX$ for some $Z \in \mathcal{C}$ and $X \in J$. Otherwise, output no. 
\end{enumerate}
\end{flushleft}
\vspace{-.3cm}
\end{framed}

\begin{theorem}[cf.~\cite{GQSS25}]\label{lem:module_iso}
Given two tuples $\mathbf A,\mathbf A'\in\M(n,R)^m$, there is
an algorithm running in $\poly(n,m,p,\tau)$ time that decides
whether there are matrices $S_0,S_0'\in\GL(n,R)$ satisfying
$\mathbf A'=S_0^T\mathbf A S_0'$.  If such matrices exist,
the algorithm outputs the pair $(S_0,S_0')$.
\end{theorem}

\begin{proof}[Proof of Theorem~\ref{thm:large_block}]
Lemma~\ref{lem:in-adjoint} shows that if a congruence
$S^T\mathbf A_1S=\mathbf A_2$ exists, then
$S=S_0g$ for some $g\in G=\mathcal A^\times$.
By Lemma~\ref{lem:index_bound},
\[
  [G:J]=|(\mathcal A/Q)^\times|
       \leq |\mathcal A/Q|\leq p^{nD}.
\]
The chosen $Z$ represent all left cosets $ZJ$ of $J$ in $G$.
For each $g=ZX$, put
$\mathbf B_Z=(S_0Z)^T\mathbf A_1(S_0Z)$.
The desired congruence is precisely
$X^T\mathbf B_ZX=\mathbf A_2$, equivalently
$f(X)(\mathbf A_2)=\mathbf B_Z$.
This is the transporter test in Step~5(a), so the congruence
algorithm is correct.

For the automorphism group put
$K=\{S:S^T\mathbf A_2S=\mathbf A_2\}$ and $K_J=K\cap J$.
Step 4 computes generators of $K_J$. If $K\cap ZJ$ is nonempty,
Step 5(b) finds $X_Z\in J$ with $k_Z=ZX_Z\in K$.
Then $K\cap ZJ=K_Jk_Z$, because $J$ is normal in $G$.
Thus $\Gamma$ generates $K$. By
Lemma~\ref{lem:upper_tri_generating_set}, the generators of
$K_J$ can be compressed to at most $\tau n^2$ elements;
at most $p^{nD}$ additional $k_Z$ are inserted.

All operations outside coset enumeration are polynomial in
$n,m,p,\tau$, by the cited lemmas and standard finite-ring
module and finite-field algebra algorithms.  Since there
are at most $p^{nD}$ cosets and a constant number of
transporter calls per coset, the total running time is
$p^{nD}\operatorname{poly}(n,m,p,\tau)$.
\end{proof}

\subsection{Alternating matrix tuple isometry algorithm}


\begin{lemma}\label{lem:matrix_kernal}
Let $\mathbf{A}$ be a matrix tuple and
$U=(U_1\ \cdots\ U_k)$ and $W=(W_1\ \cdots\ W_k)$
be invertible matrices such that
\[
 U^T\mathbf{A}W
 =\diag(U_1^T\mathbf{A}W_1,\ldots,U_k^T\mathbf{A}W_k)
\]
is a minimal left-right block diagonalization of $\mathbf{A}$.
Let $Q$ satisfy $Q-I=(Q_1\ \cdots\ Q_k)$, where each
$Q_i$ has the same dimensions as $U_i$, and assume that
\[
 UQ_i\in K_{U_i^T\mathbf{A}W_i}
 \text{for all }i\in[k].
\]
Then $Q$ is invertible, and there exists an invertible matrix $P$
such that
\[
 Q^T U^T\mathbf{A}WP=U^T\mathbf{A}W.
\]
\end{lemma}
\begin{proof}
Put $\mathbf{B}=U^T\mathbf{A}W=\diag(\mathbf{B}_1,\ldots,\mathbf{B}_k)$,
where $\mathbf{B}_i=U_i^T\mathbf{A}W_i$. As in the proof of
Theorem~\ref{thm:left_right_decomposition}, regard the transposed
tuple $\mathbf{B}^T$ as a representation
\[
 \mathcal N=\mathcal M_1\oplus\cdots\oplus\mathcal M_k,
\]
where $\mathcal M_i$ corresponds to $\mathbf{B}_i^T$.
Each $\mathcal M_i$ is indecomposable and has a local
endomorphism ring.

By the characterization of $K_{\mathbf{B}_i}$ in the proof of
Theorem~\ref{thm:left_right_decomposition}, the condition
$UQ_i\in K_{\mathbf{B}_i}$ means that there exists a morphism
$f_i:\mathcal M_i\to\mathcal N$ whose first component, expressed
in the $U$-coordinates, is $Q_i$, and such that $f_i$ is not a
split monomorphism. Since $\operatorname{End}(\mathcal M_i)$ is
local, this is equivalent to
\[
 f_i\in\operatorname{Rad}(\mathcal M_i,\mathcal N).
\]
Indeed, for every $g:\mathcal N\to\mathcal M_i$, the element
$gf_i$ is noninvertible, and hence belongs to the Jacobson
radical of $\operatorname{End}(\mathcal M_i)$.

Assemble the $f_i$ into an endomorphism
$f=(f_1,\ldots,f_k)$ of $\mathcal N$.
For each $i,j$, the component of $f_i$ from $\mathcal M_i$
to $\mathcal M_j$ is a non-isomorphism: otherwise $f_i$
would be a split monomorphism. Morphisms between
non-isomorphic indecomposable summands lie in the categorical
radical, and for isomorphic summands the non-isomorphisms
lie in the radical since their endomorphism rings are local.
By the Krull--Schmidt description of the radical of the
endomorphism ring of a direct sum, we obtain
\[
 f\in\operatorname{Rad}(\operatorname{End}(\mathcal N)).
\]
Thus $I+f$ is an automorphism of $\mathcal N$.

Write $f=(Q-I,T)$ for its two components. The endomorphism
condition for the transposed tuple is
\[
 B_j^T(Q-I)=T B_j^T\text{for every }j\in[m].
\]
Taking transposes gives
\[
 (Q-I)^T B_j=B_jT^T.
\]
As $I+f$ is invertible, both $Q$ and $I+T$ are invertible.
Consequently,
\[
 Q^T\mathbf{B}=\mathbf{B}(I+T^T).
\]
Taking $P=(I+T^T)^{-1}$ yields
\[
 Q^T U^T\mathbf{A}WP=\mathbf{B}=U^T\mathbf{A}W.
\]
This proves the lemma.
\end{proof}

\begin{lemma}\label{lem:congruence_canceling}
Let $\mathbf{A} = \diag(\mathbf{A}_1, \mathbf{A}_2)$ be an alternating matrix tuple such that every minimal left-right block of $\mathbf{A}_1$ is not equivalent to any minimal left-right block or transpose of any minimal left-right block of $\mathbf{A}_2$. 
Let $n_1$ and $n_2$ denote the dimensions of $\mathbf{A}_1$ and $\mathbf{A}_2$ respectively. 
Let $F$ and $F'$ be arbitrary $n_1 \times n_2$ and $n_2 \times n_1$ matrices respectively such that $\mathbf{A}_1 F + F'^T \mathbf{A}_2 = 0$.
Then the matrix 
\[\begin{pmatrix}I & F \\ F' & I\end{pmatrix}\] is invertible and satisfies
\[\begin{pmatrix}I & F \\ F' & I\end{pmatrix}^T \mathbf{A} \begin{pmatrix}I & F \\ F' & I\end{pmatrix} = \diag(\mathbf{A}_1', \mathbf{A}_2')\]
for some $n_1 \times n_1$ matrix $\mathbf{A}_1' = \mathbf{A}_1 W_1$ and $n_2\times n_2$ matrix $\mathbf{A}_2' = \mathbf{A}_2 W_2$ with some invertible matrices $W_1$ and $W_2$. 
\end{lemma}
\begin{proof}
We first consider the case where $\mathbf{A}_1$ and $\mathbf{A_2}$ have invertible $S_1$ and $S_2$ such that $\mathbf{A}_1 S_1$ and $\mathbf{A}_2 S_2$ are the minimal left-right block diagonalizations of $\mathbf{A}_1$ and $\mathbf{A_2}$ respectively.  
By $\mathbf{A}_1 F + F'^T \mathbf{A}_2 = 0$, we have 
\[\begin{split}
& \begin{pmatrix}I & F \\ F' & I\end{pmatrix}^T \mathbf{A} \begin{pmatrix}I & F \\ F' & I\end{pmatrix} \\
= & \begin{pmatrix}I & F \\ F' & I\end{pmatrix}^T \begin{pmatrix} \mathbf{A}_1 & \\ & \mathbf{A}_2 \end{pmatrix} \begin{pmatrix}I & F \\ F' & I\end{pmatrix}\\
= & \begin{pmatrix} \mathbf{A_1} + F'^T \mathbf{A_2} F' & \mathbf{A}_1 F  + F'^T \mathbf{A_2} \\ F^T \mathbf{A_1} + \mathbf{A}_2 F'& F^T \mathbf{A_1} F + \mathbf{A_2}\end{pmatrix} \\
= & \diag( \mathbf{A}_1 + F'^T \mathbf{A}_2 F', F^T \mathbf{A_1} F + \mathbf{A_2}). 
\end{split}
\]
There are invertible matrices $Z_1$ and $Z_2$ such that 
$\mathbf{A}_1 Z_1 = \diag(\mathbf{B}_1, \dots, \mathbf{B}_k)$ and $\mathbf{A}_2 Z_2 = \diag(\mathbf{C}_1, \dots, \mathbf{C}_\ell)$, where 
$\mathbf{B}_1, \dots, \mathbf{B}_k$ and $\mathbf{C}_1, \dots, \mathbf{C}_\ell$ are minimal left-right blocks for $\mathbf{A}_1$ and $\mathbf{A}_2$ respectively.
By $\mathbf{A}_1 F + F'^T \mathbf{A}_2 = 0$ and Lemma~\ref{lem:matrix_kernal} with $U = I$, $W = \diag(Z_1, Z_2)$, 
and \[Q = I + \begin{pmatrix}& F \\ F' & \end{pmatrix},\]  
the matrix $Q$ is invertible. 
Since every column tuple of $F'^T\mathbf{A}_2$ is a linear combination of column tuples of $\mathbf{A}_1$, there exists an invertible matrix $W_1$ such that $\mathbf{A}_1 + F'^T \mathbf{A_2}F' = \mathbf{A}_1 W_1$. 
Similarly, there is an invertible matrix $W_2$ such that $F^T \mathbf{A}_1 F + \mathbf{A_2} = \mathbf{A}_2 W_2$. 

For the general case, by Corollary~\ref{cor:standard_block_diagonalization}, there are invertible matrices $U_1$ and $U_2$ such that $U_1^T \mathbf{A}_1 U_1$ and $U_2^T \mathbf{A}_2 U_2$ are standard congruent block diagonalizations for $\mathbf{A}_1$ and $\mathbf{A}_2$ respectively.
Since $\mathbf{A}_1$ and $\mathbf{A}_2$ are alternating matrix tuples, and $\mathbf{A}_1 F + F'^T \mathbf{A}_2 = 0$, 
we have 
\[\begin{split} U_1^T \mathbf{A}_1 U_1 U_1^{-1} F = & U_1^T (- F'^T \mathbf{A}_2 )  \\ = & U_1^T (\mathbf{A}_2 F')^T \\ = & U_1^T (U_2^{-T} U_2^T \mathbf{A}_2 U_2 U_2^{-1} F')^T \\ = & U_1^T F'^T U_2^{-T} (U_2^T \mathbf{A}_2 U_2
)^T U_2^{-1},\end{split}\] which implies 
$(U_1^T\mathbf{A}_1 U_1) (U_1^{-1}F U_2) = ((U_2^T \mathbf{A}_2 U_2) (U_2^{-1} F' U_1))^T$. By the first case, we have
\[\begin{pmatrix}I & U_1^{-1}F U_2 \\ U_2^{-1} F' U_1 & I\end{pmatrix}^T \begin{pmatrix}U_1 & \\ & U_2 \end{pmatrix}^T\mathbf{A}  \begin{pmatrix}U_1 & \\ & U_2 \end{pmatrix}\begin{pmatrix}I & U_1^{-1}F U_2 \\ U_2^{-1} F' U_1 & I\end{pmatrix} = \diag(\mathbf{A}_1'', \mathbf{A}_2'')\]
for some $\mathbf{A}_1''$ and $\mathbf{A}_2''$ such that there exist invertible matrices $W_1$ and $W_2$ satisfying $\mathbf{A}_1'' = U_1^T \mathbf{A}_1 U_1 W_1$ and $\mathbf{A}_2'' = U_2^T \mathbf{A}_2 U_2 W_2$.
Consequently, let 
\[S =  \begin{pmatrix}U_1 & \\ & U_2 \end{pmatrix}\begin{pmatrix}I & U_1^{-1}F U_2 \\ U_2^{-1} F' U_1 & I\end{pmatrix} \begin{pmatrix}U_1^{-1} \\ & U_2^{-1}\end{pmatrix}.\]
We have 
\[ S^T \mathbf{A}  S 
=  \begin{pmatrix} I & F \\ F' & I\end{pmatrix}^T \mathbf{A} \begin{pmatrix} I & F \\ F' & I\end{pmatrix} 
= \diag(U_1^{-T}\mathbf{A}_1''U_1^{-1}, U_2^{-T}\mathbf{A}_2''U_2^{-1}).\]
Then the lemma follows.
\end{proof}

\begin{lemma}\label{lem:block_diagonal_invertible}
Let $\mathbf{A}$ be an alternating matrix tuple, and let $\Lambda$ be a set of minimal left-right blocks of $\mathbf{A}$. 
Let  $\mathbf{A}'= \diag(\mathbf{A}_1', \mathbf{A}_2')$ and $\mathbf{A}''= \diag(\mathbf{A}_1'', \mathbf{A}_2'')$  be two alternating matrix tuples satisfying the following conditions:
\begin{enumerate}
    \item Both $\mathbf{A}'$ and $\mathbf{A}''$ are congruent to $\mathbf{A}$. 
    \item Any minimal left-right block of either $\mathbf{A}_1'$ or $\mathbf{A}_1''$ is left-right equivalent to a minimal left-right block or the transpose of a minimal left-right block in $\Lambda$. 
    \item None of the minimal left-right blocks of either $\mathbf{A}_2'$ or $\mathbf{A}_2''$ is left-right equivalent to a minimal left-right block or the transpose of a minimal left-right block in $\Lambda$. 
\end{enumerate}
Let $n_1$ and $n_2$ denote the number of rows/columns of $\mathbf{A}_1'$ and $\mathbf{A}_2'$ respectively. 
Then $\mathbf{A}_1''$ and $\mathbf{A}_2''$ have $n_1$ and $n_2$ rows/columns respectively. 
Moreover, for any invertible matrix \[U = \begin{pmatrix} U_{1, 1} & U_{1,2} \\ U_{2,1} & U_{2,2} \end{pmatrix}\]
with $U_{1, 1} \in \M(n_1 \times n_1, R)$, $U_{1, 2} \in \M(n_1 \times n_2, R)$, $U_{2, 1} \in \M(n_2 \times n_1, R)$, and $U_{2, 2} \in \M(n_2 \times n_2, R)$ such that $U^T \mathbf{A}' U = \mathbf{A}''$, 
$U_{1, 1}$ and $U_{2,2}$ are invertible. 
\end{lemma}
\begin{proof}
Without loss of generality, we can assume that there are invertible matrices $W_1$ and $W_2$ such that $ \mathbf{A}_1'' W_1 = \diag(\mathbf{C}_1, \dots, \mathbf{C}_k)$ is a minimal left-right block diagonalization of $\mathbf{A}_1''$ and  $ \mathbf{A}_2'' W_2 = \diag(\mathbf{D}_1, \dots, \mathbf{D}_\ell)$ is a minimal left-right block diagonalization of $\mathbf{A}_2''$.
Denote 
\[U_{1, 1} = \begin{pmatrix} U_{1, 1, 1} & \dots & U_{1, 1, k} \end{pmatrix}\text{ and } U_{2, 1} = \begin{pmatrix} U_{2, 1, 1} & \dots & U_{2, 1, k}\end{pmatrix}\] such that $U_{1, 1, i}$ and $U_{2, 1, i}$ correspond to $\mathbf{C}_i$ within $U_{1, 1}$ and $U_{2, 1}$ respectively. 
By Theorem~\ref{thm:left_right_decomposition}, \[\begin{pmatrix}  U_{1, 1, i} \\ 0 \end{pmatrix} \text{ and }\begin{pmatrix} 0 \\ U_{2, 1, i}\end{pmatrix}\] are elements in $V_{\mathbf{C}_i}$ with respect to $\mathbf{A}'$ for any $i \in [k]$.
Actually, we have 
\[\begin{pmatrix}  U_{1, 1, i} \\ 0 \end{pmatrix}  \in V_{\mathbf{C}_i} \setminus K_{\mathbf{C}_i}  \text{ and }\begin{pmatrix} 0 \\ U_{2, 1, i}\end{pmatrix} \in K_{\mathbf{C}_i}\]  because otherwise $\mathbf{C}_i$ is a left-right block of $\mathbf{A}_2'$, contradicting the condition that $\mathbf{A}_{2}'$ does not contain any minimal left-right block equivalent to a block or its transpose in $\Lambda$. 

By Theorem~\ref{thm:left_right_decomposition} and Lemma~\ref{lem:matrix_kernal},
there are invertible matrices $Q$ and $P$ such that 
\[Q^T U^T \mathbf{A}' =  \begin{pmatrix}U_{1, 1} & U_{1, 2}\\ & U_{2,2}\end{pmatrix}^T \mathbf{A}'\]
and \[Q^T U^T\mathbf{A}' U P = \mathbf{A}''.\] 
Therefore, $U_{1, 1}$ is an invertible matrix. 
The case for $U_{2, 2}$ is similar. 
\end{proof}

For a matrix tuple, we say that a minimal left-right block is frequent if in a minimal left-right block diagonalization, the number of blocks in the diagonalization left-right equivalent to the given minimal left-right block is more than $n^{1 - \Delta}$. 
Otherwise, we say the minimal left-right block is infrequent. 

Let $\mathbf{A} = \diag(\mathbf{A}_1, \dots, \mathbf{A}_k)$ 
be an alternating  matrix tuple 
such that $\mathbf{A}_1$ contains all the infrequent left-right blocks, and each of $\mathbf{A}_i$ for $i > 2$ corresponds to a standard congruent block constructed from a frequent left-right block.
Let $d_i$ denote the dimension of $\mathbf{A}_i$. 
We say that a matrix $S$ satisfies the block $i$ condition if there are matrices $F \in \M(d_1 \times d_i, R)$ and $F'  \in \M(d_i \times d_1, R)$ such that $\mathbf{A}_1 F =  (\mathbf{A}_i F')^T$,
\[S=I+\begin{pmatrix}
\dots & F & \dots \\
\vdots &  & \vdots \\
F' \\
\vdots &  & \vdots 
\end{pmatrix},
\]
in which the submatrix induced by the first $d_1$ rows and the columns from $((\sum_{j=1}^{i-1} d_j) +1)$ to $(\sum_{j=1}^{i} d_j)$ is $F$,
the submatrix induced by the rows from  $((\sum_{j=1}^{i-1} d_j) +1)$ to $(\sum_{j=1}^{i} d_j)$ and the first $d_1$ columns is $F'$, and all the other entries are zero.

\begin{framed}\vspace{-.3cm}\begin{flushleft}
\textbf{Matrix Tuple Isometry Algorithm}

\textbf{Input:} Two alternating matrix tuples $\mathbf{A}$ and $\mathbf{B}$, representative blocks $\mathbf{B}_1, \dots, \mathbf{B}_\ell$ for $\mathbf{D}_{\mathbf{A}}$, and a parameter $0 < \Delta < 1$.

\textbf{Output:}
Yes or no. If yes, also a matrix $S$ such that $S^T \mathbf{A} S = \mathbf{B}$. 
\begin{enumerate}

\item Compute an arbitrary congruent block diagonalization with $ \diag(\mathbf{A}', \mathbf{A}'')$ for $\mathbf{A}$ and $\diag(\mathbf{B}', \mathbf{B}'')$ for $\mathbf{B}$ such that all the minimal left-right blocks in $\mathbf{A}''$ and $\mathbf{B}''$ are frequent with respect to $\mathbf{A}$ and $\mathbf{B}$,
and all the minimal left-right blocks in $\mathbf{A}'$ and $\mathbf{B}'$ are infrequent.


\item Let $\mathbf{A}''$ and $\mathbf{B}''$ be their arbitrary standard block diagonalizations. 

\item Let $\Gamma_1 = \{\diag(\mathbf{A}', \mathbf{A}'')\}$. Write $\diag(\mathbf{A}', \mathbf{A}'')$ as $\diag(\mathbf{A}_{1}, \dots, \mathbf{A}_{k})$ with $\mathbf{
A}_{1}$ being $\mathbf{A}'$, and $\mathbf{A}_{2}, \dots, \mathbf{A}_{k}$ being the standard congruent blocks in $\mathbf{A}''$.
Let $d_1, d_2, \dots d_{k}$ be the number of rows/columns of $\mathbf{A}_{1}, \dots \mathbf{A}_{k}$. 
\item For $i = 2, \dots, k$, 
\begin{enumerate}
\item Let $\Gamma_{i} = \Gamma_{i-1}$.
\item For an arbitrary $\mathbf{A} \in \Gamma_{i-1}$
with $\mathbf{A}  = \diag(\mathbf{A}_1, \dots, \mathbf{A}_{k})$ 
and an arbitrary $S$ satisfying the block $i$ condition,  add $S^T \mathbf{A} S$ to $\Gamma_i$. 
\item  
Keep only one representative of each blockwise isometry class:
$\diag(\mathbf{D}_1, \dots, \mathbf{D}_k)$ and $\diag(\mathbf{E}_1, \dots, \mathbf{E}_k)$ are equivalent if 
$\mathbf{D}_j$ and $\mathbf{E}_j$ are isometric for all $j\leq i$, and 
$\mathbf{D}_j = \mathbf{E}_j = \mathbf{A}_j$ for all  $j>i$.

\end{enumerate}
\item If there is a matrix tuple $\mathbf{D} = (\mathbf{D}_1, \dots, \mathbf{D}_k) \in \Gamma_k$ such that $\mathbf{D}_1$ and $\mathbf{B}'$ are isometric, and 
$\diag(\mathbf{D}_2, \dots, \mathbf{D}_k)$ and $\mathbf{B}''$ are isometric via backtracking, then output yes and a matrix $S$ such that $S^T \mathbf{A} S = \mathbf{B}$, otherwise, output no. 
\end{enumerate}
\end{flushleft}
\vspace{-.3cm}\end{framed}

\begin{lemma}\label{lem:matrix_tuple_tuple_iso_algo_correctness}
The Matrix Tuple Isometry Algorithm is correct.
\end{lemma}
\begin{proof}
By the congruence-canceling lemma, each update (1) is invertible,
preserves block diagonality, and keeps the two affected blocks
left-right equivalent to their previous values.  In particular,
every state is congruent to $\mathbf A$, so acceptance implies
$\mathbf A\cong\mathbf B$.

First ignore the removal of duplicate states, and suppose
$\mathbf A\cong\mathbf B$.  Choose
\[
 Q=\begin{pmatrix}Q_{11}&Q_{12}\\Q_{21}&Q_{22}\end{pmatrix}
 \in\GL(n,R), Q^T\mathbf C Q=\mathbf E,
\]
partitioned according to the infrequent and frequent parts.
The block-diagonal invertibility lemma gives
$Q_{11}\in\GL(d_1,R)$ and $Q_{22}\in\GL(n-d_1,R)$.
After multiplying $Q$ on the right by
$\diag(Q_{11}^{-1},I)$, we obtain
\[
 \widetilde Q=\begin{pmatrix}I&U\\F&V\end{pmatrix},
  V=Q_{22}\text{ invertible}.
\]
The upper-right block of $\widetilde Q^T\mathbf C\widetilde Q$
vanishes.  Thus, writing
$F=(F_2^T,\ldots,F_k^T)^T$ and
$UV^{-1}=(G_2,\ldots,G_k)$, we have
\[
 \mathbf A_1G_i+F_i^T\mathbf A_i=0
 (2\leq i\leq k). \tag{3}\label{eq:target}
\]

We construct local updates $S_2,\ldots,S_k$ inductively.
After step $i-1$, let
$\mathbf D^{(i-1)}=Q_{i-1}^T\mathbf C Q_{i-1}$ have first block
$\mathbf D_1^{(i-1)}=\mathbf A_1W_{i-1}$ with $W_{i-1}$ invertible,
and suppose the first $d_1$ columns of $Q_{i-1}$ have block entries
$I,F_2,\ldots,F_{i-1},0,\ldots,0$.  All unprocessed block columns
remain standard.  This holds initially with $Q_1=I$.
Since the $i$-th block is still $\mathbf A_i$, setting
\[
 (S_i)_{1,i}=W_{i-1}^{-1}G_i, (S_i)_{i,1}=F_i
\]
satisfies (1) by (3).  The congruence-canceling lemma implies that
the update is invertible and the new first block remains right
equivalent to $\mathbf A_1$.  Moreover $Q_i=Q_{i-1}S_i$ has the
required first block column with the additional entry $F_i$.
This completes the induction.

At the last step, $Q_k$ and $\widetilde Q$ have the same first
$d_1$ columns.  Therefore
\[
 Q_k^{-1}\widetilde Q=
 \begin{pmatrix}I&H\\0&V'\end{pmatrix}, V'\text{ invertible}.
\]
Writing $\mathbf D^{(k)}=\diag(\mathbf D_1,\mathbf D'')$ and
comparing blocks in
\[
 \diag(Q_{11}^{-T}\mathbf B'Q_{11}^{-1},\mathbf B'')
 =\begin{pmatrix}I&0\\H^T&V'^T\end{pmatrix}
 \diag(\mathbf D_1,\mathbf D'')
 \begin{pmatrix}I&H\\0&V'\end{pmatrix},
\]
we obtain
\[
 \mathbf D_1=Q_{11}^{-T}\mathbf B'Q_{11}^{-1},
  \mathbf D_1H=0,
  \mathbf B''=V'^T\mathbf D''V'.
\]
Hence the unpruned search reaches an accepting state.

It remains to justify discarding states.  If two states after
iteration $i$ are blockwise isometric, write
$\widetilde{\mathbf D}=P^T\mathbf DP$ for
$P=\diag(P_1,\ldots,P_i,I,\ldots,I)$ with all $P_j$ invertible.
Every next update $S$ from $\mathbf D$ corresponds to
$P^{-1}SP$ from $\widetilde{\mathbf D}$; this is again an update
of the form (1), and
\[
 (P^{-1}SP)^T\widetilde{\mathbf D}(P^{-1}SP)
 =P^T(S^T\mathbf DS)P.
\]
The two successors are blockwise isometric.  Induction shows that
discarding such duplicates cannot remove all accepting paths.
This proves completeness and correctness.
\end{proof}

\begin{lemma}\label{lem:compatible_correction_classes}
Let $\mathbf A_1$ be an alternating tuple of dimension $d_1$,
let $\mathbf E_1,\ldots,\mathbf E_s$ be alternating tuples of
dimensions $h_1,\ldots,h_s$, and assume no indecomposable
left-right summand of $\diag(\mathbf E_1,\ldots,\mathbf E_s)$
occurs in $\mathbf A_1$.  Put
\[
 H=\sum_{j=1}^s h_j, B=\sum_{j=1}^s b_jh_j.
\]
For each $j,t$, take $X_{j,t},Z_{j,t}\in\GL(h_j,R)$ so that
$X_{j,t}\mathbf E_jZ_{j,t}$ is alternating, and consider the tuples
\[
 \mathbf A_1'=\mathbf A_1+
 \sum_{j=1}^s\sum_{t=1}^{b_j}
 F_{j,t}'{}^T(X_{j,t}\mathbf E_jZ_{j,t})F_{j,t}',
 \tag{4}\label{eq:corrections}
\]
where $F_{j,t}'\in\M(h_j\times d_1,R)$ satisfies
\[
 L_{j,t}\mathbf A_1
 =(X_{j,t}\mathbf E_jZ_{j,t})F_{j,t}'
 \quad\text{for some }L_{j,t}\in\M(h_j\times d_1,R).
 \tag{5}\label{eq:compat}
\]
Let $N_{\mathrm{iso}}$ be the number of isometry classes among
(4). Then for $p=2$,
\[
 N_{\mathrm{iso}}\le
 2^{\min\left\{\tau d_1B,\ 6d_1^2,\
 4\tau d_1H^3\bigl(1+\lceil\log_2(2\tau d_1)\rceil\bigr)
 \right\}}.
 \tag{6}\label{eq:niso}
\]
For odd $p$, $N_{\mathrm{iso}}=1$.
\end{lemma}
\begin{proof}
In the proof, a left-right morphism from $\mathbf A$ to
$\mathbf E$ means a pair $z=(L,F)$ satisfying
$L\mathbf A=\mathbf EF$, with transpose adjoint
$z^*=(F^T,L^T)$.  Products and adjoints are the usual
compositions of these pairs; they are componentwise compatible
with all matrices of the tuples.  This is the same left-right
representation formalism used earlier in the paper.

Write $\mathbf E=\diag(\mathbf E_1,\ldots,\mathbf E_s)$ and
$T=\Hom(\mathbf A_1,\mathbf E)$.  Because a morphism is given by
two matrices of size $H\times d_1$,
\[
 |T|\leq p^{2\tau d_1H}.
 \tag{7}\label{eq:Tbound}
\]
For one summand $\mathbf C=X\mathbf E_jZ$, let
$z=(X^{-1}L,ZF')\in\Hom(\mathbf A_1,\mathbf E_j)$ and
\[
 A=(Z^{-T}X,\ X^TZ^{-1})\in\End(\mathbf E_j).
\]
The fact that $X\mathbf E_jZ$ is alternating implies
$A^*=A$.  The defining equation for $z$ and the transpose of
(5) show that
\[
 \mathbf A_1(z^*Az)_R=F'^T\mathbf C F',
\]
where $(\cdot)_R$ denotes the right coordinate of an endomorphism.
Extending $z$ and $A$ by zero to $\mathbf E$ gives
\[
 \mathbf A_1'=\mathbf A_1U,
 U=1+\sum_{j,t}z_{j,t}^*A_{j,t}z_{j,t},
  A_{j,t}^*=A_{j,t}.
 \tag{8}\label{eq:U}
\]

Set
\[
 K=\operatorname{span}_R\{z^*w:z,w\in T\}
 \subseteq\End(\mathbf A_1).
\]
Every element of $K$ factors through $\mathbf E$.  Because
$\mathbf A_1$ and $\mathbf E$ have disjoint indecomposable
left-right summands, the Krull--Schmidt radical criterion gives
$K\subseteq\Rad(\End(\mathbf A_1))$.  Also
$(z^*w)^*=w^*z$ and
$(z^*w)(u^*v)=z^*(wu^*)v$, so $K$ is an
adjoint-stable, nilpotent subalgebra.  In particular every
$1+x$, $x\in K$, is invertible, and $U=U^*\in1+K$.

For odd $p$, $1+K$ is a finite $p$-group.  Every self-adjoint
$U\in1+K$ has odd order, so choosing an integer $e$ with
$2e\equiv1\pmod{\operatorname{ord}(U)}$ and putting $P=U^e$
yields $P^*=P$ and $P^*P=U$.  Its invertible right coordinate
provides an isometry between $\mathbf A_1$ and $\mathbf A_1U$.
Thus $N_{\mathrm{iso}}=1$.

Now suppose $p=2$.  A submodule of $R^q$ has at most $q$
$R$-generators.  Since $\End(\mathbf E)$ embeds in a pair of
$H\times H$ matrices, its self-adjoint submodule has at most
$2H^2$ generators.  Choose self-adjoint generators
$C_{0,1},\ldots,C_{0,t_0}$ containing all $A_{j,t}$ in their span,
with $t_0\le2H^2$.
For self-adjoint $C$, write $q_C(z)=z^*Cz$ and
$\operatorname{Tr}(x)=x+x^*$.
Then
\[
 q_C(z+w)=q_C(z)+q_C(w)+\operatorname{Tr}(z^*Cw).
 \tag{9}\label{eq:polar}
\]
Modulo $\operatorname{Tr}(K)$, this is additive, hence
$R$-linear because $R=\Z/2^\tau\Z$.  Combining all terms
in (8) with the same coefficient gives
\[
 U=1+\sum_{\ell=1}^{t_0}w_{0,\ell}^*C_{0,\ell}w_{0,\ell}
 +\operatorname{Tr}(y_1),
 w_{0,\ell}\in T,\ y_1\in K.
 \tag{10}\label{eq:initial}
\]

For $a\ge1$, every element of $K^a$ is a sum of products
$z_1^*w_1\cdots z_a^*w_a$.  Writing such a product as
\[
 z_1^*(w_1z_2^*)\cdots(w_{a-1}z_a^*)w_a
 =z_1^*Aw_a,
\]
we see that $A$ is a product of $a-1$ factors of the form
$wz^*\in\End(\mathbf E)$ (and $A=I$ for $a=1$).
Choose self-adjoint generators
$C_{a,1},\ldots,C_{a,t_a}$, $t_a\le2H^2$, of the
$R$-span of the matrices
\[
 A^*(zz^*)A
 \quad(z\in T,\ A\text{ a product of }a-1
 \text{ factors }wz^*).
 \tag{11}\label{eq:coeffa}
\]
These coefficients depend only on $\mathbf A_1$, $\mathbf E$
and $a$, not on the number of copies of the $\mathbf E_j$.
For $y\in K^a$, expand $y=\sum_j y_j$ into products of
$a$ factors from $K$.  The cross terms in $y^*y$ are
$\operatorname{Tr}(\sum_{i<j}y_i^*y_j)$ with
$y_i^*y_j\in K^{2a}$.  Each diagonal term is
\[
 y_j^*y_j=v_j^*A_j^*(z_jz_j^*)A_jv_j.
\]
Expand its middle coefficient in the generators from (11).
For each fixed $C_{a,\ell}$, the polarization identity (9)
combines all the associated terms into one term
$v_\ell^*C_{a,\ell}v_\ell$ modulo
$\operatorname{Tr}(K^{2a})$: the required cross term
$v^*C_{a,\ell}w$ belongs to $K^{2a}$ by construction.
Thus
\[
 y^*y=\sum_{\ell=1}^{t_a}v_\ell^*C_{a,\ell}v_\ell
 +\operatorname{Tr}(x),
 v_\ell\in T,\ x\in K^{2a}.
 \tag{12}\label{eq:square}
\]

Suppose a self-adjoint correction has the form
\[
 V=1+\sum_jw_j^*D_jw_j+\operatorname{Tr}(y),
  y\in K^a,
\]
where each $D_j=D_j^*\in\End(\mathbf E)$.  Congruence by
$P=1-y$ is valid, and the identity
\[
 (1-y)^*(1+y+y^*)(1-y)
 =1-y^*y-\operatorname{Tr}(y^2-y^*y^2)
 \tag{13}\label{eq:traceidentity}
\]
gives, after using (12),
\[
 P^*VP=1+\sum_j(w_jP)^*D_j(w_jP)
 +\sum_{\ell=1}^{t_a}v_\ell^*(-C_{a,\ell})v_\ell
 +\operatorname{Tr}(y'),
  y'=-x-y^2+y^*y^2\in K^{2a}.
 \tag{14}\label{eq:doubling}
\]
Thus each step replaces a residual trace in $K^a$ by one in
$K^{2a}$, adding at most $2H^2$ compatible vectors.

The endomorphism algebra of $\mathbf A_1$ acts faithfully on the
two vertex spaces, of total cardinality $2^{2\tau d_1}$.
The descending radical filtration of this module shows that
$\Rad(\End(\mathbf A_1))^{2\tau d_1}=0$.
Hence $K^{2\tau d_1}=0$.  Starting with (10) and applying (14)
for $a=1,2,4,\ldots$ eliminates the residual trace after at most
$\lceil\log_2(2\tau d_1)\rceil$ steps.  The resulting normal
forms use at most
$2H^2(1+\lceil\log_2(2\tau d_1)\rceil)$ vectors from $T$,
with fixed coefficients.  By (7), they number at most
\[
 2^{4\tau d_1H^3(1+\lceil\log_2(2\tau d_1)\rceil)}.
 \tag{15}\label{eq:logcount}
\]

The original $F_{j,t}'$ give at most $2^{\tau d_1B}$
corrections.  For the additional bound independent of $\tau$,
observe that any two self-adjoint units $U,V\in1+K$ satisfying
$U\equiv V\pmod{8\End(\mathbf A_1)}$ are congruent.  Indeed,
$W=U^{-1}V=1+8x$ is self-adjoint for the adjoint
$Y\mapsto U^{-1}Y^*U$; the $2$-adic binomial expansion gives
a polynomial square root $P$ of $W$ modulo $2^\tau$.
Then $P^*UP=UP^2=V$.
Since $\End(\mathbf A_1)$ is represented by two
$d_1\times d_1$ matrices, reduction modulo $8$ has at most
$2^{6d_1^2}$ values.  Combining these three bounds proves (6).
All normalizing matrices act through invertible endomorphisms of
$\mathbf A_1$ and preserve the other diagonal blocks.
\end{proof}

\begin{lemma}\label{lem:matrix_tuple_tuple_iso_algo_time}
The Matrix Tuple Isometry Algorithm has running time
\[
 p^{O\left(n^{2-\Delta}+\tau n^{1+\Delta}
 +\min\{n^2,\,\tau n^{1+3\Delta}(1+\log(2\tau n))\}\right)}
 \poly(p,\tau,n,m).
\]
\end{lemma}
\begin{proof}
For the first step of the algorithm, 
to compute $\mathbf{A}'$ and $\mathbf{A''}$, 
one can enumerate an arbitrary standard congruent block constructed from a left-right block of at most $n^{\Delta}$ rows and columns in time $p^{\tau \cdot n \cdot n^{\Delta}}$, and recurse on the matrix tuple excluding this standard congruent block. 
At the end, the matrix tuple is decomposed into a congruent block diagonalization with standard congruent blocks constructed from left-right blocks of at most $n^{\Delta}$ rows and columns, and a single congruent block containing all the left-right blocks with more than $n^{\Delta}$ rows or columns. 
We further identify the standard congruent blocks corresponding to frequent left-right blocks, and assign these blocks to $\mathbf{A}''$ and the rest to $\mathbf{A}'$.
Construct $\mathbf{B}'$ and $\mathbf{B}''$ similarly. 
The running time of this step is then $p^{\tau \cdot n \cdot n^{\Delta}} \cdot \poly(n, m, p, \tau)$. 

By the definition of the frequency threshold and the standard
congruence-block classification in the paper, every frequent block
has $d_i=O(n^\Delta)$, and
\[
 \sum_{i=2}^k d_i^2\leq
 \left(\max_{i\geq2}d_i\right)\sum_{i=2}^k d_i
 =O(n^{1+\Delta}).
\]
Each first block in $\Gamma_i$ is a compatible correction of
$\mathbf A_1$ using the original small blocks
$\mathbf A_2,\ldots,\mathbf A_i$.  Indeed, the $j$-th small
block is unchanged until step $j$, and the updated first block
is $\mathbf A_1+\sum_{j=2}^iF_j'^T\mathbf A_jF_j'$.
The compatibility equation can be transported back to the
original $\mathbf A_1$ because all intermediate first blocks
are right equivalent to $\mathbf A_1$.

Each frequent minimal left-right type occurs more than
$n^{1-\Delta}$ times.  Summing the dimensions of one
representative per frequent left-right type and its possible
single/paired congruence shapes gives $H=O(n^\Delta)$.
Lemma~\ref{lem:compatible_correction_classes} therefore bounds
the number of isometry classes of the first block by
\[
 2^{O\left(\min\{n^2,\tau n^{1+3\Delta}
                      (1+\log(2\tau n))\}\right)}.
\]
By the congruence-canceling lemma, each processed small block
remains right equivalent to its original block, so it has at most
$p^{\tau d_j^2}$ possible values.  After deduplication,
\[
 |\Gamma_i|\leq 
 p^{O\left(\tau n^{1+\Delta}
 +\min\{n^2,\tau n^{1+3\Delta}
                      (1+\log(2\tau n))\}\right)}.
\]
Each step enumerates at most $p^{2\tau d_1d_i}$ compatible
pairs, and testing blockwise isometry costs
\[p^{O(n^{2-\Delta}+\tau n^{1+\Delta}+n\log n)}
\poly(p,\tau,n,m)\] by the cited subroutines.
Even comparing candidates against all retained states only
multiplies the state-count exponent by a constant.  There are at
most $n$ stages.  This proves the displayed running-time bound.
\end{proof}

\begin{proof}[Proof of Theorem~\ref{thm:matrix_tuple_isometry}]
Correctness follows from
Lemma~\ref{lem:matrix_tuple_tuple_iso_algo_correctness}, and
running time from Lemma~\ref{lem:matrix_tuple_tuple_iso_algo_time}.
For $\Delta=1/4$ the exponent is
\[O\left(n^{7/4}+\tau n^{7 / 4}
 +\min\{n^2,\tau n^{7 / 4}(1+\log(2\tau n))\}\right)
 =O(\tau n^{7/4} \log (\tau n)).
\]
\end{proof}
\subsection{Automorphism group computation}

In this section, we compute a generating set for the automorphism group of $\mathbf{A}$ and prove Theorem~\ref{thm:matrix_tuple_automorphism}. 

\begin{lemma}\label{lem:auto_group_basic}
Let $\mathbf{A} = \diag(\mathbf{A}_1, \mathbf{A}_2)$ be a non-degenerate alternating matrix tuple in $\M(n, R)^{m}$. 
Let $n_1$ be the dimension of $\mathbf{A}_1$. 
Let $G$ be the group generated by 
\begin{equation}\label{equ:special_matrix_1}S = \begin{pmatrix}I \\ & Z\end{pmatrix}\end{equation}
for some matrix $Z$ s.t. $S^T \mathbf{A} S = \mathbf{A}$, and 
$H$ be the group generated by  
\begin{equation}\label{equ:special_matrix_2}S = \begin{pmatrix}I & X \\ & Z\end{pmatrix}\end{equation}
for some matrices $X$ and $Z$ s.t. $S^T \mathbf{A} S = \mathbf{A}$. 
There is an algorithm to compute a set $\Gamma$ of size $O(\tau \cdot n_1 \cdot (n - n_1) \cdot \log p)$ in time $\poly(p^{\tau \cdot n_1 \cdot (n - n_1)}, m)$ such that $\Gamma \cup G$ generates $H$. \end{lemma}
\begin{proof}
Every $S$ in $G$ is of the form in Equation~(\ref{equ:special_matrix_1}) and satisfies $S^T \mathbf{A} S = \mathbf{A}$, 
and every $S$ in $H$ is of the form Equation~(\ref{equ:special_matrix_2}) and satisfies $S^T \mathbf{A} S = \mathbf{A}$.

For an arbitrary matrix $S$ of the form in Equation~(\ref{equ:special_matrix_2}) s.t. $S^T \mathbf{A} S = \mathbf{A}$, 
since 
\[\begin{pmatrix} I^T \\ X^T & Z^T \end{pmatrix} \begin{pmatrix}\mathbf{A}_1 \\ & \mathbf{A}_2 \end{pmatrix}\begin{pmatrix} I & X \\ & Z \end{pmatrix} = 
\begin{pmatrix} \mathbf{A}_1 & \mathbf{A}_1 X \\X^{T} \mathbf{A}_1 & X^T \mathbf{A}_1 X +  Z^T \mathbf{A}_2 Z \end{pmatrix} 
= \begin{pmatrix}\mathbf{A}_1 \\ & \mathbf{A}_2 
\end{pmatrix},
\]
we have $\mathbf{A}_1 X = 0$ and $\diag(I, Z)$ belongs to $G$. 
Consequently, $\diag(I, Z^{-1})$ is also in $G$. 
Since $G$ is a subgroup of $H$, 
\[\begin{pmatrix} I & X \\ & I \end{pmatrix}\]
is also in $H$. 
For any 
\[\begin{pmatrix} I & X_1 \\ & I \end{pmatrix}, \begin{pmatrix} I & X_2 \\ & I \end{pmatrix},\] 
we have \[\begin{pmatrix} I & X_1 \\ & I \end{pmatrix}\cdot \begin{pmatrix} I & X_2 \\ & I \end{pmatrix}  =  \begin{pmatrix} I & X_1 +  X_2 \\ & I \end{pmatrix}.\]

Let $F$ be the group generated by 
\[S = \begin{pmatrix}I & X \\ & I\end{pmatrix}\]
satisfying $S^T \mathbf{A} S = \mathbf{A}$. 
Let $\Gamma$ be a generating set of $F$. Then $\Gamma\cup G$ generates $H$. 

To compute a generating set of $F$, we enumerate all the elements of $F$, and then compute a generating set of size $O(\log |F|)$. 
Since $|F|$ is at most $(p^\tau)^ {n_1 \cdot n_2}$, 
$\Gamma$ is of size $O(\tau \cdot n_1 \cdot n_2 \cdot \log p)$. 
The running time is upper bounded by $\poly(p^{\tau \cdot n_1 \cdot (n - n_1)}, m)$
\end{proof}

\begin{lemma}\label{lem:automorphism}
Let $\mathbf{A} = \diag(\mathbf{A}_1, \mathbf{A}_2)$ be a non-degenerate alternating matrix tuple in $\M(n, R)^{m}$. 
Let $n_1$ and $n_2$ be the dimensions of $\mathbf{A}_1$ and $\mathbf{A}_2$ respectively. 
Let $G$ be the group generated by 
\begin{equation}S = \begin{pmatrix}I \\ & Z\end{pmatrix}\end{equation}
for some matrix $S$ s.t. $Z^T \mathbf{A}_2 Z = \mathbf{A}_2$, and 
$J$ be the group generated by all the $S$ s.t. $S^T \mathbf{A} S = \mathbf{A}$. 
Given a generating set $\Gamma_0$ of $G$, 
there is an algorithm to compute a set $\Gamma$ of size $O(\tau \cdot n_1 \cdot n \cdot \log p)$ in time $p^{O(\tau \cdot n_1 \cdot n + \tau n^{7/4} \log (\tau n))} 
\cdot \poly(|\Gamma_0|, m)$ such that $\Gamma \cup G$ generates $J$. 
\end{lemma}
\begin{proof}
Let $H$ be the group generated by  
\begin{equation}S = \begin{pmatrix}I & X \\ & Z\end{pmatrix}\end{equation}
for some matrices $X$ and $Z$ s.t. $S^T \mathbf{A} S = \mathbf{A}$. 
We consider the following algorithm.
\begin{enumerate}
\item For each $W \in \M(n_1, R)$ and $Y \in \M((n - n_1) \times n_1, R)$, compute an arbitrary matrix 
\[ S = \begin{pmatrix} W & X \\ Y & Z \end{pmatrix}\]
for some $Y \in \M((n-n_1) \times n_1, R)$ and $Z \in \M(n - n_1, R)$ such that $S \in J$, if one exists. 

\item Let $\Lambda$ be the set of all the matrices computed in the first step,  $\Xi$ be an empty set and $\Gamma$  be a set of matrices such that $\Gamma_0 \cup \Gamma$ generates $H$, where $H$ is defined as in Lemma~\ref{lem:auto_group_basic}. 

\item Repeat the following process until $\Lambda = \Xi$. 
\begin{enumerate}
\item Choose an arbitrary matrix in $\Lambda \setminus \Xi$ and add it to $\Gamma$. 
\item 
For every matrix $S\in \Lambda$ not in $\Xi$ s.t. $S$ is generated by $H$ and $\Gamma$, 
add $S$ to $\Xi$. 
\end{enumerate}
\item Return $\Gamma$.
\end{enumerate}

We first show that $\Gamma$ and $G$ generate $J$. 
After step 2 of the algorithm, $\Gamma$ and $G$ generate $H$. 
By step 3 of the algorithm, a matrix is added to $\Xi$ only if it is generated by $\Gamma$ and $G$. 
So at the end of the algorithm, every matrix of $\Lambda$ is generated by $\Gamma$ and $G$. 
The algorithm always terminates because we can always add a matrix of $\Lambda$ to $\Gamma$. 
To show that $\Gamma$ and $G$ generate $J$, it is sufficient to show that $\Lambda$ and $H$ generate $J$, which follows from Fact~\ref{fact:product_gen_set}, which shows that every matrix in $J$ is a product of a matrix in $\Lambda$ and a matrix in $H$. 
\begin{fact}\label{fact:product_gen_set}
Let 
\[S = \begin{pmatrix} W & X \\ Y & Z \end{pmatrix} \text{ and } S' = \begin{pmatrix}  W & X' \\ Y & Z' \end{pmatrix} \]
be two matrices in $J$. 
There is a matrix $Q$ in $H$ such that $S' = S Q$. 
\end{fact}
\begin{proof}
Write $S^{-1}$ as \[\begin{pmatrix}W'' & X'' \\ Y'' & Z'' \end{pmatrix}.\]
We have $W'' W + X'' Y = I$, $W'' X + X'' Z = 0$, $Y'' W + Z'' Y = 0$, and $Y'' X + Z'' Z = I$. 
Hence, 
\[S^{-1} S' = \begin{pmatrix}W'' & X'' \\ Y'' & Z'' \end{pmatrix}\begin{pmatrix}  W & X' \\ Y & Z' \end{pmatrix}  = \begin{pmatrix}I & W'' X' + X'' Z' \\ 0 & Y'' X' + Z'' Z'\end{pmatrix}\]
Since $S^{-1}$ is also in $J$, $S^{-1} S'$ is in $J$ too, and thus in $H$. 
Taking $Q$ to be $S^{-1} S'$ proves the fact.
\end{proof}

Now we bound the size of $\Gamma$. 
Before step 3 of the algorithm, by Lemma~\ref{lem:auto_group_basic}, $\Gamma$ is of size at most  $O(\tau \cdot n_1 \cdot (n - n_1) \cdot \log p)$.
We claim that after each iteration of step 3 of the algorithm, the size of $\Xi$ is at least doubled. 
By Fact~\ref{fact:product_gen_set}, after each iteration of step 3, every element can be generated by $\Gamma$ and $G$ can be represented as a product of an element in $\Xi$ and an element of $H$. 
Hence, the elements of $\Xi$ represent all the cosets of $H$ for the group generated by $\Gamma$ and $G$. 
Since the size of the group generated by $\Gamma$ and $G$ at least doubles after each iteration of step 3, the size of $\Xi$ is at least doubled. 

Hence, the number of iterations of step 3 is at most $\log |\Lambda|$, which is $O(\tau \cdot n_1 \cdot n \cdot \log p)$. So the size of $\Gamma$ is upper bounded by $O(\tau \cdot n_1 \cdot n \cdot \log p)$ at the end. 

Finally, we bound the running time of the algorithm.
In the first step of the algorithm, for each $W$ and $Y$, let \[U = \begin{pmatrix} W \\ Y\end{pmatrix}.\] 
If $U^T \mathbf{A} U \neq \mathbf{A}_1$, no matrix in $J$ has first $n_1$ columns $U$. 
Otherwise, let $M = \{v \in R^n : U^T \mathbf{A} v = 0\}$, which is computed by solving a linear system over $R$ in time $\poly(n, m, \tau, \log p)$. 
Since a matrix over $R$ is invertible if and only if it is invertible modulo $p$, by Gaussian elimination over $\F_p$ on $U$ and the generators of $M$ modulo $p$, 
we either find $X', Z'$ with every column of \[\begin{pmatrix} X' \\ Z'\end{pmatrix}\] in $M$ such that 
\[S' = \begin{pmatrix}W & X' \\ Y & Z'\end{pmatrix}\]
is invertible, or conclude that no such $X', Z'$ exist. 
In the latter case, no matrix in $J$ has first $n_1$ columns $U$, because the last $n - n_1$ columns of such a matrix would give such $X', Z'$. 
In the former case, \[U^T \mathbf{A} \begin{pmatrix} X' \\ Z'\end{pmatrix} = 0,\] and since $\mathbf{A}$ is alternating, $S'^T \mathbf{A} S' = \diag(\mathbf{A}_1, \mathbf{A}_2')$ for some $\mathbf{A}_2'$. 
There exists an automorphism of $\mathbf{A}$ whose first $n_1$ columns equal $U$ if and only if $\mathbf{A}_2'$ and $\mathbf{A}_2$ are congruent. 
Indeed, the ``if'' direction is shown below. For the ``only if'' direction, let $\begin{pmatrix} U & C \end{pmatrix} \in J$. 
Then the columns of $C$ are in $M$, and $C^T \mathbf{A} C = \mathbf{A}_2$. 
Write \[C = U T + \begin{pmatrix} X' \\ Z'\end{pmatrix} G,\] where $G$ is invertible because $\begin{pmatrix} U & C \end{pmatrix}$ is invertible. 
Since the columns of both $C$ and \[\begin{pmatrix} X' \\ Z'\end{pmatrix}\] are in $M$, we have $\mathbf{A}_1 T = U^T \mathbf{A} U T = 0$. 
Expanding $C^T \mathbf{A} C$ and using \[U^T \mathbf{A} \begin{pmatrix} X' \\ Z'\end{pmatrix} = 0,\] its transpose, and $T^T \mathbf{A}_1 T = 0$, we obtain $\mathbf{A}_2 = C^T \mathbf{A} C = G^T \mathbf{A}_2' G$. 
If such a matrix $E$ exists, choose one satisfying $E^T \mathbf{A}_2' E = \mathbf{A}_2$. Then  
\[S = \begin{pmatrix} W & X' E\\ Y & Z'E \end{pmatrix}\] satisfies $S^T \mathbf{A} S = \mathbf{A}$, and hence belongs to $J$. 
Hence, the running time of the first step of the algorithm is $p^{\tau \cdot n_1 \cdot n } \cdot (p^{O(\tau n^{7/4} \log (\tau n))} \cdot \poly(n) + \poly(n, m, p^\tau))$ by Theorem~\ref{thm:matrix_tuple_isometry}.

We perform the following process to implement step 3(b) of the algorithm. 
Repeat the following process until $\Xi$ does not change: For any two elements $S_1$ and $S_2$ in $\Xi \cup \Gamma \cup \Gamma_0$,  
if no element of $\Xi$ has the same first $n_1$ columns as $S_1 S_2$, then add the element of $\Lambda$ whose first $n_1$ columns agree with those of $S_1 S_2$ (computed from $S_1 S_2$) to $\Xi$. 
This process implements Step 3(b) of the algorithm because the first $n_1$ columns of $S_1 S_2$ are obtained by multiplying $S_1$ by the first $n_1$ columns of $S_2$. 
And this process has a running time of $\poly((|\Gamma| + |\Gamma_0| + |\Lambda|), p^\tau, n)$. 
Then the lemma follows.
\end{proof}

\begin{framed}\vspace{-.3cm}\begin{flushleft}
\textbf{Matrix Tuple Automorphism Algorithm}

\textbf{Input:} An alternating matrix tuple $\mathbf{A}$.

\textbf{Output:} A generating set of the automorphisms of $\mathbf{A}$. 
\begin{enumerate}
\item Set $\Delta = 1/ 2$. 

\item Compute an arbitrary congruent block diagonalization of the form $ \diag(\mathbf{A}_1, \dots, \mathbf{A}_k)$
such that $\mathbf{A}_k$ contains all the infrequent left-right blocks, and each of $\mathbf{A}_i$ for $1 \leq i \leq k - 1$ is a standard congruent block constructed from a frequent left-right block.


\item Compute a generating set $\Gamma_k$ of the automorphism group of $\mathbf{A}_k$. 

\item For $i = k -1, \dots, 1$, compute a generating set $\Gamma_i$ for the automorphism group of $\diag(\mathbf{A}_{i}, \dots, \mathbf{A}_k)$. 

\item Return $\Gamma_1$. 
\end{enumerate}
\end{flushleft}
\vspace{-.3cm}\end{framed}

\begin{proof}[Proof of Theorem~\ref{thm:matrix_tuple_automorphism}]
The correctness of the algorithm and the bound on the size of the generating set follow from Theorem~\ref{thm:large_block} and Lemma~\ref{lem:automorphism}.

We bound the running time of the algorithm. Step 3 of the algorithm runs in $O(p^{n^{1.5}}\cdot \tau \cdot n^2)$ time by Theorem~\ref{thm:large_block}. 
Since every $\mathbf{A}_i$ for $1 \leq i \leq k - 1$ is constructed from a frequent left-right block, every $\mathbf{A}_i$ for $1 \leq i \leq k - 1$ is of dimension at most $n^{1/2}$. 
Hence, every iteration of Step 4 takes  
$p^{O(\tau \cdot  n^{1.5} + \tau n^{7/4} \log (\tau n) + n^{1.5})}  \cdot \poly(m)$ by Lemma~\ref{lem:automorphism}. 
Then we get the overall running time bound.
\end{proof}

\end{document}